\documentclass[twocolumn,groupedaddress]{revtex4}

\newcommand{\beq}{\begin{equation}}
\newcommand{\eeq}{\end{equation}}
\newcommand{\beqa}{\begin{eqnarray}}
\newcommand{\eeqa}{\end{eqnarray}}
\newcommand{\ket} [1] {\vert #1 \rangle}
\newcommand{\bra} [1] {\langle #1 \vert}

\usepackage{graphicx}  
\usepackage{dcolumn}   
\usepackage{bm}        
\usepackage{amssymb}   
\usepackage{amsmath}   
\usepackage[utf8]{inputenc}
\usepackage[english]{babel}
\usepackage{mathtools}
\usepackage{mathrsfs}   
\usepackage{tikz}
\usetikzlibrary{matrix,arrows.meta,decorations.markings,calc}
\usepackage{tcolorbox}
\usepackage{color}

\begin{document}

\title{Symmetry reduction induced by anyon condensation: a tensor network approach}
\author{Jos\'e Garre-Rubio, Sofyan Iblisdir, David P\'{e}rez-Garc\'{i}a}
\affiliation{Departamento de An\'alisis Matem\'atico, Universidad Complutense de Madrid, 28040 Madrid, Spain \\
Instituto de Ciencias Matem\'aticas, C/ Nicol\'as Cabrera, Campus de Cantoblanco, 28049 Madrid, Spain}
  
\begin{abstract}

 Topological ordered phases are related to changes in the properties of their quasi-particle excitations (anyons). We study these relations in the framework of projected entanglement pair states (\textsf{PEPS}) and show how condensing and confining anyons reduces a local gauge symmetry to a global on-site symmetry. We also study the action of this global symmetry over the quasiparticle excitations. As a byproduct, we observe that this symmetry reduction effect can be applied to one-dimensional systems as well, and brings about appealing physical interpretations on the classification of phases with symmetries using matrix product states (\textsf{MPS}). The case of $\mathbb{Z}_2$ on-site symmetry is studied in detail.

\end{abstract}

\maketitle

\section{Introduction} 

Classifying quantum phases of many-body systems is a central issue in condensed matter physics. During the last decade or so, the problem has required to go beyond Landau theory of symmetry breaking \cite{Landau}: phase transitions in topologically ordered systems, such as a quantum Hall sample, simply cannot be correctly accounted for with \emph{local} order parameters \cite{QHEWen}. Beyond this impossibility lies a rich variety of non-local phenomena such as exotic statistics of excitations, protected edge modes, or ground state degeneracy on manifolds with non-trivial topology.

The two-dimensional topological phases we are considering are described by their elementary excitations: quasi-particles called anyons completely characterized by their fusion and braiding properties. Condensation and confinement of these anyons have been put forth in \cite{HopfBais} to relate different topological phases. The basis is the formation of a condensate of bosonic quasiparticles which will correspond to a new vacuum. Anyons having trivial statistics with the condensed anyons will characterize the new topological phase. Otherwise, quasiparticles which are not compatible with the new vacuum (non-trivial braiding with the condensed anyons) will not be able to move around freely in the system; they will become confined. Also quasiparticles previously distinguishable by the braiding with the condensed anyons will be identified. Because of the condensation, confinement and identification this mechanism reduces the topological content of the parent phase. This reduction has been originally formulated mathematically as the 'symmetry breaking' of a discrete gauge theory \cite{Mark, Bais1,Bais2,Bais3,Bais4}. 

This paper is devoted to the study of these phenomena with projected entangled pair states (\textsf{PEPS}) in a 2D square lattice. This class of states and its properties are characterized locally, by a five-rank tensor placing in each site with a physical index and four {\it virtual} indices that take care of the correlations present in the system. PEPS has turned out to be relevant to the description of quantum matter obeying an area law \cite{PEPSarealaw}, or exhibiting topological order \cite{PEPSdescp,topor}. Indeed \textsf{PEPS} have been shown to provide an adequate formalism to analyze anyon condensation and topological phase transitions \cite{ShadowsAnyons,CondTNS,conRan,newNorbert,Huan}. This is because \textsf{PEPS} allow to represent the characteristic properties of the anyons present in the model, identify the possible different behavior of these quasiparticles (condensation, confinement and identification) and because they describe systems beyond renormalization fixed points of topological phases.

\begin{figure}[ht!]
\begin{center}
\includegraphics[scale=0.39]{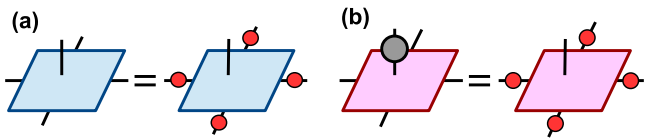}
\caption{ (a) The parent tensor (blue) is invariant under the virtual action of a representation of the group $G$.  This gauge symmetry endows the \textsf{PEPS} with topological order. (b) Some operators acting on the virtual level of the restricted tensor (pink) are equivalent to an action on the physical index of the tensor; they correspond to a global on-site symmetry. The elements of $G$ which still are gauge symmetry of the restricted tensor form a subgroup, that we name $G_{\rm topo}$, and characterize the topological order of the restricted \textsf{PEPS}.} \label{fig:tendual}
\end{center}
\end{figure}

Our work participates in this line of research: we focus on the effect of symmetries in particular pairs of \textsf{PEPS} and relate their anyonic excitations. It is known \cite{PEPSdescp} that topological order in \textsf{PEPS} can be characterized by a gauge symmetry acting on the virtual indices of the local tensor that defines the \textsf{PEPS} (see Fig. \ref{fig:tendual}(a)). In this paper we will restrict to the case in which the topological order comes from a group $G$, the so called quantum doubles models \cite{Kitaev}. We then modify this initial ({\it parent}) tensor and construct a new tensor --the {\it restricted} tensor-- in which the gauge symmetry, and hence the topological content, is reduced to a subgroup of $G$. Since the elements of the gauge symmetry group correspond to flux excitations in the \textsf{PEPS} \cite{PEPSdescp}, intuitively the restricted tensor will have less fluxes and then less capability to distinguish by braiding operations among the different charges. One expects then some charges of the parent model to be effectively {\it condensed} in the restricted system. Explicitly we find that some parent charges no longer belongs to the non-trivial topological sector; they have to be identified with the vacuum under interferometric experiments. 

We point out that the name 'anyon condensation' analyzed here does not correspond to a phase transition in the usual sense, i.e. interpolating two Hamiltonians varying a parameter. As in the original work of Bais {\it et. al.} \cite{HopfBais, Bais1, Bais2, Bais3, Bais4, Mark} and many subsequent ones \cite{Bombin,Titus, Kong}, we just restrict ourselves to comparing the behavior of anyons in a pair of models which are renormalization fixed points, and hence can be seen as the two extreme points of an interpolating path.

Moreover, we will show analytically how the restricted tensor {\it ungauges} part of the symmetry coming from $G$ so that the virtual operators induce now a non-trivial action in the physical index (see Fig. \ref{fig:tendual}(b)). This is known to correspond to a global on-site symmetry in the \textsf{PEPS} \cite{PEPSsym}.  The parent phase will then correspond to a purely topological model and the restricted phase belongs to a so-called symmetry enriched topological phase: intrinsic topological order plus a global on-site symmetry (see \cite{SETQ} and references therein). Indeed, we will see that the global symmetry will in general have a non-trivial action on the anyons of the restricted \textsf{PEPS}, e.g. permuting anyon types. Finally, we will analyze how anyon confinement and symmetry reduction are inextricably linked. These relations have been studied  in \cite {SETWan1, SETWan2} using a generic language of topological orders. 

Besides, we will see how anyon condensation, and its effects on the symmetries of the systems, sheds new light on the classification of one-dimensional symmetry protected topological phases (\textsf{SPT}) in terms of matrix product states (\textsf{MPS}) \cite{phasesMPS}. Namely, we will translate the approach taken in 2D to the language of \textsf{MPS} and subsequently we will obtain all the possible phases in 1D systems with symmetries and degenerate ground state. Also the structure classifying this setup will appear very naturally and we will give an explicit method to construct the state and operators of each phase.

An important aspect of our construction is the relation between the groups that will describe the topological content of each \textsf{PEPS} of the pair. If $G$ describes the parent topological phase we will start from, the phase of the restricted \textsf{PEPS} will be indexed by a normal subgroup $G_{\rm topo}$ of $G$, and exhibit a global on-site symmetry represented by the quotient group $G/G_{\rm topo}$; the mathematical counterpart of the physical picture is encoded in the short exact sequence:
\begin{equation}\label{eq:exactseq}
1\rightarrow G_{\rm topo} \rightarrow G \rightarrow G_{\rm sym} \rightarrow 1,
\end{equation}
where $G_{\rm sym}\cong G/G_{\rm topo}$. This sequence characterizes the group extension of the group $G_{\rm topo}$ by the group $G_{\rm sym}$. 

The paper is organized as follows: In Section \ref{sec:mathTNS}, we introduce the two families of tensor network states which are used in this study: \textsf{MPS} and \textsf{PEPS}, and describe their relevant properties to this work. In Section \ref{approach}, we describe our approach for relating topological phases in \textsf{PEPS} and we analyze the symmetry properties of the involved phases. In Section \ref{sec:phasesMPS}, we give a short review of the current classification of phases focusing on the case of one-dimensional bosonic systems treated \cite{phasesMPS}. Next, we present our approach, and recover all possible representations of an on-site global symmetry at the virtual level of the \textsf{MPS}. We also show some concrete example with the global on-site $\mathbb{Z}_2$ symmetry. We conclude and discuss open questions in Section \ref{sec:conc}.

\section{The mathematical formalism: \textsf{MPS} and \textsf{PEPS}} \label{sec:mathTNS}

Matrix product states are pure states that describe quantum spin chains and are fully characterized by a tensor $A$ (associated to each site). This tensor description reduces the number of parameters from exponential to linear in the system size (i.e. the number of particles). The key assumption that enables this reduction is that the entanglement present in the system is bounded and follows a nearest neighbor pattern. The \textsf{MPS} representation reflects the local character of the physical interactions and encapsulates the relevant part of the full Hilbert space describing the system.

Given a chain of $N$ particles, each described by a $d$-dimensional Hilbert space, $\mathbb{C}^{d}$, an \textsf{MPS} is a state of the form
\begin{equation}\label{eq:def-MPS-ti-pbc}
|\mathcal{M}(A) \rangle= \sum_{i_1, \cdots ,i_N=1}^d {\rm tr}[A^{i_1}\cdots A^{i_N}] | i_1,\cdots,i_N \rangle.
\end{equation}
where $A^i$ denotes a $D \times D$ matrix associated with a particular state $| i \rangle \in \mathbb{C}^{d}$. The tensor associated to an \textsf{MPS} denoted by $A^i_{\alpha, \beta}$ has rank three. Actually, the expression (\ref{eq:def-MPS-ti-pbc}) does not represent the most general form of an \textsf{MPS}; it restricts to the case of periodic boundary conditions and translational invariance \cite{MPSrep}. We will work with these assumptions. 

\begin{figure}[ht!]
\begin{center}
\includegraphics[scale=0.50]{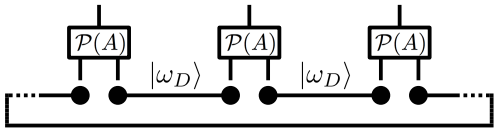}
\caption{Construction of an \textsf{MPS} with periodic boundary condition by placing maximally entangled pairs and applying the map $\mathcal{P}$. Let us note that the map acts on virtual particles belonging to different maximally entangled pairs.
} \label{fig:vbpictureMPS}
\end{center}
\end{figure}

A way to picture how to construct an \textsf{MPS}  is to consider the following process. One first constructs a product state $|\omega\rangle^{\otimes N}$ of virtual degrees of freedom (d.o.f.) or generalized spins with maximally entangled pairs $|\omega_D \rangle= \sum_{i=1}^D  |i,i \rangle$ to describe the amount of entanglement between two sites. Then, we apply the map $\mathcal{P}(A):\mathbb{C}^{D}\otimes\mathbb{C}^{D} \to \mathbb{C}^{d}$:
\begin{equation}\mathcal{P}(A)=\sum_{i=1}^{d} \sum_{\alpha, \beta =1} ^D (A^i)_{\alpha \beta} |i\rangle_p \langle \alpha \beta|_v \notag \end{equation}
on each site, i.e. on neighboring virtual spins corresponding to parts of different maximally entangled pairs, as illustrated in Fig. \ref{fig:vbpictureMPS}. The subscripts $p$ and $v$ denote the physical and virtual d.o.f. respectively. This process translates the $N$ site chain described initially by maximally entangled states into the $|\mathcal{M}(A) \rangle$ state. 
These states are important because every ground state of a one-dimensional gapped local Hamiltonian can be approximated by an \textsf{MPS} \cite{Hast1,Hast2,Arad}. A simple example of an \textsf{MPS} is the GHZ state: $|\Psi_{{\rm GHZ}} \rangle= \sum_i^n| i,i,\cdots,i \rangle$, where $(A^i)_{\alpha \beta}=\delta_{i \alpha} \delta_{i \beta}$.

The matrices of an \textsf{MPS} can be brought into a canonical form, which is unique, and these matrices generally have a block-diagonal representation \cite{MPSrep}. This block diagonal structure restricts the support of the map $\mathcal{P}$ to a direct sum space: $({\rm ker} \mathcal{P})^{\perp} = \bigoplus^{\mathcal{A}}_{\alpha=1}  \mathcal{H}_{\alpha} \subset \mathbb{C}^{D}\otimes\mathbb{C}^{D}$. The uniqueness of the canonical form allows to prove that if a quantum state admits two different \textsf{MPS} representations, their matrices are unitarily related. For every \textsf{MPS} whose canonical form has one block, one can build a local gapped Hamiltonian (the so-called parent Hamiltonian) whose unique ground state is exactly the \textsf{MPS}. In the case where the canonical form has several blocks, the ground state of the associated Hamiltonian is not unique and is spanned by the \textsf{MPS} constructed from the individual blocks.  Another property of this parent Hamiltonian, called frustration-freeness, is that its \textsf{MPS} ground state also minimizes the energy of each of its local terms.

The symmetries of an \textsf{MPS} can be characterized in terms of the virtual d.o.f. \cite{MPSsym}. A global on-site symmetry of an \textsf{MPS} is a local operation which leaves the state invariant up to a global phase:  $u^{\otimes N} |\mathcal{M}(A) \rangle=e^{i\phi} |\mathcal{M}(A) \rangle$ (case of unique ground state). As a result of the uniqueness of the canonical form and the equivalence between tensors describing the same state, the physical symmetry can be replaced by a local transformation in the virtual d.o.f. This means that  there exists a unitary $U=P(\oplus_{\alpha}V_{\alpha})$, where $P$ is a permutation matrix of the blocks and $V_{\alpha}$ is a unitary acting on the $\alpha$-th block, such that

\begin{equation}\label{eq:MPSsym}
\sum_j u_{ij} A^j=WUA^iU^{\dagger},
\end{equation}
where $W=\oplus_{\alpha}e^{i\theta_{\alpha}}\mathbb{I}_{\alpha}$ is an arbitrary phase in each block.

Although \textsf{MPS} were first introduced for numerical purposes (the main motivation was the minimization of the energy of a target Hamiltonian), they have become a very successful formalism to analyze quantum many body systems, since they capture the properties of gapped systems with nearest neighbor entanglement patterns.
 
\begin{figure}[ht!]
\begin{center}
\includegraphics[scale=0.34]{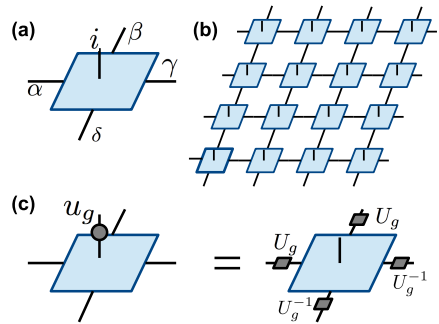}
\caption{(a) Diagram picture of the tensor $A^i_{\alpha \beta \gamma \delta}$ of a \textsf{PEPS} for a square lattice. (b) The \textsf{PEPS} constructed via the concatenation of the virtual d.o.f. of the tensors (without boundary conditions). (c) A single on-site unitary (which is a representation of the global symmetry) applied to the physical system is equivalent in an injective \textsf{PEPS}, as far as the tensor level is concerned, to the action of four unitaries, representing a group element, in the virtual d.o.f.
} \label{fig:tensorsPEPS}
\end{center}
\end{figure}

\textsf{PEPS} are a natural 2D analogue of \textsf{MPS} for two-dimensional lattices (or graphs in general). To construct the state, the tensor associated with each site has a number of virtual indices equal to the coordination number of the site. Taking for simplicity a square lattice, the tensors that define the state have five indices, $A^i_{\alpha \beta \gamma \delta}$, one corresponding to the physical hilbert space and four indices associated to the virtual d.o.f. (sketched in Fig. \ref{fig:tensorsPEPS}(a) ). The trace and multiplication operations considered in one dimension are generalised to a contraction of all pairs of nearest neighbour virtual d.o.f.  $|\Psi(A)\rangle= \sum_{i_1,\cdots,i_{N}} \mathcal{C}(\{A^i_{\alpha \beta \gamma \delta}\})|i_1\cdots i_{N} \rangle$. As for \textsf{MPS}, these states arise as ground states of local gapped Hamiltonians and are unique ground states if the \textsf{PEPS} tensors satisfy a condition called injectivity \cite{uniPEPS}.

The characterization of local and spatial symmetries of injective \textsf{PEPS} has been done in \cite{PEPSsym} at the level of the tensor. If an injective \textsf{PEPS} has a global on-site symmetry the local operator acting on the tensor is equal to the tensor under the action of some operator in the virtual d.o.f., as depicted in Fig. \ref{fig:tensorsPEPS}(c).\\

\begin{figure}[ht!]
\begin{center}
\includegraphics[scale=0.8]{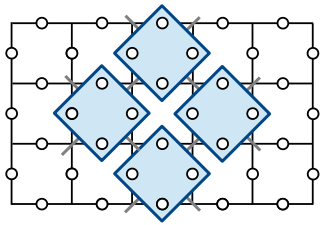}
\caption{Relation between the lattice model of \cite{Kitaev} and the tensor construction (\ref{eq:gisopeps}). The white dots represent the generalized spins living on the edges. Each tensor is placed on alternating plaquettes of the lattice, gathering the four surrounding spins. The physical index of the tensor corresponds to the tensor product of four local Hilbert spaces and this fact is reflected in Eq.(\ref{eq:tensordesc}). Diagonal grey edges represent \textsf{PEPS} virtual bonds.  
} 
\label{fig:latticetensor}
\end{center}
\end{figure}

A family of \textsf{PEPS} of crucial interest for our work are the so-called $G$-isometric \textsf{PEPS}:
\begin{equation}\label{eq:gisopeps}
A=\frac{1}{|G|}\left (\sum_{ g \in G}  L_g \otimes L_g \otimes L^{\dagger }_g\otimes L^{\dagger }_g \right ),
\end{equation}
where $L_g$ is the left regular representation of the group $G$ \cite{PEPSdescp}. The relation between the physical labels and the virtual d.o.f. of the tensor is as follows
\begin{equation} \label{eq:tensordesc}
A^i_{\alpha \beta \gamma \delta}\equiv  A^{(a,b,c,d)}_{\alpha \beta \gamma \delta}=\sum_{g\in G} [L_g]_{\alpha a}  [L_{g}]_{\beta b}[L^{\dagger}_g]_{\gamma c}  [L^{\dagger}_g]_{\delta d}.
\end{equation}
The physical index is now four-fold: $'i'\cong(a,b,c,d)$, see caption of Fig. \ref{fig:latticetensor} for explanation. 

\begin{figure}[ht!]
\begin{center}
\includegraphics[scale=0.35]{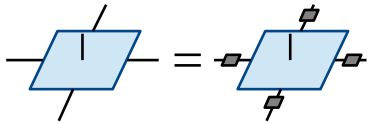}
\caption{Gauge symmetry of the tensor in the virtual d.o.f. Within the family of $G$-isometric \textsf{PEPS}, the invariance is achieved with tensor product operators: $\{ L_g\otimes L_g\otimes L^{\dagger }_g\otimes L^{\dagger }_g: g\in G \}$. More generally the gauge symmetry is realised by Matrix Product Operators \cite{MPOtop}. } \label{fig:gaugePEPS}
\end{center}
\end{figure}

The family of $G$-isometric \textsf{PEPS} is invariant under the action of (a representation of) the group $G$ acting purely on the virtual d.o.f. (as illustrated in Fig. \ref{fig:gaugePEPS}). 

It can be shown that the state (\ref{eq:gisopeps}) is a renormalization fixed point of the topological phases described by the Quantum Double model of the group $G$ \cite{PEPSdescp, Kitaev} and that a parent Hamiltonian can be constructed for it. The most important features of this parent Hamiltonian are (i) its ground state degeneracy depends on the topology of the lattice considered, (ii) its elementary excitations exhibit anyonic statistics. These excitations are in correspondence with irreducible representations (irreps) of the Quantum Double of $G$. The \textsf{PEPS} (\ref{eq:gisopeps}) is also the ground state of a Hamiltonian proposed and studied in \cite{Kitaev}, where physical d.o.f. are placed on the edges of a square lattice (Fig. \ref{fig:latticetensor}). 

The excitations are quasi-particles that exhibit topological interaction: braiding and fusion analogous to the Aharonov-Bohm effect \cite{Preskillnotes}. There are three different types of anyons in the Quantum Double of $G$: charges represented by the irreps of $G$, fluxes described by conjugacy classes of elements of $G$ and dyons which are characterized by the irreps of the centralizer of each conjugacy class. Such models/phases are interesting, not only because of their exotic topological behavior, but also by their potential application in (topological) quantum computing \cite{Kitaev}.

One can also define the analogue object in 1D, the so-called $G$-isometric \textsf{MPS}, described by the tensor 
\begin{equation}
\sum_{g\in G} L_g \otimes L^{\dagger }_g\; .
\end{equation}

On the one hand, $G$-isometric \textsf{MPS} are intimately connected to the {\it isometric forms} for \textsf{MPS} \cite{PEPSdescp} used in the 1D classification of phases and, on the other hand, they have the same mathematical structure as their 2D analogue. We will come back to this point in Section \ref{sec:phasesMPS}.

\section{Symmetry changes through anyon condensation in $\mathbf{G}$-isometric \textsf{PEPS}} \label{approach}

Anyon condensation has been proposed to relate two orders involved in a topological symmetry breaking \cite{HopfBais}. It is based on the formation of a condensate of bosonic quasiparticles of the parent phase which can be addressed mathematically as the 'symmetry breaking' of a gauge theory. The quasiparticles that braid trivially with the condensed excitations characterize the new topological phase. But the quasiparticles which are not compatible with trivial fusion and braiding with the condensed anyons cannot move around freely in the system; they become confined. 

It is observed in \cite{SETWan1,SETWan2} that the mathematical formalism describing the confined anyons can be associated to the action of a global symmetry on the new topological phase. This relates a symmetry enriched topological phase with a purely topological phase, where the former is the result of an anyon condensation of the latter. The purpose of this work is to show how this picture can be materialized in a lattice model very naturally in the \textsf{PEPS} formalism.

It has been shown recently in \cite{BoconTNS} that anyon condensation in \textsf{PEPS} can be associated to a breaking of the local symmetry of the virtual d.o.f. Following this idea, we will start with a $G$-isometric \textsf{PEPS} 
\beq \label{eq:parentten}
A_G=\frac{1}{|G|}\left (\sum_{ g \in G}  L_g\otimes L_g\otimes L^{\dagger}_g\otimes L^{\dagger}_g \right ),
\eeq
which we call parent tensor. We will also consider a normal subgroup of $G$, $G_{\rm topo}$, and construct the following tensor 
\begin{equation} \label{eq:restensor}
A^{\rm{res}}_G=\frac{1}{|G_{\rm topo}|}\left (\sum_{ g \in G_{\rm topo}}  L_g\otimes L_g\otimes L^{\dagger }_g\otimes L^{\dagger }_g \right )
\end{equation}
which we call the restricted tensor. We will show how, partially breaking the local invariance of a parent tensor, we will end up with a new tensor enriched with a non-trivial virtual representation of the broken gauge symmetry. The \textsf{PEPS} constructed with these tensors, representing the ground state of the model,  will inherit the global and local symmetry relations between them.

We will be able to identify the remaining topological order, flux confinement, the associated charge condensation and the splitting of quasiparticles excitations in the \textsf{PEPS} constructed with the restricted tensor. This identification is done comparing the behavior of the operators corresponding to anyons in the parent model when they are placed in the restricted state. Moreover we will connect the new symmetry features and the different behaviors of the anyons in both models. Finally we will analyze the effect of the broken gauge symmetry over the unconfined anyons of the restricted phase.

\subsection{Symmetry reduction: from local to global}

The state associated with the tensor $A^{\rm{res}}_G$ is in the same phase as the Quantum Double of $G_{\rm topo}$. This is because we can express the restricted tensor as
\begin{equation}
A^{\rm{res}}_G\cong A_{G_{\rm topo}}\otimes\mathbb{I}, \notag
\end{equation}
where $A_{G_{\rm topo}}$ is the tensor of the $G_{\rm topo}$-isometric \textsf{PEPS}. We can perform this decomposition since the action of $\{ L_k: k \in G_{\rm topo}\}$ can be decomposed as  $L^{G_{\rm topo}}_k\otimes \mathbb{I}^{G_{\rm sym}}$ (see Appendix \ref{ap:ext}). Therefore the state associated to the tensor $A^{\rm{res}}_G$ is in the same phase as the $G_{\rm topo}$-isometric \textsf{PEPS} (if we do not impose symmetries) with the subsequent properties studied in \cite{PEPSdescp}.

In the \textsf{PEPS} representation of these models, a gauge invariance of the tensor in the virtual d.o.f. is equivalent to a local gauge symmetry of the state formed with the tensor. Let us explicitly construct these operators for the models that we are considering. We define the operator $$S_g\equiv L_g\otimes L_g\otimes L^{\dagger }_g\otimes L^{\dagger }_g, $$ where $g\in G$. It is easy to see that these operators realized a gauge invariance of the parent tensor for all $g$: 
\beq
(S_g)_p \; A_G=A_G=A_G \; (S_g)_v \; \forall g \in G, \notag
\eeq
where the subscript $v$ denotes the action on the virtual d.o.f., and $p$ the action on the physical d.o.f. Thus
\begin{equation}
 (S^{\mathcal{R}}_g\otimes \mathbb{I}^{{\rm rest}})|\Psi(A_G)\rangle = |\Psi(A_G)\rangle \quad \forall  g\in G, \notag
 \end{equation} 
where $S^{\mathcal{R}}_g$ is the operator $S_g$ acting in any region $\mathcal{R}$ and $|\Psi(A_G)\rangle$ is the state constructed with the tensor $A_G$ using periodic boundary condition. That is, the representation of $G$ given by the operators $S_g$ is a gauge symmetry of the state $|\Psi(A_G)\rangle$. The representation on the virtual d.o.f. of the operators $S_g$ is the identity. Trivially, $|\Psi(A_G)\rangle$ is also left invariant when the same $S_g$ is applied on each lattice site. The restricted tensor has a similar gauge symmetry: 
$$(S_k)_p A^{\rm{res}}_G=A^{\rm{res}}_G=A^{\rm{res}}_G(S_k)_v \quad \forall k\in G_{\rm topo},$$
\begin{equation}\label{eq:gaugeres}
 (S^{\mathcal{R}}_k\otimes \mathbb{I}^{{\rm rest}})|\Psi(A^{\rm{res}}_G)\rangle = |\Psi(A^{\rm{res}}_G)\rangle \quad \forall k\in G_{\rm topo}.
 \end{equation} 
 When $g\in G - G_{\rm topo}$, $S_g$ will no longer leave $A^{\rm{res}}_G$ invariant: the restricted tensor breaks the gauge invariance of the parent tensor to a normal subgroup $G_{{\rm topo}}$ of $G$.\\

\begin{tcolorbox}
Part of the gauge symmetry of the state $|\Psi(A_G)\rangle$ is degraded to just a global symmetry of the state $|\Psi(A^{\rm{res}}_G)\rangle$. The physical representation associated with such a global symmetry of the restricted tensor belongs to the quotient group $G_{\rm sym}\cong G/G_{\rm topo}$.
\end{tcolorbox}
\vspace{5mm}

\begin{figure}[ht!]
\begin{center}
\includegraphics[scale=0.35]{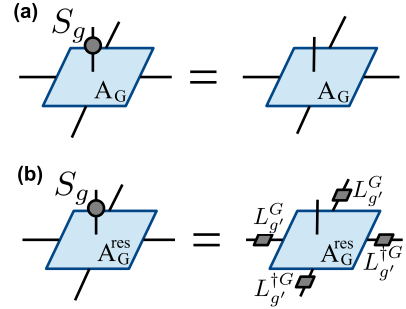}
\caption{(a) The action of the operator $S_g$ leaves invariant the tensor $A_G$. (b) The action of the operator $S_g$ over $A^{\rm res}_G$ change the tensor as depicted. }
\label{fig:symAG}
\end{center}
\end{figure}

Let us prove the above statement. Consider the action of $S_g$ on $A^{\rm{res}}_G$:
\begin{align}\label{eq:symtensor}
(S_g)_p A^{\rm{res}}_G  &=  \sum_{ k \in G_{\rm topo}}  L_{kg}\otimes L_{kg}\otimes L^{\dagger }_{kg}\otimes L^{\dagger }_{kg} \notag \\
&= \sum_{ \tilde{g} \in [g]}  L_{\tilde{g}}  \otimes L_{\tilde{g}}  \otimes L^{\dagger }_{\tilde{g}} \otimes L^{\dagger }_{\tilde{g}}  \notag \\
& = \sum_{ k \in G_{\rm topo}}  L_{g'k}\otimes L_{g'k}\otimes L^{\dagger }_{g'k} \otimes L^{\dagger }_{g'k}    \notag \\
& = A^{\rm{res}}_G \left (S_{g'} \right )_v \quad {\rm for \; all} \;  g'\in [g], 
\end{align}
where we have used the fact that $G_{\rm topo}$ is a normal subgroup. The operator $S_g$ acting on the physical Hilbert space is translated into the operator $S_{g'}$ on the virtual d.o.f., as illustrated in Fig. \ref{fig:symAG}(b), where $g'$ is any element of the coset of $g$ in $G/G_{\rm topo}$; $[g]=\{gk ; \;  k \in G_{\rm topo} \}$. 
Then, when we concatenate the perturbed tensors the virtual operators cancel out in each edge and the state remains invariant. 

The $G$ gauge symmetry has been reduced to a $G_{\rm topo}$ gauge symmetry plus a $G/G_{\rm topo}$ global symmetry. Since $G_{\rm topo}$ is normal in $G$, $G/G_{\rm topo}$ is a group which we will denote $G_{\rm sym}$ from now on. The $G_{\rm sym}$ global symmetry over this state means:

\begin{equation}\label{eq:Gsymtensor_1}
(S_g)_p A^{\rm{res}}_G =(S_{g'})_p A^{\rm{res}}_G \quad {\rm if} \;  [g']= [g],
\end{equation}
$$(S_g)_p (S_{g'})_p A^{\rm{res}}_G =(S_{g g'})_p A^{\rm{res}}_G.$$

Combining (\ref{eq:gaugeres}) and (\ref{eq:Gsymtensor_1}), the most general form of symmetry reads
\begin{equation}\label{eq:symstate}
\bigotimes_{x\in \Lambda}  S_{g_x} |\Psi(A^{\rm{res}}_G)\rangle = |\Psi(A^{\rm{res}}_G)\rangle  \;  {\rm if} \; \exists g \in G:\; g_x\in[g] \; \forall x\in \Lambda,
\end{equation}
where $\Lambda$ denotes the set of lattice sites. Let us finally show that the global symmetry forms a representation of the quotient group. This is done first using the projection from $G$ to $G_{\rm sym}$, $g\mapsto [g]$ and then proving that the symmetry  operators $S_g$ are a representation of the quotient group.  It is sufficient to consider the action of $S_g$ on the relevant Hilbert space ($\mathcal{H}_p$) defined as:
$$\mathcal{H}_p= \left \{ \;
\parbox[c]{0.18\textwidth}{ \includegraphics[scale=0.25]{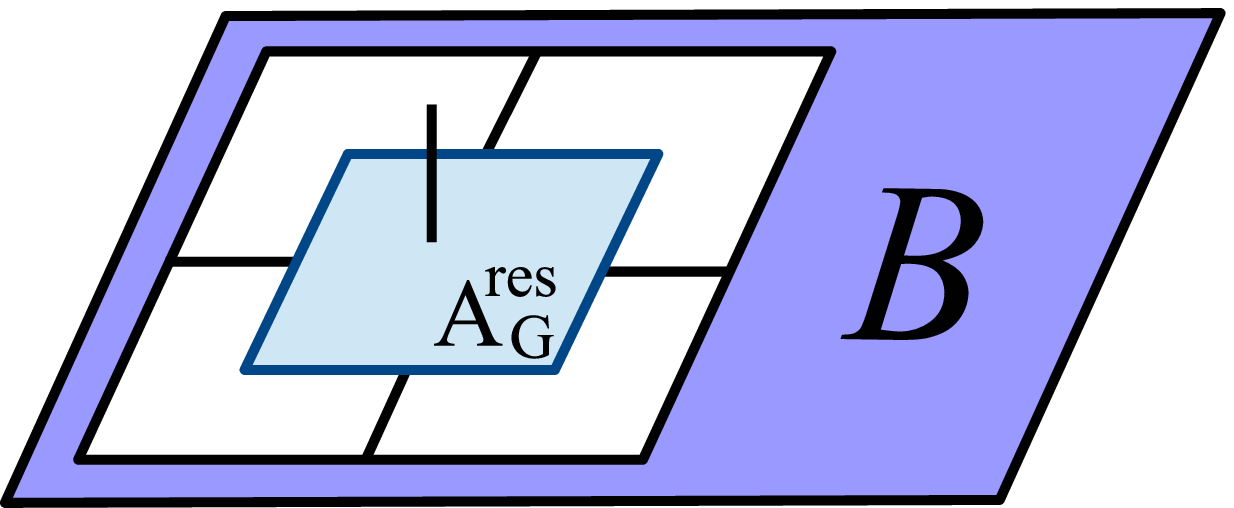}}\Bigg|B \right \}\subset \mathbb{C}^d,$$
where $B$ runs over all boundary conditions and $d=|G|^4$. We will restrict to this space, defining $\tilde{S} _g=S_g\big|_{\mathcal{H}_p}$ because  $S_g(\mathcal{H}_p)\subset \mathcal{H}_p$. The operator $\tilde{S} _g$ is uniquely defined by $G_{\rm sym}$ since
$\tilde{S}_g=\tilde{S}_{g'} \; {\rm if} \;  [g']= [g]$. Moreover these operators are a representation of $G_{\rm sym}$ because $\tilde{S}_g  \tilde{S} _{g'}= (S_g S_{g'}) \big|_{\mathcal{H}_p}= S_{gg'} \big|_{\mathcal{H}_p}=\tilde{S}_{gg'}$. Then, the operators $S_g$ are a representation of the quotient group $G_{\rm sym}\cong G/G_{\rm topo}$  over the restricted tensor.

The restriction Eq.(\ref{eq:parentten}) $\to$ Eq.(\ref{eq:restensor}), together with the action of the operators $S_g$ can be viewed as a local tensor (partial) {\it ungauging}. This means that we end up with a new tensor with a (non-trivial) gauge subgroup $G_{\rm topo}$ of $G$ and that the rest of the parent gauge group $G/G_{\rm topo}$ is not lost: it becomes now a global on-site symmetry under $S_g$. The full gauge symmetry $G$ of the parent tensor is recovered by the gauge  group $G_{\rm topo}$ and the global symmetry group $G_{\rm sym}$  of the restricted tensor via the short exact sequence of Eq.(\ref{eq:exactseq}).  This procedure can be regarded as the inverse of the tensor gauging studied in \cite{GaugePEPS}, in which the authors modify a tensor with a global symmetry to end up with a tensor with purely gauge symmetry.
We remark that we are comparing the same symmetry operators over different tensors; this is possible because the dimensions of the indices of both tensors coincide. This fact allows us to compare another type of operators, namely those which create quasiparticle excitations. This will be analyzed in the next subsections for the cases of pure fluxes and pure charges (the case of dyons is discussed in Appendix \ref{ap:dyonic}).

\subsection{Comparing pure flux excitations of the parent and restricted models: confinement}

\begin{figure}[ht!]
\begin{center}
\includegraphics[scale=0.35]{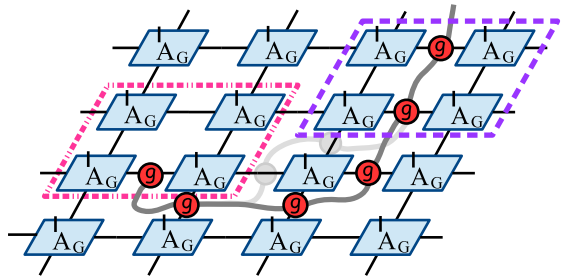}
\caption{A subset of the lattice, in a background of $A_G$ tensors, having a flux excitation string. The red dots represent the operator $L_g$ acting on the virtual d.o.f. of the tensors. Each colored square drawn represents the place where a local hamiltonian acts. Only the pink (dot-dashed line) plaquette is an excitation of the local hamiltonian $h$: adds $+1$ to the total energy of the state.}
\label{fig:string}
\end{center}
\end{figure}

We will study the effect of inserting flux excitations of the parent Hamiltonian of $A_G$ on an $A_G^{\rm res}$ background. A pair of flux excitations of the Quantum Double of $G$ is described, in the \textsf{PEPS} formalism, by an open string of operators $L_g$ placed on the {\it virtual} d.o.f. of the lattice constructed with $A_G$. An example of this configuration is shown in Fig. \ref{fig:string}. Such strings can be moved and deformed freely on an $A_G$ background, thanks to the $G$-invariance of the tensor $A_G$ \cite{PEPSdescp}. But, on an $A_G^{\rm res}$ background, there is an energy cost associated with a string of $L_g$ operators whenever $g\in G- G_{\rm topo}$. The fluxes which do not belong to $G_{\rm topo}$ become confined: the excited state containing confined fluxes has an energy that grows with the length of the string, i.e., it exhibits string tension. We are actually going to see that the confinement of some fluxes is intimately related to the change of the symmetry described above.

Let us sketch the construction of \cite{PEPSdescp} in order to use it for our purposes. Given a state on the form of a $G$-isometric \textsf{PEPS} with tensor $A_G$, the parent Hamiltonian is constructed by summing identical four-body local terms $h$, $H=\sum_{i \in \Lambda} h_i$, which act on a $2\times 2$ region of the lattice. The operator $h$ is the projector onto the orthonormal subspace generated by all virtual conditions applied to the concatenation of four tensors in a $2\times 2$ set, $h=\mathbb{I}-\Pi_{\mathcal{S}_{2\times2}}$. Pictorially we represent this subspace as
$$\mathcal{S}_{2\times2}= \left \{ \;
\parbox[c]{0.16\textwidth}{ \includegraphics[scale=0.15]{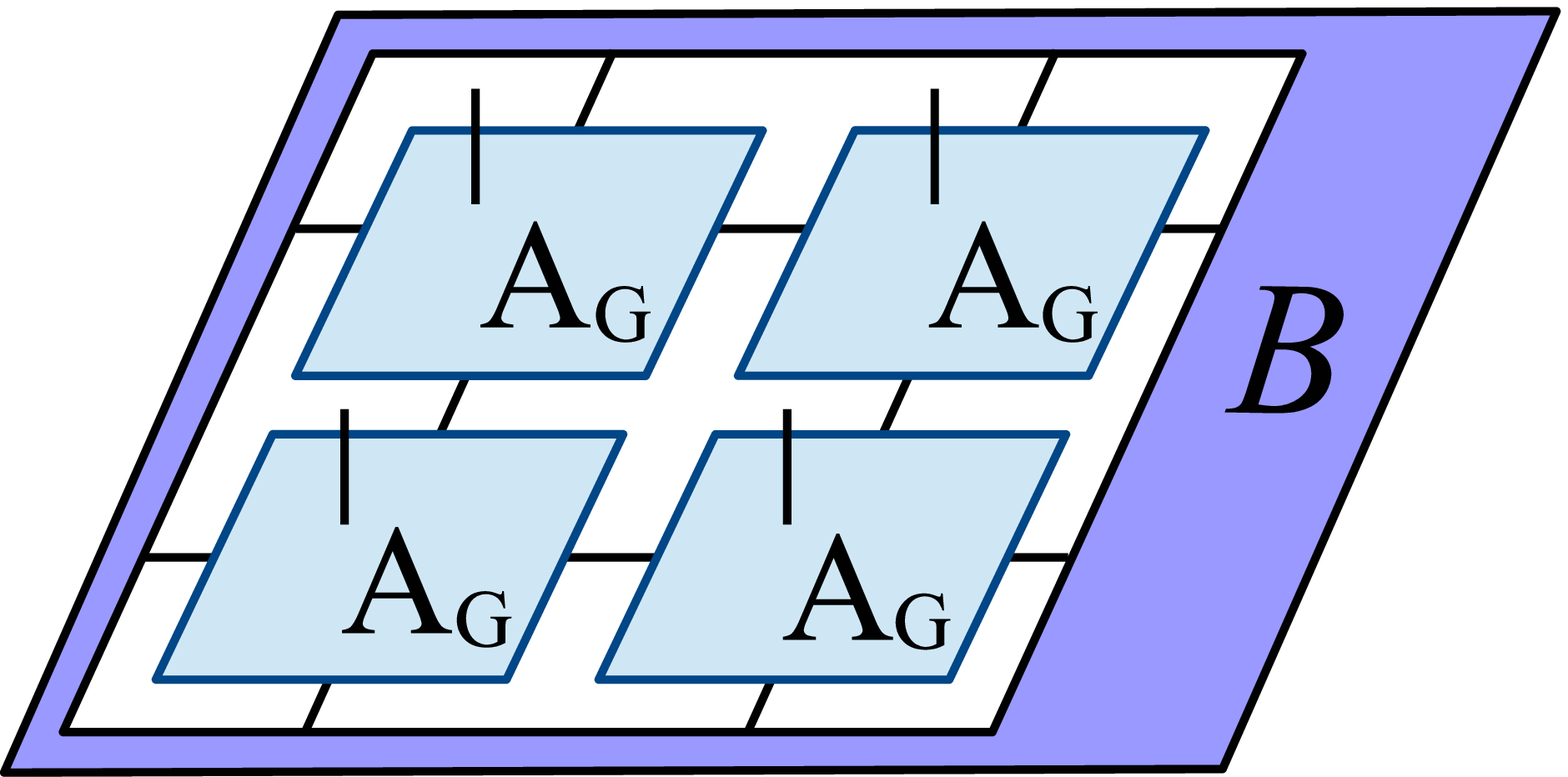}}\Bigg|B\in(\mathbb{C}^{|G|})^{\otimes 8} \right \}.$$
Therefore, $h$ has the subspace $\mathcal{S}_{2\times2} $ as its ground state:
$$h\;\parbox[c]{0.15\textwidth}{ \includegraphics[scale=0.15]{S2x2parent.eps} } =0.$$

Suppose we have a pair of fluxes in the lattice, see Fig. \ref{fig:string}, the plaquette at each end point of the string corresponds to an excitation of the local hamiltonian, $h$, of $A_G$. This is because the end point plaquette cannot be moved using the $G$-invariance of the tensor and it does not belong to the kernel of $h$. But the rest of the string can be deformed freely through the lattice. Thus the plaquette terms are not distinguished locally from the ground state; the energy of the string does not depend on the distance between the two end points.

\begin{figure}[ht!]
\begin{center}
\includegraphics[scale=0.35]{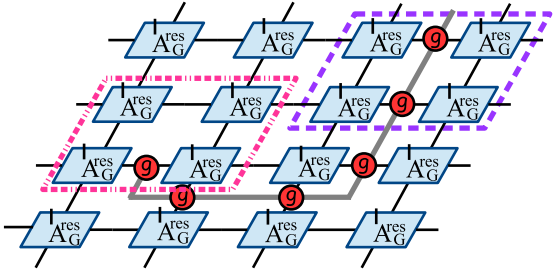}
\caption{A subset of the lattice, in a background of $A^{\rm res}_G$ tensors, having a string of $L_g$ operators acting on the virtual d.o.f. Each colored square drawn represents the place where a local hamiltonian acts. In this situation both plaquette are excitations of the local hamiltonian $h^{\rm res}$: adds $+1$ to the total energy of the state.}
\label{fig:stringres}
\end{center}
\end{figure}

We now consider a string of operators $L_g$, with $g\in G - G_{{\rm topo}}$, in a background of $A^{\rm res}_G$ as depicted in Fig. \ref{fig:stringres}.

This string cannot be deformed using the $G_{{\rm topo}}$-invariance of the tensors. Instead, the element $g$ of the operator $L_g$ can be sent to $k'gk$ (and then to the operator $L_{k'gk}$) where $k,k'\in G_{{\rm topo}}$ applying the $G_{{\rm topo}}$-invariance of the tensors for every virtual edge. This transformation cannot change the coset $g$ belongs to because $k'gk\in[g]$ ($G_{{\rm topo}}$ is normal in $G$). Thus, an element $g\notin G_{{\rm topo}}\equiv[e]$ cannot be transformed into one belonging to the trivial coset $[e]$, so the operator cannot be moved freely from its position. This fact shows that some strings can be detected locally in contrast with a string formed with operators $L_k$, where $k \in G_{\rm topo}$, that can be deformed freely.

We now show that a state modified with a string operator as the one depicted in Fig. \ref{fig:stringres} is an eigenstate of the local term of the parent Hamiltonian of $A^{\rm{res}}_G$. Therefore this modified state is an excitation of the parent Hamiltonian. For this purpose, let us define the parent Hamiltonian of $A^{\rm res}_G$ in analogy with the $G$-isometric \textsf{PEPS}:  $H^{\rm res}= \sum_{i \in \Lambda} h^{\rm res}_i$ where the local projector is $h^{\rm res}=\mathbb{I}-\Pi_{\mathcal{S}^{\rm res}_{2\times2}}$ and the subspace $\mathcal{S}^{\rm res}_{2\times2}$ is constructed as

$$\mathcal{S}^{\rm res}_{2\times2}= \left \{ \;
\parbox[c]{0.19\textwidth}{ \includegraphics[scale=0.18]{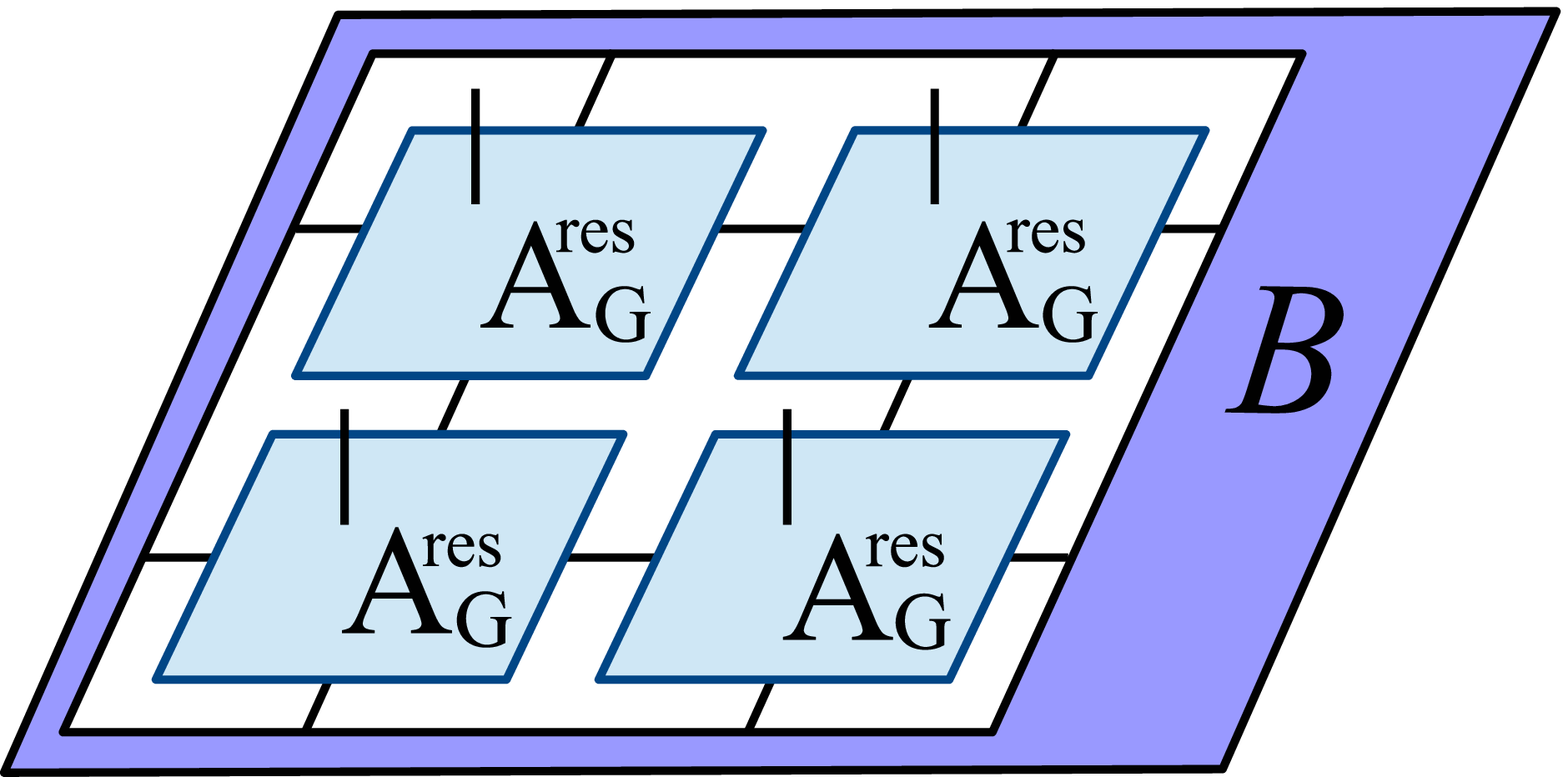}}\Bigg|B\in(\mathbb{C}^{|G|})^{\otimes 8} \right \}.$$

We calculate explicitly the scalar product between an arbitrary element of $\mathcal{S}^{\rm res}_{2\times2}$ and a $2\times2$ subset of the lattice containing part of the $g$-string. 

If the result is zero, the state with a string is locally orthogonal to $\mathcal{S}_{2\times2}$. So, decorating an $A_G^{\rm res}$ background with a string of operators $L_g, g \in G - G_{\rm topo}$ gives rise to an eigenstate with eigenvalue one for each local term $h^{\rm res}$ whose support overlaps with the support of the string (recall that it is a projector):
$$h^{\rm res}\; \parbox[c]{0.14\textwidth}{ \includegraphics[scale=0.25]{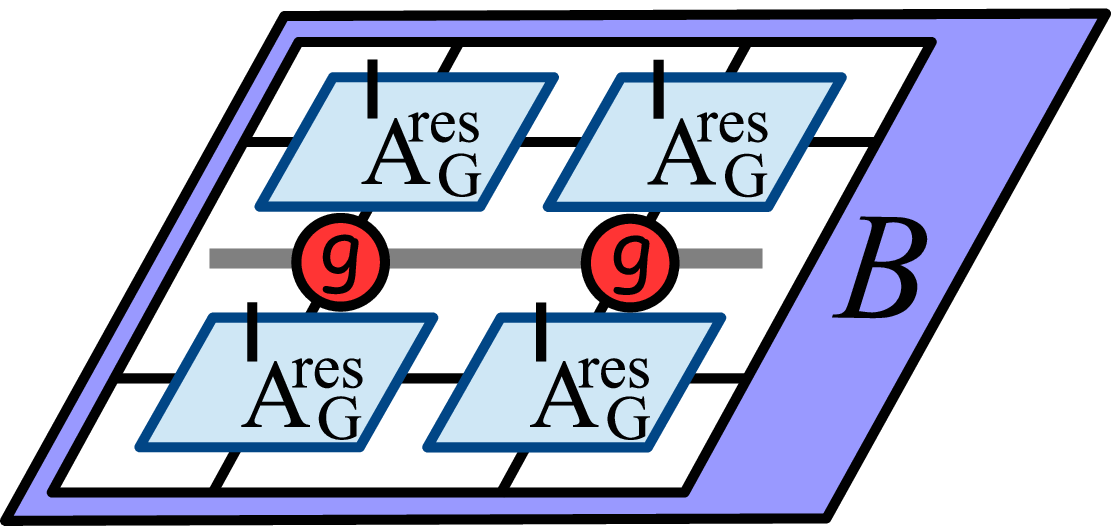}} =\parbox[c]{0.14\textwidth}{ \includegraphics[scale=0.25]{S2x2twodefect2.eps}} , \;g \in G - G_{\rm topo}. $$

 We can express the scalar product pictorially as following:
\begin{widetext}

\beq \label{eq:scapro}
 \left \langle \parbox[c]{0.145\textwidth}{ \includegraphics[scale=0.14]{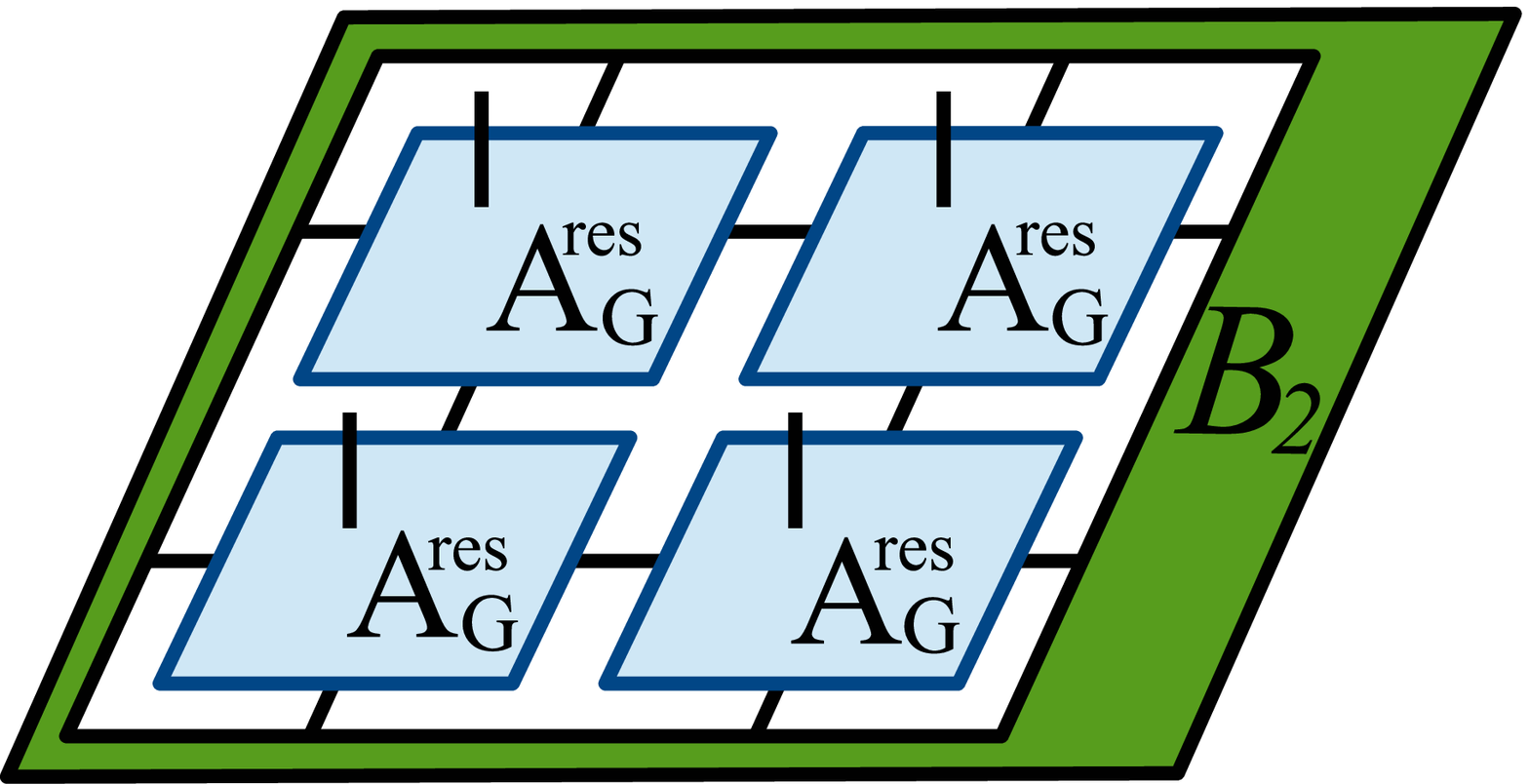}} 
\Bigg | \;
\parbox[c]{0.155 \textwidth}{ \includegraphics[scale=0.25] {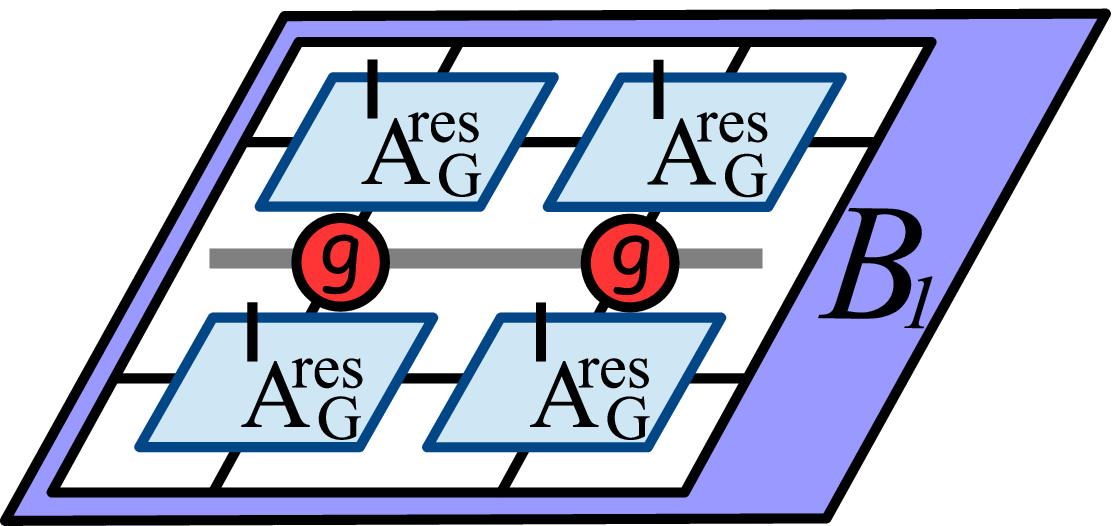}} \right \rangle =
\parbox[c]{0.2\textwidth}{ \includegraphics[scale=0.23]{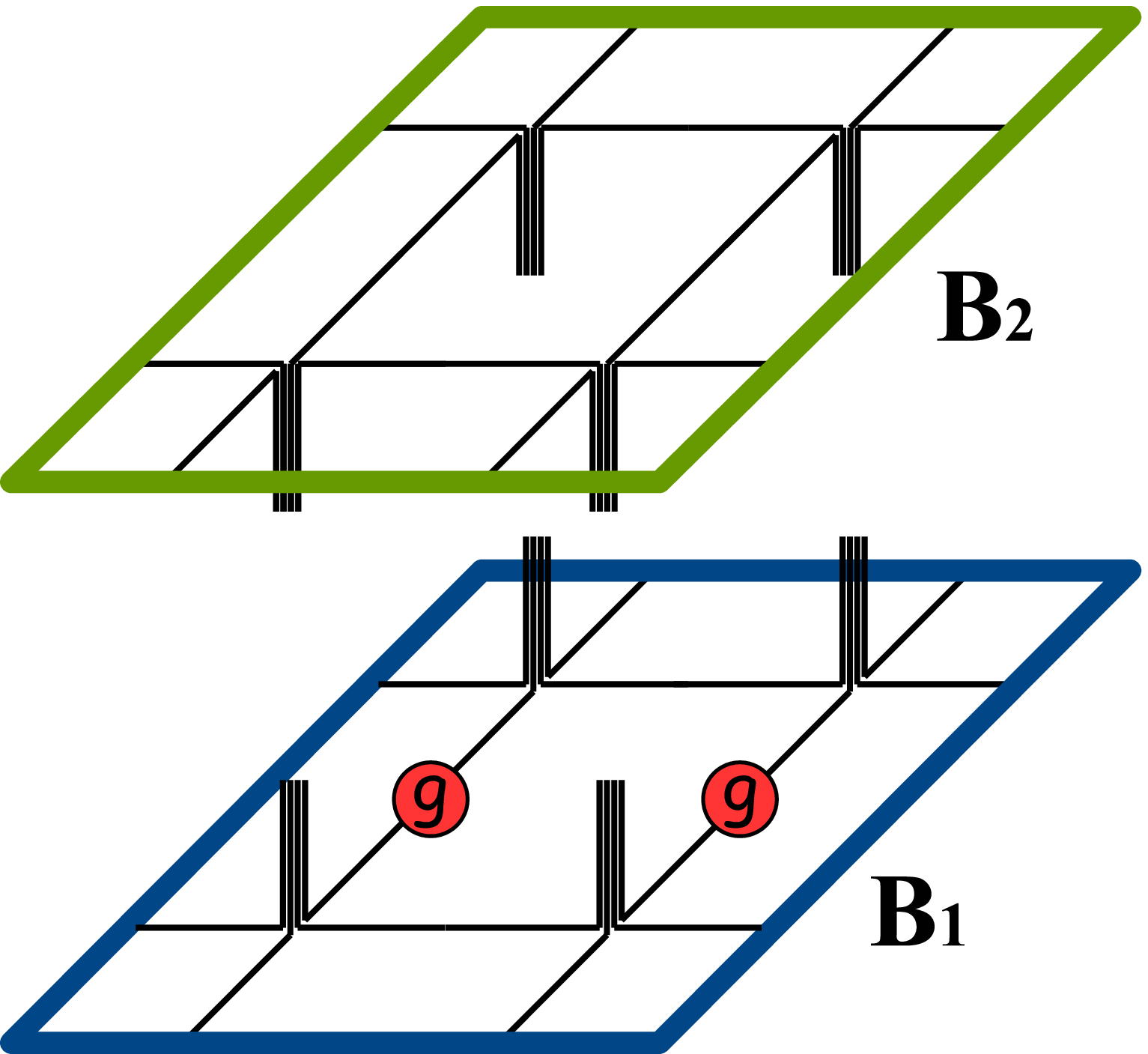} }=
\parbox[c]{0.19\textwidth}{ \includegraphics[scale=0.23]{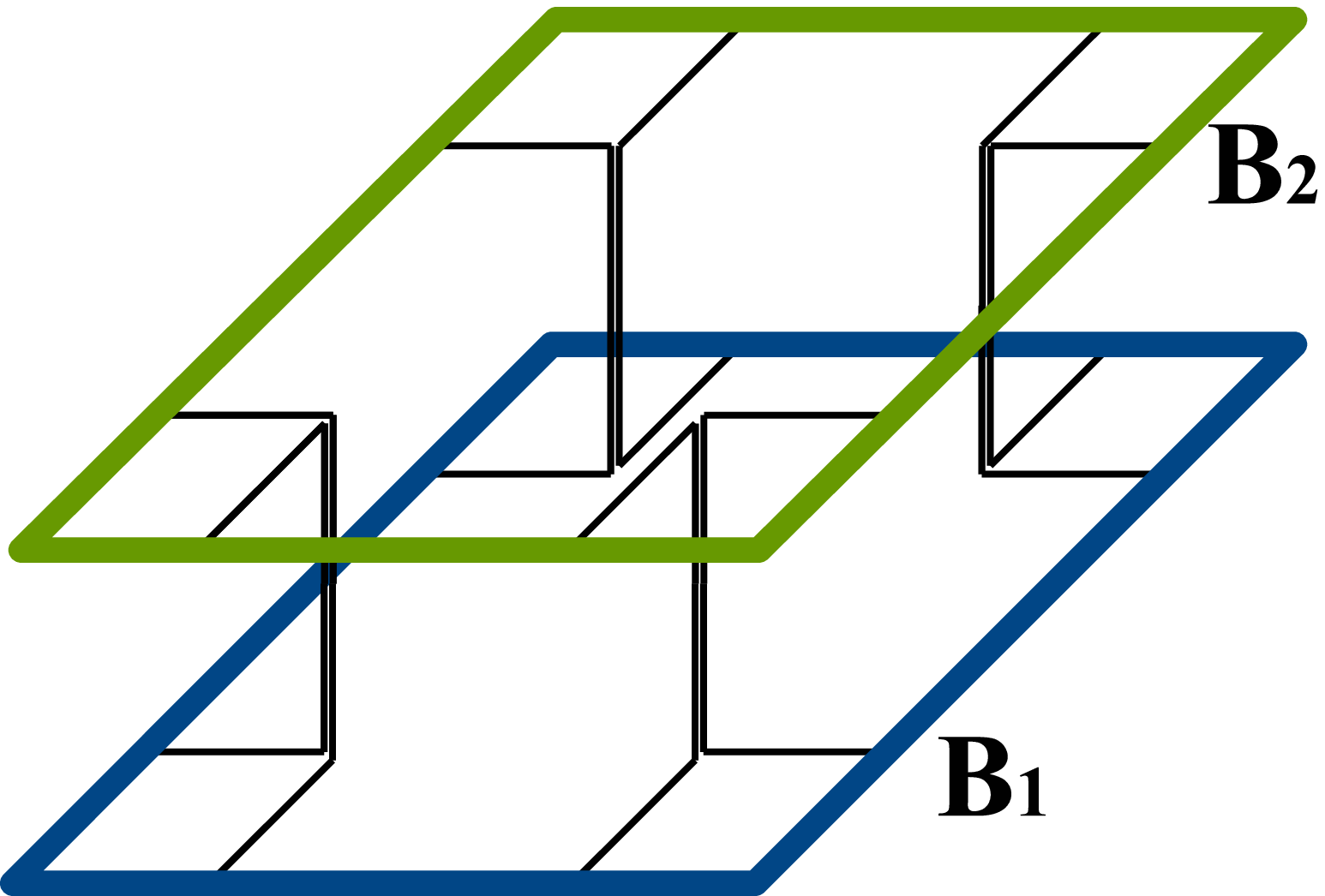} } \times \;
\parbox[c]{0.11\textwidth}{ \includegraphics[scale=0.23]{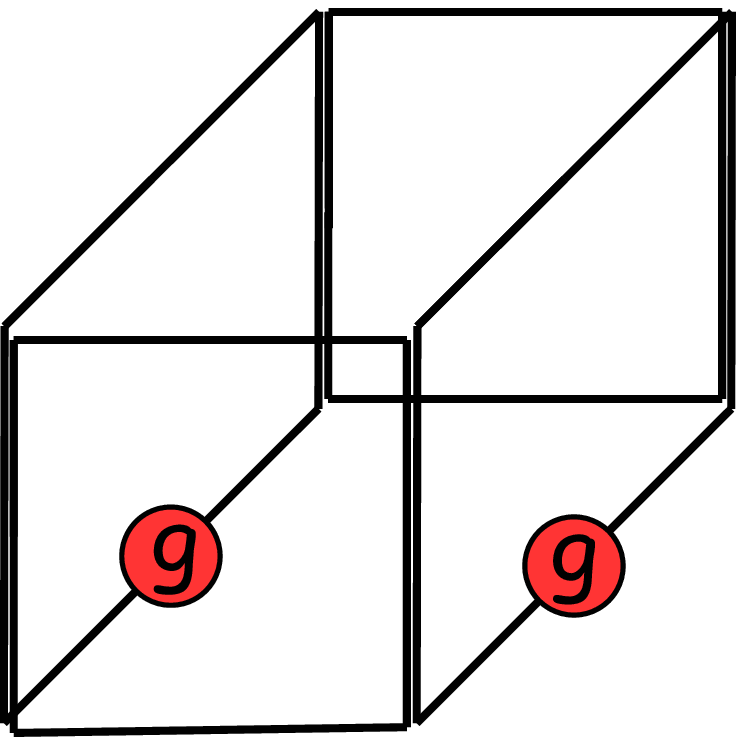} },
\eeq
\end{widetext}
where we have separated the boundary term and the inside term in virtue of the tensor product form of $A^{\rm{res}}_G$ (see Eq.(\ref{eq:restensor})). For the sake of simplicity, we have omitted the sum running over the elements of $G_{{\rm topo}}$ in all tensors forming the construction. The scalar product of Eq.(\ref{eq:scapro}) is proportional to 
\begin{align}
\sum^{G_{\rm topo}}_{a,b,c,d} \mathcal{C}(B_1,B_2,a,b,c,d)&\chi_{L}(a^{-1}b) \chi_{L}(b^{-1}gc) \times \notag\\
& \chi_{L}(c d^{-1}) \chi_{L}(a^{-1}gd), \notag
\end{align}
 where each element of the sum comes from the individual tensors,  $\mathcal{C}(B_1,B_2,a,b,c,d)$ is the boundary term and the traces comes from the loops of the last drawing containing the operator of the string $L_g$. The left regular representation obeys $\chi_{L}(g)=|G|\delta_{e,g}$ so the scalar product computed is proportional to $\sum^{G_{{\rm topo}}}_{k} \mathcal{C} \delta_{k,g}$. Finally we conclude that this value is zero if and only if $g\notin [e]\equiv G_{{\rm topo}}$.
The fact that the state with these operators placed in the virtual d.o.f. is an eigenstate with eigenvalue one of the local hamiltonian $h^{\rm res}$ and that a string of this kind cannot be deformed (using the $G_{\rm topo}$ invariance) through the lattice is equivalent to a string tension. The string tension is manifested in the fact that there exits an energy dependence on the string's length of the excitation. That is:
\begin{equation*} 
H^{\rm res} \left(  |\Psi_l^*(A^{\rm{res}}_G,L_g)\rangle\right) \propto l |\Psi_l^*(A^{\rm{res}}_G,L_g)\rangle,\quad {\rm with}  \; g\notin [e],
\end{equation*}
where $|\Psi_l^*(A^{\rm{res}}_G,L_g)\rangle$ is the state constructed with the tensor $A^{\rm{res}}_G$ and placing a string flux of length $l$ with element $g\in G$ in the virtual d.o.f. When $g$ belongs to $G_{{\rm topo}}$ the scalar product is not zero so we cannot conclude that it is an excitation. In fact the operator in the virtual d.o.f. can be cancelled out using the $G_{{\rm topo}}$-invariance of the tensor and it then corresponds to a string not locally detectable. 

Using the same kind of argument we can verify that the operator, on the virtual d.o.f., corresponding to an end of the string gives rise to an eigenvector of $H^{\rm res}$:
$$h^{\rm res}\; \parbox[c]{0.14\textwidth}{ \includegraphics[scale=0.25]{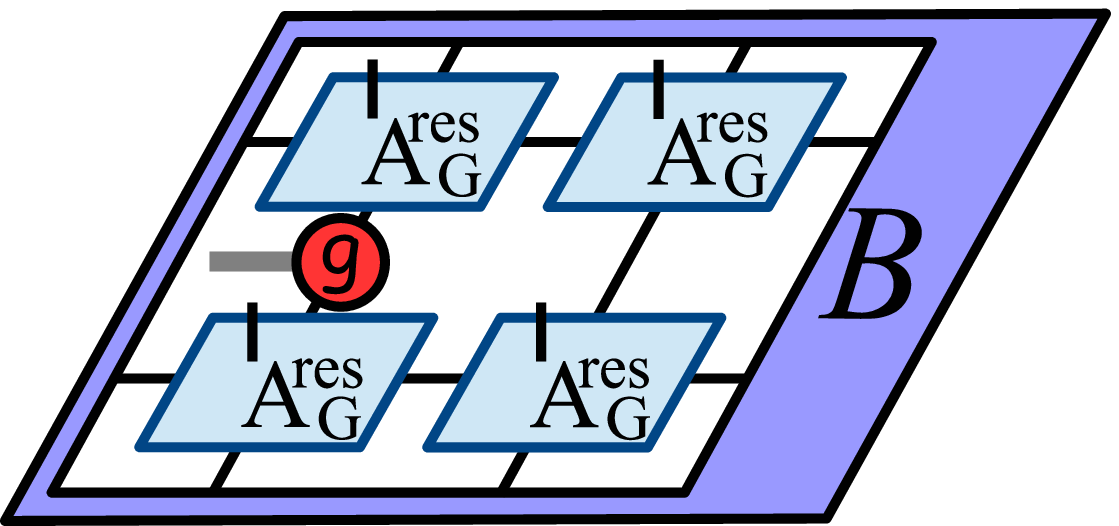}} =\parbox[c]{0.14\textwidth}{ \includegraphics[scale=0.25]{S2x2onedefect.eps}}, g\in G-G_{\rm topo} .$$\\

\begin{tcolorbox}
 The flux excitation string of the state formed with $A_G$ is also an excitation of $A^{\rm{res}}_G$. The fluxes associated with elements of $G$ that do not belong to $G_{{\rm topo}}$ become confined in the $A^{\rm{res}}_G$ model. The string of confined fluxes cannot be extended freely through the lattice of $A^{\rm{res}}_G$ and the energy penalty of the excitation depends on the length of the string. 
\end{tcolorbox}

\vspace{5mm}

This shows that we have two very different flux string excitations in the state formed with the tensor $A^{\rm{res}}_G$. The open string created by placing operators $L_k$ with $k\in G_{{\rm topo}}$ in the virtual d.o.f. has the freedom of being deformed in the entire chain except at the ending plaquettes. Therefore the energy penalty of this excitation comes from the two end points, regardless of the length of the string. As opposed to this case we have the string constructed with elements $g\in G- G_{{\rm topo}}$ which cannot be deformed freely; the operators $L_g$ are confined in its position of the lattice. Therefore the energy penalty of this chain depends on the length of the string. We say that all the flux-type particles of the parent model, conjugacy classes of $G$, which do not belongs to $G_{\rm topo}$ become confined fluxes in the state formed with $A^{\rm{res}}_G$.

The unconfined fluxes of the parent model split into a superposition of fluxes of the restricted model. This is because a conjugacy class of $G$ belonging to $G_{\rm topo}$ can be decomposed into multiple conjugacy classes of $G_{\rm topo}$. Different internal states of the same type of flux of the parent model, $g_1,g_2\in C^G[g] \subset G_{\rm topo}$, should be considered now internal states of a different type of flux of the restricted model, $g_1\in C^{G_{\rm topo}}[g]$, $g_2\in C^{G_{\rm topo}}[g']$ with $C^{G_{\rm topo}}[g] \neq C^{G_{\rm topo}}[g']$.\\

\begin{tcolorbox}
 The energy of the restricted state modified by $S^{\otimes \mathcal{M}}_g$, over a compact region $ \mathcal{M}$, depends on the length of the boundary of the region $ \mathcal{M}$, when $g\in G-G_{\rm topo}$. The virtual representation of this action is a confined loop defect which only depends on the quotient group $G_{\rm sym}\cong G/G_{{\rm topo}}$. 
 \end{tcolorbox}
 \vspace{5mm}
 
\begin{figure}[ht!]
\begin{center}
\includegraphics[scale=0.35]{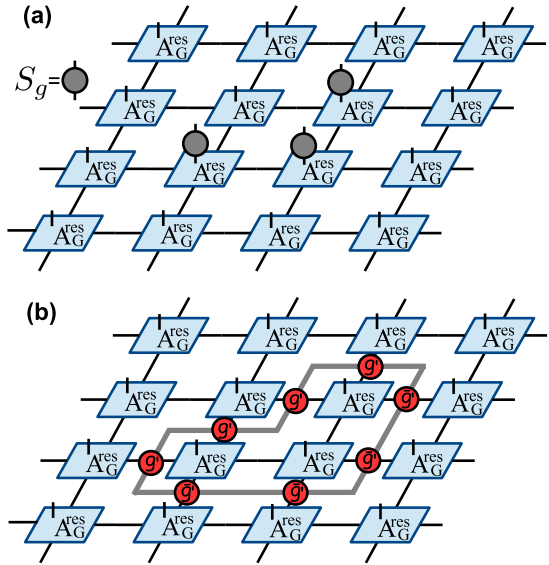}
\caption{(a) We apply the operator $S^{\otimes 3}_g$ on a background of $A^{\rm res}_G$ tensors; the result is depicted on part (b). (b) The confined loop of $L_{g'}$ operators affects 7 plaquettes, corresponding each one to an excitation of the local hamiltonian $h^{\rm res}$}
\label{fig:symconf}
\end{center}
\end{figure}

Let us prove the above statement. The flux confinement and the emergence of a non-trivial representation of the on-site symmetry studied before (see Eq.(\ref{eq:symtensor}) and (\ref{eq:symstate})) are features intimately related. When we apply the operator $S_g$ with $g\in G-G_{{\rm topo}}$ on a connected subset of the lattice ($\mathcal{M}$) in a background of $A^{\rm{res}}_G$, we create a closed loop of $L_{g'}$ virtual operators around the affected region (see Fig. \ref{fig:symconf} for an example). The state with this closed string is an excitation of the parent Hamiltonian of $A^{\rm{res}}_G$ because it is formed with the same virtual operators as the confined fluxes. Then the virtual representation of the unitary $S_g$ over a connected region is a confined closed string. Therefore the energy of the state with the confined closed string depends on the length of the loop:
 \begin{equation}
H^{\rm res} \left(S_g^{\otimes \mathcal{M}}|\Psi(A^{\rm{res}}_G)\rangle\right) \propto |\partial \mathcal{M}| \left(S_g ^{\otimes \mathcal{M}}|\Psi(A^{\rm{res}}_G)\rangle \right),\notag
\end{equation}
where $g\notin [e]$. We notice that we can obtain the same defect acting on the complementary region of $\mathcal{M}$. The fact that these defects cannot be deformed freely through the lattice imply that the state does not remain invariant under the action of $S_g$ over $\mathcal{M}$. Unlike the parent model situation, the action of closed loops of operators on the ground state may increase energy \cite{Kitaev}. 
In order to leave the restricted state invariant, i.e., remove the confined closed string, we have to act over the complementary of $\mathcal{M}$. That is, we need to act on the whole lattice to preserve the state, ending up with a global on-site symmetry of the group $G_{\rm sym}$. 
The $G_{{\rm topo}}$-invariance of the tensors $A^{\rm{res}}_G$ does not change the coset, of $G$ by $G_{{\rm topo}}$, the confined loop belongs to. This leads us to conclude that the virtual representation of the symmetry is of  the group $G_{{\rm sym}}\cong G/G_{{\rm topo}}$.

\subsection{Comparing pure charge excitations of the parent and restricted models: condensation}

Let us now study the properties of the charge excitations of $A_G$ when they are placed on a background formed by $A^{\rm{res}}_G$ tensor. As explained in Section \ref{sec:mathTNS}, the charge-type excitations of the Quantum Double of $G$ are identified with the irreps of the group $G$. More precisely, charge excited states can be given a \textsf{PEPS} representation, starting from that of the $G$-isometric \textsf{PEPS}, and inserting operators $\sum^G_g\chi_\nu(pg)|g\rangle\langle g|$ on some virtual d.o.f., where $\chi_\nu$ is the character of the irrep $\nu$ of $G$  \cite{PEPSdescp}. Unlike flux excitations, charge particles are not associated with a string (in this \textsf{PEPS} representation) and they are point defects in the lattice constructed with $G$-isometric \textsf{PEPS}. Therefore, although they are always created in pairs, we can safely just consider how one particle of the pair affects the background. These operators also correspond to excitations when they are placed in the state constructed with $A^{\rm{res}}_G$. In order to prove it we compute the scalar product between a state with this operator in the virtual d.o.f. and one arbitrary element of $\mathcal{S}_{2\times2}$ of $A^{\rm{res}}_G$ as we did in the case of flux excitations. We obtain a result proportional to 
\begin{align}
\sum^{G_{{\rm topo}}}_{a,b,c,d} \mathcal{C}(&B_1,B_2,a,b,c,d)  \chi_{L}(a^{-1}b)\chi_{L}(bc^{-1})\times \notag\\
&  {\rm tr} [L_{c  d^{-1}}  \sum^G_g\chi_\nu(pg)|g\rangle\langle g| ]\chi_{L}(ad^{-1}), \notag
\end{align}
which reduces to
\begin{align}
{\rm tr}[  \sum^G_g\chi_\nu(pg)|g\rangle\langle g| ]=& \sum^G_g\chi_\nu(pg)=\sum^G_g\chi_\nu(g)= \notag\\
&\sum^G_g\chi_\nu(g)\chi_1(g)=|G|\delta_{\nu,1} \notag 
\end{align}
Therefore the restricted state with one charge operator placed on the virtual d.o.f. is orthogonal to the ground state of the parent Hamiltonian associated with $A^{\rm{res}}_G$. As for fluxes above, one can further prove that it is an eigenstate with eigenvalue 1 of the local parent Hamiltonian and then, as there is no string associated, of the parent Hamiltonian.

Similarly, with each irrep $\sigma$ of $G_{\rm topo}$, a charge operator acting on the $G_{\rm topo}$-isometric \textsf{PEPS} can be associated: $\sum^{G_{\rm topo}}_k\chi_{\sigma}(qk)|k\rangle\langle k|$ where $\chi_{\sigma}$ is the character of $\sigma$ and where $\{\ket{k}: k=1 \ldots |G_{\rm topo}|\}$ denote a basis of the space where the left regular representation of $G_{\rm topo}$ lives, $\mathbb{C}[G_{\rm topo}]$. The corresponding charge operator in the $A^{\rm{res}}_G$ \textsf{PEPS} is $\sum^{G_{\rm topo}}_k\chi_{\sigma}(qk)|k\rangle\langle k| \otimes \mathbb{I}^{G_{\rm sym}}$ in analogy with the flux operator $L_k=L^{G_{\rm topo}}_k\otimes \mathbb{I}^{G_{\rm sym}}$ (acting on $\mathbb{C}[G_{\rm topo}]\otimes \mathbb{C}[G_{\rm sym}]$). We rewrite this operator and we denote it as follows: 
\begin{equation}\label{eq:defQres}
Q^{\sigma}_q=  \sum^{G_{\rm topo}}_k\chi_{\sigma}(qk)  \left( \sum_{[g]}^{G_{\rm sym}}|k,[g]\rangle\langle k,[g]| \right).
\end{equation}
Placing $Q^{\sigma}_q$ on an $A_G^{\rm res}$ background, we get an excited eigenstate of the parent Hamiltonian of $A_G^{\rm res}$.

The confinement of some fluxes is intimately related to the condensation of charge particles. An anyon condensation is a situation where a topological excitation cannot be distinguished from the vacuum with topological interactions (i.e. using braiding operations). Since some fluxes are confined in the $A_G^{\rm res}$ background, there are less fluxes 'available' (the remaining unconfined) to braid with the charge particles to topologically distinguish amongst them.
Let us consider an elementary charge excitation of the parent model and we try to identify the class of this charge in the restricted model. To perform this experiment we have to create a charge-pair excitation belonging to the class of the irrep $\nu$ of $G$, braid one charge of the pair with a flux characterized by the element $k\in G_{\rm topo}$ and then fuse the charge modified by the braiding with the other charge of the pair. The probability the charge pair fuses back to the vacuum is given by \cite{Preskillnotes, PEPSdescp} 
\begin{equation}
{\rm Prob(vacuum)}=\left | \frac{\chi_\nu(k)}{|\nu|} \right |^2.\notag
\end{equation}
This interferometric process can be conceived as a method to identify the irrep associated with the charge particle. To completely identify the irrep $\nu$ of $G$ with an $A_G^{\rm res}$ background, we would need to braid charges with confined fluxes also. We forbid this operation because we are restricting to topological interactions; we do not allow  processes whose energy cost depend on lengths. If two irreps of $G$ are the same when restricted to elements of $G_{\rm topo}$; they have to be consider equivalent in the $A_G^{\rm res}$ background. Then if one irrep, say $\mu$, is the identity for all elements of $G_{{\rm topo}}$, we obtain that the probability to fuse it with the vacuum is one. Therefore this irrep associated to a charge is not modified by the braiding of any of the unconfined fluxes, and we now have to identify it with the trivial topological charge (the vacuum); we will call this phenomenon 'charge condensation'. 

We now claim that there is always a charge excitation of the parent model that is condensed in the $A_G^{\rm res}$ background. This charge excitation does not need to be elementary, i.e. associated with an irrep of $G$. It can be a composite of elementary charges (related to a reducible representation of $G$). We know that given any normal subgroup $G_{\rm topo}$ of $G$ there is always a representation $\rho$ of $G$ such that the kernel of the corresponding character is exactly $G_{\rm topo}$ \cite{Char}. Therefore, if we perform an interferometric experiment with $\rho$ we obtain
\begin{equation}
{\rm Prob(vacuum)}=\left | \frac{\chi_\rho(k)}{|\rho|} \right |^2=1, \notag
\end{equation}
for all $k\in G_{\rm topo}$, that is, the charge associated to $\rho$ has condensed.

Also the charges of the model $A_G$ split into a superposition of charges of the group $G_{{\rm topo}}$. As mentioned before, we can only allow to braid them with the fluxes corresponding to the elements of $k\in G_{\rm topo}\subset G$ and the irrep of $G$ restricted to these elements becomes:
$$\chi^G_{\nu}(k)\cong\sum_{\sigma} m_{\sigma} \chi^{G_{{\rm topo}}}_{\sigma}(k),$$
where $m_{\sigma}$ is the multiplicity of the irrep $\sigma$ of $G_{{\rm topo}}$ that appears in the decomposition of  the irrep $\nu$ of $G$.

\subsection{Symmetry action on anyons and symmetry breaking}

\begin{figure}[ht!]
\begin{center}
\includegraphics[scale=0.4]{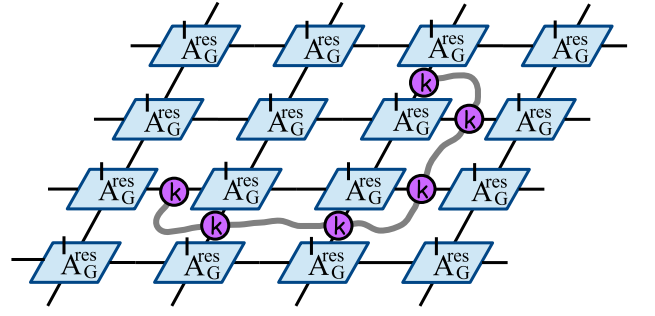}
\caption{A subset of the lattice with an open string of virtual operators $L_k^G$, representing a flux excitation.}
\label{fig:openstringk}
\end{center}
\end{figure}

We analyze here the effect of the symmetry operators over fluxes and charges of the restricted model. We will see how the symmetry permutes the different ground state sectors in the torus.
We start with an open string of operators $L_k$ with $k\in G_{{\rm topo}}$ on an $A^{\rm{res}}_G$ background. We illustrate this setup on Fig. \ref{fig:openstringk}.
This corresponds to a flux excitation of the model, where the flux string can be deformed freely by the $G_{{\rm topo}}$-invariance of the tensors (except at the plaquette where the string terminates). Now we apply the operator $S_g, g \in G$ at each lattice site, which corresponds to a symmetry of the state $|\Psi(A^{\rm{res}}_G)\rangle$, on the lattice with a string flux excitation. It is clear that the fixed end points initially in the element $k$ with $k\in G_{{\rm topo}}$ is sent to the element $k'$ where $k'=gkg^{-1}$ of $G_{{\rm topo}}$.

\begin{figure}[ht!]
\begin{center}
\includegraphics[scale=0.4]{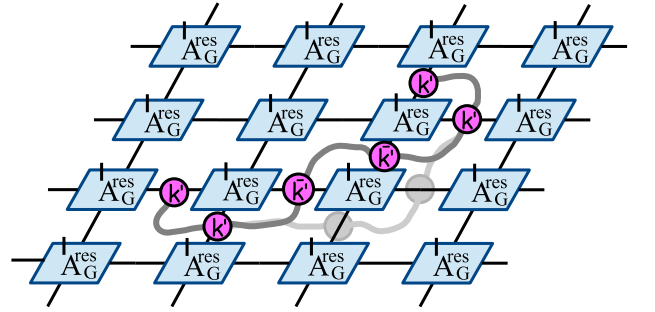}
\caption{The previous flux excitation modified by the action of the symmetry operator $S^N_g$. The virtual operators defining the excitation are now $L_{k'}$ with $k'=gkg^{-1}$. Then, the flux-type particle belongs now to the conjugacy class $C[k']$ which represent a change in type of particle after the action of the symmetry.}
\label{fig:openstringk2}
\end{center}
\end{figure}
Let us notice that not only the end points are modified by the symmetry operators, the elements of the chain are also sent to $k'\in G_{\rm topo}$ and it also can be moved freely through the lattice as it is shown in Fig. \ref{fig:openstringk2}. Therefore the symmetry operators also correspond to a symmetry in the space of flux excitations sending the conjugacy class $C[k]$ of $G_{{\rm topo}}$ to a different conjugacy class $C[k']$ with $k'=gkg^{-1}$ when $G$ is non-abelian. This operation, which we will denote as the map $\rho_g(k) = gkg^{-1} $, corresponds to a representation of the group $G_{{\rm sym}}$ which permutes the flux type particles (conjugacy classes of $G_{{\rm topo}}$). Let us prove this. Given a conjugacy class of $G_{\rm topo}$ the action $\rho_g$ only depends on the coset $[g]$ of $g$. Let us take $k$ and $g'\in [g]$, we can write $g'=k'g$ for some $k'\in G_{\rm topo}$ and then $\rho_{g'}(k)=g'kg'^{-1}= k'\rho_g(k) k'^{-1}$. Also if $\tilde{k}\in C[k]$ it follows that $\rho_{g'}(\tilde{k})=g'\tilde{k}g'^{-1}= g'k_1k k^{-1}_1g'^{-1}= k_2g'k g'^{-1} k^{-1}_2= k_2 k'\rho_g(k) (k_2k')^{-1}$ for some $k_1,k_2\in G_{\rm topo}$. In the same way it can be shown that $\rho_g \circ \rho_h(k')$ belongs to the same conjugacy class as $\rho_{gh}(k)$ for any $k'\in C[k]$. We point out that the particle-type of the flux, i.e the conjugacy class of $G_{\rm topo}$, can be interferometrically detected and it can also be measured in the plaquette using the \textsf{PEPS} representation \cite{PEPSdescp}. The permutation of restricted fluxes is related to the splitting of the parent fluxes; the permuted fluxes are those coming from the same parent flux and they appear in its splitting. 

The previous construction depends on the group extension Eq.(\ref{eq:exactseq}) chosen, as explained in Appendix \ref{ap:ext}. When the normal subgroup $G_{{\rm topo}}$ of $G$ is abelian, the group and the conjugacy classes are equivalent, then the action $\rho$ is a homomorphism from $G_{\rm sym}$ to Aut$(G_{\rm topo})$. If $G_{{\rm topo}}$ is non-abelian the action is a homomorphism from $G_{\rm sym}$ to Out$(G_{\rm topo})$. \\

We now place one charge particle (see Eq.(\ref{eq:defQres})), corresponding to the irrep $\sigma$ of $G_{\rm topo}$, on the lattice. Then we act with $S_g, g \in G$ at each lattice site. The effect on the charge operator is the following:
\beq\label{eq:charge-sym}
L_{g'} Q^{\sigma}_q L^{\dagger}_{g'}= \sum^{G_{\rm topo}}_k\chi_{\sigma}(qk)  \left( \sum_{[h]}^{G_{\rm sym}} L_{g'}|k,[h]\rangle\langle k,[h]|L^{\dagger}_{g'} \right).
\eeq
We now use that $G$, as a set, can be rewritten as a cartesian product and the multiplication can be expressed in the following form
$$ L_{g'}|k,[h]\rangle =|k_g\rho_{g}(k) \omega([g],[h]),[g][h]\rangle,$$
where we have denoted $|g'\rangle$ as $|k_g,[g]\rangle$ and $\omega([g],[h])\in H^2(G_{\rm sym},G_{\rm topo})$. We can rewrite (\ref{eq:charge-sym}) as
\begin{align}
\sum^{G_{\rm topo}}_k\chi_{\sigma}(q \rho_{g^{-1}}(k_g^{-1}k))  \Big( \sum_{[h]}^{G_{\rm sym}} & |k \omega([g],[h]),[g][h]\rangle \notag \\
&\langle k \omega([g],[h]),[g][h] | \Big).\notag
\end{align}
We now use that $q \rho_{g^{-1}}(k_g^{-1}k)= \rho_{g^{-1}}[ \rho_{g}(q) k_g^{-1}k ]= g^{-1}[ \rho_{g}(q) k_g^{-1}k ] g$. Then 
$$ \chi_{\sigma}(q \rho_{[g]^{-1}}(k_g^{-1}k))=   \chi_{\sigma}( g^{-1}[ \rho_{[g]}(q) k_g^{-1}k ] g). $$
Clifford's Theorem \cite{Clifford} establishes that given an irrep $\Pi_{\sigma}(k)$ of $G_{\rm topo}$, the operator $\Pi_{\sigma}(g^{-1}kg)$ corresponds to $\Pi_{\sigma'}(k)$ where $\sigma'=\sigma' (\sigma, g^{-1})$ is another irrep of $G_{\rm topo}$. Therefore the action of the symmetry operator includes a permutation of the particle type of the charge:
\begin{align}
L_{g'} Q^{\sigma}_q L^{\dagger}_{g'}=&
\sum^{G_{\rm topo}}_k\chi_{\sigma'}( \rho_{g}(q) k_g^{-1}k ) \times \notag \\
 \sum_{[h]}^{G_{\rm sym}} & |k \omega([g],[h]),[g][h]\rangle  \langle k \omega([g],[h]) ,[g][h]|.\notag
\end{align}
If the cocycle is trivial $L_{g'} Q^{\sigma}_q L^{\dagger}_{g'}= Q^{\sigma' }_{\rho_{g}(q) k_g^{-1}}$. 
We notice that the action of the symmetry operator has been analyzed only on one charge of the pair. It can be shown that the symmetry changes similarly the associated conjugate charge, allowing to detect physically the permutation of the particle type \cite{PEPSdescp}. We can also show the following:
\beq \label{eq:SF}
L_{g'}L_{h'} Q^{\sigma}_q L^{\dagger}_{g'}  L^{\dagger}_{h'}=  L_{gh'} Q^{\sigma}_{qk_{gh}k_g^{-1} k_h^{-1} \bar{\omega}([g],[h])} L^{\dagger}_{gh'}, 
\eeq
when $\rho$ is trivial. The virtual action of the operators $S_gS_h$ and $S_{gh}$ over $Q^{\sigma}_q$ is related by a braiding to the flux $\omega([g],[h])$ up to gauge redundancies ($k_g k_hk_{gh}^{-1}$). Eq.(\ref{eq:SF}) evokes the phenomenon of symmetry fractionalization discussed in \cite{SETQ}.
Let us summarize:\\

\begin{tcolorbox}
The symmetry operators of the ground space also correspond to a symmetry on the space of quasiparticles excitations. The symmetry action over charges and fluxes is a representation of $G_{\rm sym}$ and it permutes the particles types. The action on charges also depends on a cocycle of $H^2( G_{\rm sym},G_{\rm topo})$. These two functions determine the extension group $G$.
\end{tcolorbox}
\vspace{5mm}

\begin{figure}[ht!]
\begin{center}
\includegraphics[scale=0.35]{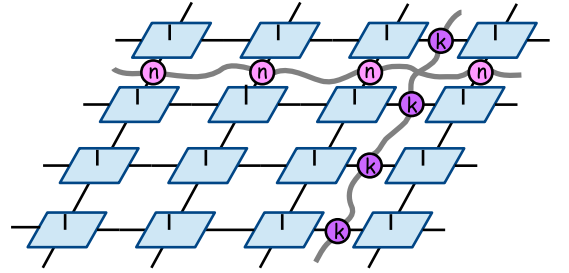}
\caption{The ground state corresponding to the pair conjugacy class of $(k,n)$ is depicted.}
\label{fig:GStorus1}
\end{center}
\end{figure}

Let us now study the effect of the global symmetry on the ground space of the restricted model. The parent Hamiltonian of $A_{G}^{\rm res}$ has a degenerate ground space on the torus. This subspace is spanned by the states constructed with the tensor $A_{G}^{\rm res}$ and placing $(k,n)$-closed boundary condition, where $kn=nk$ \cite{PEPSdescp}. This boundary condition is constructed placing a tensor product string of $L_n$ operators in the horizontal part of the boundary and a string with $L_k$ operators in the vertical boundary and then closing the boundaries periodically, as illustrated in Fig. \ref{fig:GStorus1}.

The number of different inequivalent closures corresponds to the different anyons sectors in the quantum double of $G_{\rm topo}$. The equivalence relation of closures is given by $(k,n)\sim (k',n')$ if there exists $t\in G_{\rm topo}$ such that $(k,n)= (tk't^{-1},tn't^{-1})$. Let us denote each ground state constructed with the tensor $A_{G}^{\rm res}$ and with closure type $(k,l)$ satisfying $kn=nk$ as $|\Psi(A_{G}^{\rm res}|(k,n))\rangle$. If we apply the symmetry operator $S_g$, where $g\in G$, at each lattice site we obtain:
\begin{equation}
S^{\otimes N}_g|\Psi(A_{G}^{\rm res}|(k,n))\rangle=|\Psi(A_{G}^{\rm res}|(g'kg'^{-1},g'ng'^{-1}))\rangle,\notag
\end{equation}
which also belongs to the ground space because $k'n'=n'k'$ where $k'=g'kg'^{-1}$ and $n'=g'ng'^{-1}$. Therefore the class $(k,l)$ is sent to $(\rho_g(k),\rho_g(n))$, this action is a representation of $G_{\rm sym}$ that acts as a permutation of the different ground states. 

\subsection{ An example: $\mathbf{G_{topo}= \mathbb{Z}_n }$, $\mathbf{G_{sym}= \mathbb{Z}_2 }$}
Let us consider the dihedral group $G=D_{2n}$, with $n$ odd, $\{ r,s | r^n=s^2=e, sr^ks=r^{-k}\}$. There are $1+1+(n-1)/2$ conjugacy classes: $[e], [s]=\{sr^{k}\}_{k=0}^{n-1}, [r^k]=\{r^k,r^{n-k}\}\;k=1,\ldots, n-1$ which correspond to fluxes in the quantum double of $D_{2n}$: $QD(D_{2n})$. There are two one-dimensional representation of $D_{2n}$; the trivial one and the irrep $Z$ given by $Z(r^k)=1$ and $Z(sr^{k})=-1 \; \forall k=0,\ldots,n-1$. There are $n-1$ two-dimensional irreps $\Pi_\nu$, labelled by $\nu=1,\cdots, n-1$ given by
\beq
\Pi_\nu(r^k)=\left(\begin{array}{cc}q^{ k\nu}& 0 \\ 0 & q^{- k\nu} \end{array}\right), \; \Pi_\nu(s)=\left(\begin{array}{cc}0&1 \\ 1 & 0 \end{array}\right),\notag
\eeq
where $q=e^{\frac{2\pi i}{n}}, \; k=0,\cdots, n-1$.
The normal subgroup $\{ r^k \}_{k=0}^n$ of $D_{2n}$ is isomorphic to the abelian group $\mathbb{Z}_n$. The fluxes of $QD(\mathbb{Z}_n)$ are the $n$ conjugacy classes given by single elements: $\{r^k\}$ for $k=0,\ldots,n-1$. The charges correspond to the $n$ one-dimensional irreps, given by $\pi_\sigma(r^k)= q^{k\sigma}$ where $\sigma=0,\ldots,n-1$. The restriction of fluxes from $D_{2n}$ to $\mathbb{Z}_n $ can be associated to a confinement of the flux $[s]$ and a condensation of the charge $Z$ (because $Z(r^k)=1$) of $QD(D_{2n})$. The splitting of charges of $QD(D_{2n})$ is given by $\Pi_\nu \cong \pi_\sigma \oplus \pi_{-\sigma}$ for $\nu=\sigma$. This example is also shown in \cite{HopfBais}. The associated global symmetry comes from the quotient group $\mathbb{Z}_2\cong D_{2n} / \mathbb{Z}_n$. The non-trivial action of the symmetry on the anyons of $QD(\mathbb{Z}_n)$ is given by:
\beq
r^k\mapsto sr^ks=r^{-k}, \; \pi_\sigma(r^k)\mapsto  \pi_\sigma(sr^ks)=  \pi_{-\sigma}(r^k). \notag
\eeq
This example corresponds to the non-trivial extension group of $\mathbb{Z}_n$ by $\mathbb{Z}_2$. One also can consider the restriction from $QD(\mathbb{Z}_n\times \mathbb{Z}_2 )$ (the trivial extension) to $QD(\mathbb{Z}_n)$ where the corresponding condensation and confinement can be identified \cite{HopfBais} but the symmetry action is trivial on the anyonic sector.

\section{The 1D \textsf{SPT}  classification} \label{sec:phasesMPS}

We now wish to show how the notion of group extension discussed in the previous section is also useful to analyse 1D systems with symmetries. Namely, it allows for a transparent derivation of the classification of 1D phases in the \textsf{MPS} formalism. We start with a short summary of this classification \cite{phasesWen,phasesMPS, phasesKitaev}. Our main focus will be on symmetry protected topological (\textsf{SPT}) phases, that is, phases that remain distinct if we impose that the paths that connect them should respect a given symmetry. An example of a system with non-trivial \textsf{SPT} phase is the celebrated spin-1 Haldane chain \cite{Haldane}.   

\subsection{Overview of \textsf{SPT}  classification in 1D} \label{secclass}
Formally, two systems are said to be in the same quantum phase if they can be connected by a smooth path of Hamiltonians, which is local, bounded-strength and uniformly gapped \footnote{One requieres that the gap of the Hamiltonian along the path was uniformly lower bounded by a constant in the size of the system. This requirement ensures that the gapped is preserved in the thermodynamic limit.}. Along this path the physical properties of the system will change smoothly. If at some point the gap closes, it will result in a change of the global properties and usually a phase transition will occur. When a symmetry is imposed on Hamiltonians and the paths connecting them, phase diagrams become richer; two systems are then said to be in the same phase if the previous path exists and if, moreover, there exists a representation of the symmetry which commutes with the Hamiltonian along the entire path. 

The classification of quantum gapped phases is restricted to the task of classifying \textsf{MPS}. This is because it has been proven 
that the family of \textsf{MPS} approximate efficiently ground states of gapped quantum Hamiltonians \cite{Hast1,Hast2,Arad}. And for any of these ground states the associated parent Hamiltonian can be constructed  (see Section \ref{sec:mathTNS}). The classification is restricted to the so-called {\it isometric form} of an \textsf{MPS}: those \textsf{MPS} which are renormalization fixed points. In \cite{phasesMPS} a gapped path of Hamiltonians is built connecting {\it any} \textsf{MPS} with its corresponding isometric form. Therefore, the task of classifying phases is restricted to the classification of isometric \textsf{MPS} without loss of generality. The final step is to identify the obstructions to design gapped paths of Hamiltonians between the different isometric forms. 

The main conclusions of \cite{phasesMPS} are: (i) without symmetries, all systems with the same ground state degeneracy are in the same phase, where the representative states are the product state for the unique ground state case and the GHZ state for the degenerate case. (ii) When on-site linear symmetries are imposed to the systems, the different phases are classified, in the unique ground state case, by the second cohomology group $H^2(G,U(1))$ of  the symmetry group $G$ over $U(1)$. This classification is best understood if one considers the virtual d.o.f. of the \textsf{MPS}: the unitary representation $u_g$ realising the physical global symmetry translates into an action $V_g \otimes V^{\dagger}_g$ on these virtual d.o.f., where $V_g$ is a projective representation of $G$ (see Appendix \ref{ap:coho}). When a symmetry is imposed, the possible phases that can be obtained are labeled by the equivalence classes of the representation $V_g$; $H^2(G,U(1))$ precisely identifies them. 

In the non-injective case, the \textsf{MPS} map $\mathcal{P}$ of a system with degenerate ground state is supported on a "block-diagonal" space
\begin{equation}
\mathcal{H}=\bigoplus^{\mathcal{A}}_{\alpha=1} \mathcal{H}_{\alpha},\notag
\end{equation}
which is the known block structure of the matrices forming the tensor of the \textsf{MPS}. The phases are determined by a representation of the symmetry group in terms of permutations between the blocks of the \textsf{MPS} (see Section \ref{sec:mathTNS}), and the $\mathcal{A}$-fold cartesian product of $H^2(G,U(1))$ with itself.

The map $\mathcal{P}$ is injective when restricted to each subspace $\mathcal{H}_{\alpha}$, so a unique canonical form can be constructed allowing to characterize the action of symmetries.

An on-site global symmetry of an \textsf{MPS} under a linear unitary representation of the group $G$, which we will call $u_g$, is given by the following action on the virtual d.o.f. of the tensor \footnote{The complex phases $\theta_\alpha$ that appear in Eq.(\ref{eq:MPSsym}) do not play any role in the classification because we are restricting ourselves to finite group representations; these phases are therefore rational, and can be absorbed under blocking \cite{phasesMPS}.  }:

\begin{equation}\label{eq:symblocks}
u_g\mathcal{P}= \mathcal{P}\left ( P_g\left [ \bigoplus^{\mathcal{A}}_{\alpha=1}V^{\alpha}_g\otimes \bar{V}^{\alpha}_g\right] \right).
\end{equation}

The operators $\{ V^{\alpha }_g: g \in G \}$ form a (projective) unitary representation acting on $\mathcal{H}_{\alpha}$, as in the case of a unique ground state. The operators $\{ P_g: g \in G \}$ form a representation of $G$ that acts as a permutation between the subspaces $\mathcal{H}_{\alpha}$. This representation is in general reducible: $\{P_g \mathcal{H}_{\alpha}: g \in G\} \neq \mathcal{H}$. As a result, the subspaces $\mathcal{H}_\alpha$ can be lumped into larger subspaces of $\mathcal{H}$, $\mathcal{H}_{\mathfrak{a}}$, that are irreducible under the action of $G$: $\{P_g \mathcal{H}_{\mathfrak{a}}: g \in G\}= \mathcal{H}_{\mathfrak{a}}$. From the splitting $\mathcal{H}=\bigoplus_{\mathfrak{a}} \mathcal{H}_{\mathfrak{a}}$, a decomposition of the operators $P_g$ into irreducible representations $P_g^{\mathfrak{a}}$ can be derived, which we can use to re-express Eq.(\ref{eq:symblocks}): 
\begin{equation}\label{eq:symblocks_2}
u_g\mathcal{P}=\mathcal{P}\left (\bigoplus_{ \mathfrak{a} }P^{\mathfrak{a}}_g\left [ \bigoplus_{\alpha\in \mathfrak{a} }V^{ \alpha }_g\otimes \bar{V}^{\alpha }_g\right]\right).
\end{equation}
Interestingly, the decomposition in terms of $\mathcal{H}_{\mathfrak{a}}$ is generally unstable under perturbations, even those that preserve the symmetry \cite{phasesMPS}. 

We now study each summand $P^{\mathfrak{a}}_g\left ( \bigoplus_{\alpha\in \mathfrak{a} }V^{\alpha }_g\otimes \bar{V}^{\alpha }_g\right)$ of Eq.(\ref{eq:symblocks_2}) and explain how it relates to the concept of \emph{induced} representation \cite{Simon}. For a fixed summand index $\mathfrak{a}$, we pick a reference block $\alpha_0\in \mathfrak{a}$ and we define the subgroup:
$$ H:=\{h\in G : \; P^{\mathfrak{a}}_h (\mathcal{H}_{\alpha_0})=\mathcal{H}_{\alpha_0}  \}\subset G.$$
We can split $G$ in disjoint cosets $k_{\beta}H$ labelled by the blocks $\beta \in \mathfrak{a}$ for a (non-unique) choice of $k_{\beta}\in G$ chosen such that $P^{\mathfrak{a}}_{k_{\beta}} (\mathcal{H}_{\alpha_0})=\mathcal{H}_{\beta}$ (let us notice that this is possible because $P_g$ is irreducible in this subset $\mathfrak{a}$). Then we can use that for every $g$ and $\alpha$, there exist unique $h\in H$ and $\beta$ such that
\begin{equation}\label{eq:galphah}
g k_{\alpha}=k_{\beta}h.
\end{equation}
It can be proven that the action on each summand is unitarily equivalent to
\begin{equation}
P^{\mathfrak{a}}_g\left ( \bigoplus_{\alpha \in \mathfrak{a} }V^{\alpha_0 }_h\otimes \bar{V}^{\alpha_0 }_h\right),\notag
\end{equation}
where $h$ is determined by $g,\alpha$ in Eq.(\ref{eq:galphah}). One can also show that two systems are in the same phase if the permutation representation $P_g$ are the same and if for each irreducible subset $\mathfrak{a}$, the projective representation $V^{\alpha_0 }_h$ has the same cohomology class. Since the permutation is effectively encoded in $H$, a phase is characterized by the choice of $H$ together with one of its cohomology classes.

\subsection{Non-trivial virtual representation from restriction in \textsf{MPS}}

\begin{figure}[ht!]
\begin{center}
\includegraphics[scale=0.3]{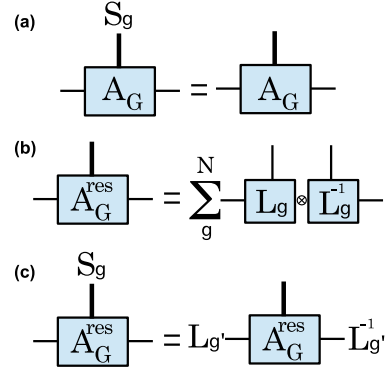}
\caption{(a) The local symmetry of the tensor $A_G$ is represented. (b) The tensor $A^{{\rm res}}_G$ of equation (\ref{redtens}) is depicted. The physical Hilbert space is decomposed into a tensor product form. (c) The action of the operators $S_g=L_g\otimes L_{g^{-1}}$ is translated into the virtual d.o.f. with a freedom on the choice of the element $g'$ within coset $[g]$.} 
\label{Aresym}
\end{center}
\end{figure}

Let us start with a $G$-isometric \textsf{MPS} with tensor $ A_G=|G|^{-1}\left (\sum_{ g \in G}  L_g\otimes L^{\dagger}_g \right )$, where $L_g$ denotes again the left regular representation of $G$ \cite{PEPSdescp}. This \textsf{MPS} has the local symmetry (see Fig. \ref{Aresym}(a)):
\begin{equation}
(L_g\otimes L^{\dagger}_g)_p A_G \; =A_G (L_g\otimes L^{\dagger}_g)_v =A_G \quad \forall g \in G. \notag
\end{equation}
Its parent Hamiltonian has a degenerate ground subspace whose dimension is equal to the number of inequivalent irreps of $G$. In analogy to the 2D case, we wish to study the restricted tensor
\begin{equation} \label{redtens}
 A^{{\rm res}}_G=\frac{1}{|N|}\left (\sum_{ g \in N}  L_g\otimes L^{\dagger }_g \right ), 
\end{equation} 
where $N$ is a normal subgroup of $G$. Unlike the 2D case, there is no topological content associated with $N$.

The operators $\{S_g=L_g\otimes L_{g^{-1}}: g\in G\}$ represent a global on-site symmetry of both periodic boundary condition \textsf{MPS} constructed with $A_G$ and $A^{{\rm res}}_G$, as shown in Fig. \ref{Aresym}. As in 2D (see Eq.(\ref{eq:symtensor})), these operators no longer represent a local symmetry of the tensor $A^{{\rm res}}_G$ for all the elements $g\in G$. In fact, $S_g$ is a local symmetry of $A^{{\rm res}}_G$ if and only if $g \in N$. 

Given $g,g' \in G$, as already seen in the more general 2D case, whenever $g N=g' N$, the actions of $S_g$ and $S_{g'}$ on $|{\mathcal{M}(A^{{\rm res}}_G)} \rangle$ are identical; $S_g$ is actually a representation of $G_{{\rm sym}}\cong G/N$. Part of the local symmetry of the state $|\mathcal{M}(A_G)\rangle$ is degraded to a global symmetry in the state $|\mathcal{M}(A^{\rm{res}}_G)\rangle$. The local $G$-symmetry is reduced to a local $N$-symmetry plus a global on-site $G_{{\rm sym}}$ symmetry. Applying $S_g$ on $A^{\rm{res}}_G$ translates into a \emph{non}-trivial representation of $G_{{\rm sym}}$ on the virtual d.o.f., in contrast to applying $S_g$ on $A_G$. 

We now study the effect of $S_g$ on $A^{{\rm res}}_G$. To do so, we will exploit the block diagonal structure of $A^{{\rm res}}_G$ and we will express $\{ L_g: g \in G \}$ as its direct sum decomposition in terms of irreps of $G$. It will turn out that the block structure of the virtual matrices of $A^{{\rm res}}_G$ is related to the irreps of $N$. For our purposes, we will be led to study how, given a proper normal subgroup $N \subset G$, and an irrep of $G$, $\Pi_\nu$, the irreps of $N$ contained in $\Pi_\nu$ gives a particular structure to the matrices $\Pi_\nu$. This issue has been analysed by Clifford in \cite{Clifford}. Using his results, we will: 1) obtain all the possible phases in 1D with symmetries and degenerate ground state, 2) show how the notion of induced representation appears naturally in the 1D phase classification, 3) exhibit an explicit method to construct the state and operators of each phase 4) associate the restriction $G \to N$ to an appealing physical mechanism in 2D (see previous section). 

To begin with, let us write $A_G$ a bit more explicitly: 
\begin{equation}\label{eq:generictensorA}
A_G= \frac{1}{|G|} \sum_{g}^G \sum_{\alpha,\beta,i,j}[L_g]_{\alpha i} \overline{[L_g]}_{\beta j }|\alpha ) \langle i| \otimes|j \rangle (\beta|,
\end{equation}
where  $|i\rangle,|j\rangle,|\alpha)$ and $|\beta)$ are basis elements of the vector space that supports $L$. The basis of the space describing the virtual d.o.f. of the tensor $A_G$ are represented with round brackets in Eq.(\ref{eq:generictensorA}) whereas regular brackets are used for the physical Hilbert space.
Let us denote 
\beq 
L_g \cong \bigoplus_\nu \mathbb{I}_{m_\nu} \otimes \Pi_\nu(g), \notag
\eeq
the decomposition into irreps of the left regular representation, which acts on $\mathbb{C}^{|G|} \cong \mathcal{H} = \bigoplus_\nu \mathcal{K}_\nu \otimes \mathcal{H}_\nu$. That is, $\Pi_\nu$ are the irreps of $G$, and $m_\nu$ their multiplicities; $\Pi_\nu$ acts on $\mathcal{H}_\nu$ and $\mathcal{K}_\nu$ is the multiplicity space associated with $\Pi_\nu$. Some Clebsch-Gordan matrix allows to write 
\begin{align}
A_G &\cong \frac{1}{|G|}  \sum_{\substack{g \in G\\ \alpha,\beta,i,j}}   \left[ C \big(\bigoplus_{\nu} \mathbb{I}_{m_{\nu}} \otimes \Pi_{\nu}(g)\big) C^{\dagger}  \right]_{\alpha i}    \times \notag \\
&\overline{\left[   C \big(\bigoplus_{\nu'} \mathbb{I}_{m_{\nu'}} \otimes \Pi_{\nu'}(g)  \big) C^{\dagger} \right]}_{\beta j } |\alpha )\langle i| \otimes|j \rangle (\beta|.\notag
\end{align}

If we express $A_G$ using orthonormal bases $\{ |k^{(\nu)},l^{(\nu)}\rangle\}$ for each subspace $\mathcal{K}_{\nu}\otimes \mathcal{H}_{\nu}$, and the orthogonality relations of irreps, we obtain:
\begin{align}
 A_G \cong \sum_{\nu}  \frac{1}{d_{\nu}} \sum_{ \substack{ k^{(\nu)}_1,k^{(\nu)}_3 \\ l^{(\nu)}_1,l^{(\nu)}_2 }}&
|k^{(\nu)}_1,l^{(\nu)}_1)\langle k^{(\nu)}_1,l^{(\nu)}_2 |  \notag \\
\otimes &  |k^{(\nu)}_3,l^{(\nu)}_2 \rangle (k^{(\nu)}_3,l^{(\nu)}_1 |. \notag
\end{align}
This tensor exhibits an obvious block diagonal form, in line with \cite{PEPSdescp},
\begin{equation*}
A_G\cong \bigoplus_{\nu}  \frac{1}{d_{\nu}}  A_G[\nu],
\end{equation*}
where each block $A_G[\nu]$ admits a very simple diagrammatic representation:
\beq \label{eq:tendiagram}
A_G[\nu]= 
\parbox[c]{0.22\textwidth}{ \includegraphics[scale=0.25]{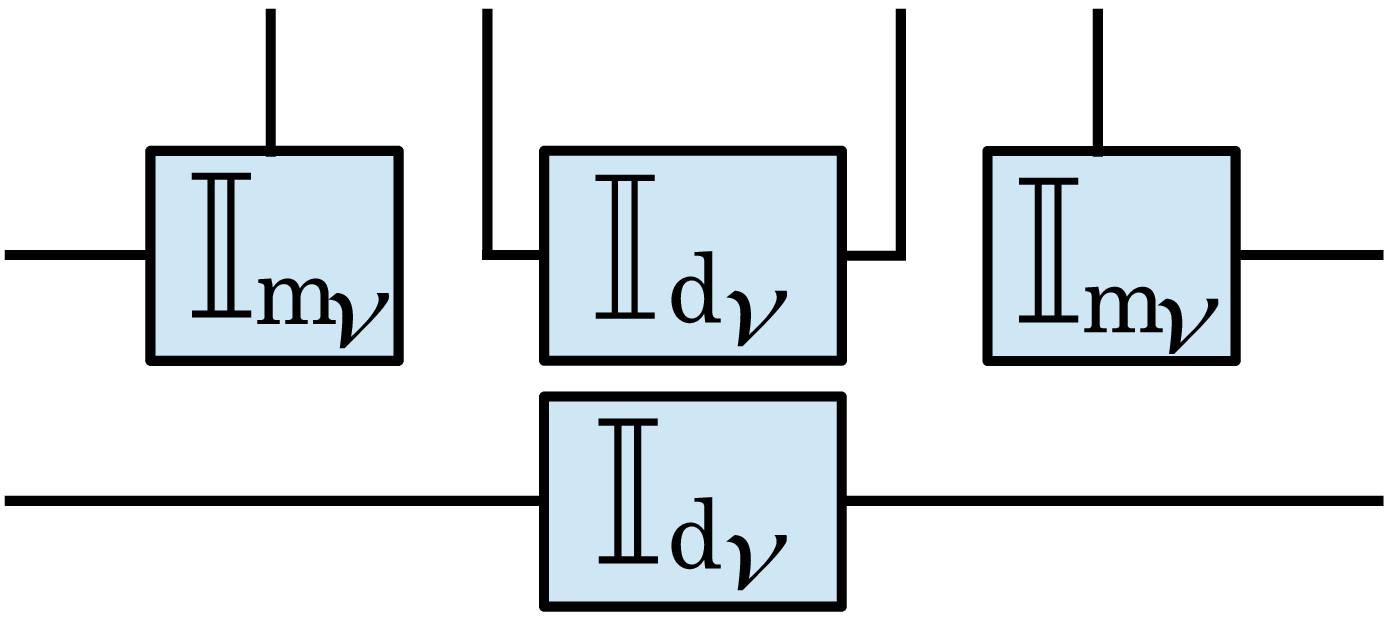}}.
\eeq
Note that the left regular representation is such that $m_\nu=d_\nu$.
Eq.({\ref{eq:tendiagram}) shows that if the irreps of two different groups, $G_1$ and $G_2$, have the same dimensions (and multiplicities), the corresponding tensors are identical modulo a local (Clebsch-Gordan) transformation. For example the left regular representations of $\mathbb{Z}_4$ and $\mathbb{Z}_2\times \mathbb{Z}_2$ both decompose into four one-dimensional irreps. Another example of this situation is that of the quaternionic group $\mathbb{Q}_8$ and the order 4 dihedral group $D_4$: in each case, the left regular representation is made of two inequivalent one-dimensional irrep and two equivalent copies of a two-dimensional irrep. \\

Given an irrep $\gamma$ of $G$ and an element $h\in G$, we consider the following operator:
\begin{align}
S_{\gamma}(h)&=\left(\bigoplus_{\nu \neq \gamma} \mathbb{I}_{m_{\nu}}\otimes \mathbb{I}_{d_{\nu}} \oplus \mathbb{I}_{m_{\gamma}}\otimes \Pi^G_{\gamma}(h)\right) \notag\\ 
& \otimes  \left(\bigoplus_{\nu \neq \gamma} \mathbb{I}_{m_{\nu}}\otimes \mathbb{I}_{d_{\nu}} \oplus \mathbb{I}_{m_{\gamma}}\otimes \bar{\Pi}^G_{\gamma}(h)\right).\notag
\end{align}
Modulo an appropriate change of basis, the action of $S_{\gamma}(h)$ reads:
$$\tilde{A}_G=S_{\gamma}(h) \left( C\otimes C)A_G( C^{\dagger} \otimes C^{\dagger} \right ).$$
It is clear that $\tilde{A}_G[\nu]={A}_G[\nu]$ for $\nu\neq \gamma$, but the block $\gamma$ is modified as

\begin{equation*}
\begin{split}
\tilde{A}&_G[\gamma]  = \frac{1}{d_{\gamma}}\sum_{k^{(\gamma)},k'^{(\gamma)}} 
 \sum_{ \substack{ m^{(\gamma)},n^{(\gamma)} \\ l^{(\gamma)},l'^{(\gamma)}}}
 [\Pi^G_{\gamma}(h)]_{n^{(\gamma)},l'^{(\gamma)}} \overline {[\Pi^G_{\gamma}(h)]}_{m^{(\gamma)},l'^{(\gamma)}  }  \\
\times & |k^{(\gamma)},l^{(\gamma)})\langle k^{(\gamma)},n^{(\gamma)} | \otimes|k'^{(\gamma)},m^{(\gamma)} \rangle (k'^{(\gamma)},l^{(\gamma)} |.
\end{split}
\end{equation*}

Since $\sum_{l'} \left([\Pi^G_{\gamma}(h)]_{n,l'} \overline {[\Pi^G_{\gamma}(h)]}_{m,l' }\right)=\delta_{m,n}$, the tensor remains invariant. That $S_{\gamma}(h)$ is a symmetry can be straightforwardly seen diagrammatically:
\begin{align}\label{eq:locsympa}
\tilde{A}_G[\gamma]= &
\parbox[c]{0.2\textwidth}{ \includegraphics[scale=0.25]{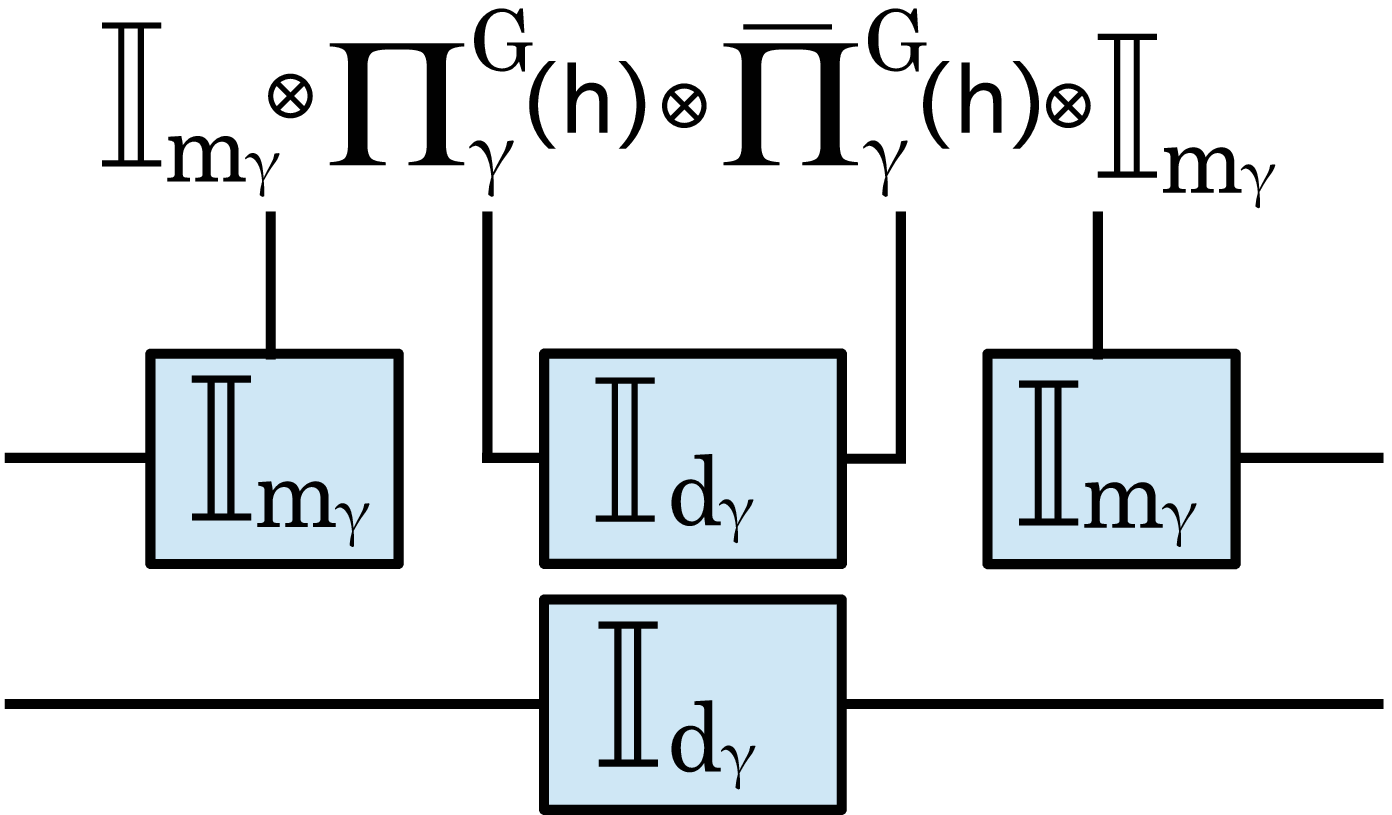}} \notag \\
= &\parbox[c]{0.2\textwidth}{ \includegraphics[scale=0.25]{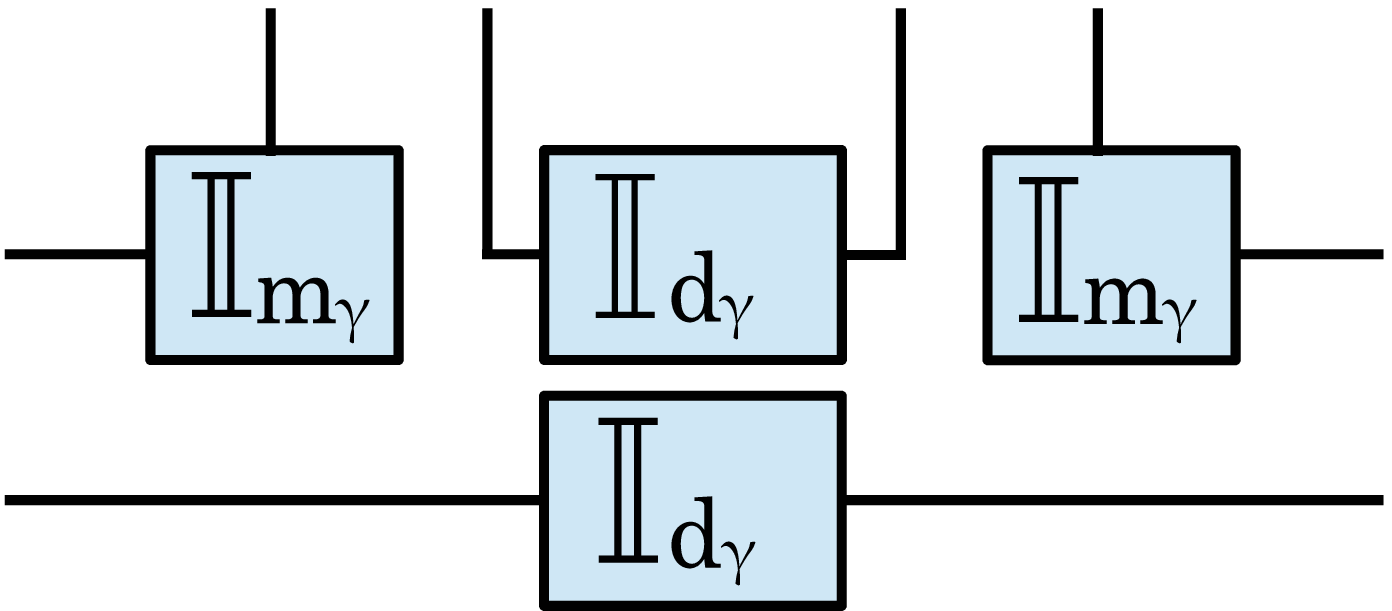}}= A_G[\gamma].
\end{align}
The operators $S_{\gamma}(h) \:\forall h\in G$, represent a local symmetry of the state constructed with the tensor $A_G$. 
Similarly the diagrams show that $S_{\gamma}(h) \:\forall h\in G$ have a trivial action on the virtual d.o.f. of $A_G$.\\

\begin{tcolorbox}
For $G$-isometric \textsf{MPS},
$$[S_{\gamma}(h)]_pA_G= A_G=[\Pi^G_{\gamma}(h)]_v A_G [\Pi^G_{\gamma}(h^{-1})]_v,$$
 $$ \left (S_{\gamma}(h)^{\otimes \mathcal{R}} \right) |\mathcal{M}(A_G) \rangle = |\mathcal{M}(A_G) \rangle,$$
where the subscripts $p$ and $v$ stand for the physical Hilbert space and virtual d.o.f. respectively, and where $\mathcal{R}$ is any region of the lattice.
\end{tcolorbox}
\vspace{5mm}

We now turn to the restricted \textsf{MPS} (\ref{redtens}) and study its symmetries. For that, we begin with the decomposition of $ \Pi_\nu(n)$ where $ n \in N$ into irreps of $N$: 
\begin{equation}\label{eq:Irdec}
\Pi_\nu(n)\cong \bigoplus_{\sigma \subset \nu} \mathbb{I}_{\widehat{\mathcal{K}}_\sigma} \otimes \pi_\sigma(n),
\end{equation} 
where $\{\pi_\sigma(n): n \in N\}$ denote the irreps of $N$ and $\Pi_\nu(n)$ acts on  $\mathcal{H}_\nu= \bigoplus_{\sigma \subset \nu} \widehat{\mathcal{K}}_\sigma \otimes \widehat{\mathcal{H}}_\sigma$. As before $\widehat{\mathcal{K}}_\sigma$ is a multiplicity space, i.e. the number of copies of $\pi_\sigma(n)$ contained in $\Pi_\nu$ is equal to $\textrm{dim} \; \widehat{\mathcal{K}}_\sigma$.  Clifford's Theorem \cite{Clifford} states that if $\sigma, \sigma' \subset \nu$, then (i) $\textrm{dim} \; \widehat{\mathcal{H}}_\sigma=\textrm{dim} \; \widehat{\mathcal{H}}_{\sigma'}=d_{\sigma(\nu)}$, (ii) the multiplicity spaces $\widehat{\mathcal{K}}_\sigma$ and $\widehat{\mathcal{K}}_{\sigma'}$  are isomorphic: $\textrm{dim} \widehat{\mathcal{K}}_\sigma= \textrm{dim} \widehat{\mathcal{K}}_{\sigma'}=\ell_{\nu}$ (iii) there exists $g \in G$ (that depends on $\sigma,\sigma'$) such that $\pi_{\sigma'}(n)=\pi_{\sigma}(g n g^{-1}), \forall n \in N$, i.e. $\pi_\sigma$ and $\pi_{\sigma'}$ are related by conjugation. 

Let $\{ \ket{l^{(\sigma)},q^{(\sigma)}} \}$ denote an orthonormal basis of $\widehat{\mathcal{K}}_\sigma \otimes \widehat{\mathcal{H}}_\sigma$. Using again irrep orthogonality relations, one can readily show that 
\begin{align}
A^{\rm res}_G &=  \frac{1}{|N|}\sum_{n \in N} L (n)\otimes L^{\dagger}(n) \notag \\
& \cong   \sum_{\nu,\mu}  \; \mathbb{I}_{\mathcal{K}_\nu}\otimes A^{\rm res}_G(\nu,\mu) \otimes \mathbb{I}_{\mathcal{K}_\mu} \notag 
\end{align}
where
\begin{align} \label{eq:blockdesc}
A^{\rm res}_G(\nu,\mu)= &
 \sum_{\sigma(\nu) \sim \rho(\mu)}  \frac{1}{d_{\sigma(\nu)}}  
\sum_{\substack{ q^{(\sigma)},q^{(\rho)}\\l^{(\sigma)},l^{(\rho)} }} | l^{(\sigma)},q^{(\sigma)} ) \notag\\
\times & \langle  l^{(\sigma)},q^{(\rho)}|  \otimes    |l^{(\rho)},q^{(\rho)}\rangle  ( l^{(\rho)},q^{(\sigma)}|,
 \end{align}
where $\sigma \sim \rho$ denotes that both representations are equivalent. We observe that if $\nu$ and $\mu$ do not contain any common irrep of $N$, $A^{\rm res}_G(\nu,\mu)=0$. We also point out that, in virtue of Clifford's Theorem, if $\nu$ and $\mu$ have one common irrep of $N$, then \emph{any} irrep of $N$ contained in $\nu$ is also contained in $\mu$ and vice versa. They are so-called associate.

We can represent Eq.(\ref{eq:blockdesc}) diagrammatically:
\beq \label{eq:diares}
A^{\rm res}_G(\nu,\mu)=  \sum_{\sigma(\nu) \sim \rho(\mu)}  \frac{1}{d_{\sigma(\nu)}}
\parbox[c]{0.2\textwidth}{ \includegraphics[scale=0.25]{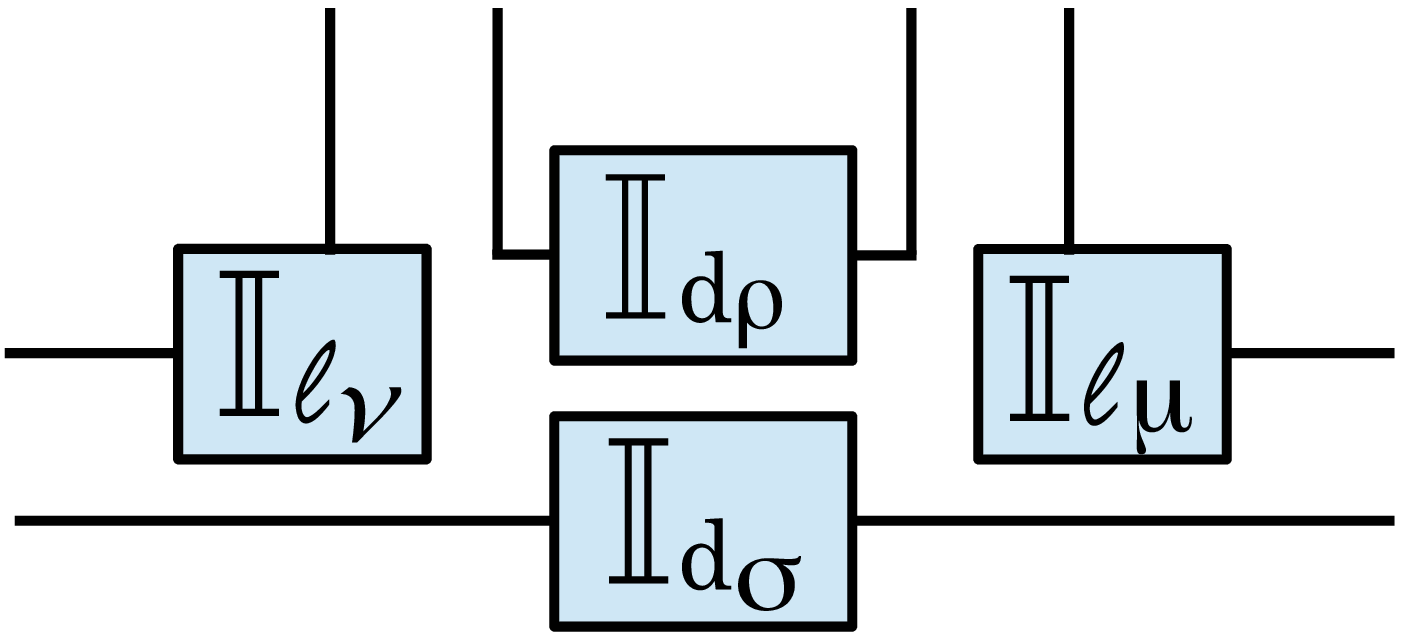}}
\eeq

In analogy with our discussion of $A_G$, we will construct symmetry operators in terms of irreps of $G$. For the sake of clarity, it is desirable that these irreps reflect the decomposition theory of the restricted tensor (\ref{eq:diares}). Clifford's theorem shows us exactly how to do that. Let $\{ \pi_{\sigma_{i}} : i=1 \ldots r_\nu\}$ represent the set of inequivalent irreps of $N$ that appear in the decomposition theory of $\{ \Pi_\nu(n): n \in N \}$. We can express the matrix $\Pi_\nu(g)$ in terms of submatrices $T_{ij}(g): \widehat{\mathcal{K}}_{\sigma_j} \otimes \widehat{\mathcal{H}}_{\sigma_j} \mapsto \widehat{\mathcal{K}}_{\sigma_i} \otimes \widehat{\mathcal{H}}_{\sigma_i}$ as:
\beq\label{eq:pi_nu_cliff}
\Pi_{\nu}(g)=\left( \begin{array}{ccc}T_{11}(g)&\cdots&T_{1r}(g)\\ \vdots &\ddots& \vdots\\ T_{r1}(g) &\cdots & T_{rr}(g)\end{array}\right).
\eeq
It is clear that $T_{ij}(n)=\delta_{i,j}  \mathbb{I}_{\widehat{\mathcal{K}}_{\sigma_i}} \otimes \pi_{\sigma_i}(n) \forall n \in N$. For fixed $i$ and $g$, it can be shown that there is one and only one value of $j$ for which $T_{ij}(g) \neq 0$. That is, $\Pi_\nu(g)$ has a permutation form, for example:
$$\Pi_{\nu}(g)=\left( \begin{array}{cccc} 0&*&0&0\\  0&0&*&0 \\ 0&0 &0& *\\ *&0 &0 & 0\end{array}\right),$$ interchanging the subspaces $\{ \widehat{\mathcal{K}}_{\sigma_i} \otimes \widehat{\mathcal{H}}_{\sigma_i} :i =1 \ldots r_\nu\}$, and acting non-trivially on them.

To elucidate the content of the operators $T_{ij}$, it is convenient to introduce the subgroup $G' \subset G$ that leaves $\widehat{\mathcal{K}}_{\sigma_1} \otimes \widehat{\mathcal{H}}_{\sigma_1}$ invariant: 
\begin{equation*}
G' \equiv \{g'\in G ; T_{11}(g') \neq 0\},
\end{equation*}
and we choose a set of elements $\{ \check{g}_2,\cdots, \check{g}_{r_\nu} \}$, such that $  T_{i1}(\check{g}_i) \neq 0$. These elements index the cosets of $G'$ in $G$:
$$G=G'+\check{g}_2 G'+\cdots+\check{g}_{r_\nu} G'. $$
We want to describe each matrix $T_{ij}(g)$ in terms of the representation $T_{11}(g')$ of $G'$. We notice that if one takes an element $g\in G$ and some element $\check{g}_i$, there exists a unique $\check{g}_j$ and $g'\in G'$ such that $g \check{g}_j=\check{g}_i g'$. Building on this observation, it can be shown that 
\begin{equation}\label{eq:indurep} T_{ij}(g) \cong  \tilde{T}_{ij}(g) \equiv
\Big\{
	\begin{array}{ll}
		T_{11}(\check{g}_i ^{-1}g \check{g}_j)  & \mbox{if } \check{g}^{-1}_i g \check{g}_j\in G' \\
		0 & \mbox{otherwise } 
	\end{array}.
\end{equation}
Hence $\Pi_\nu$ can be expressed as an induced representation of an irrep of $G'$. Moreover one can prove that $T_{11}(g' )$ admits a tensor product decomposition:
\beq \label{eq:VC}
 T_{11}(g')=V_\nu(g')\otimes C_\nu(g'),
\eeq
 where $V_\nu(g'): \widehat{\mathcal{K}}_{\sigma_1} \to \widehat{\mathcal{K}}_{\sigma_1}$ and $C_\nu(g'): \widehat{\mathcal{K}}_{\sigma_1} \to \widehat{\mathcal{K}}_{\sigma_1}$ are irreducible representations of $G'$. Since $T_{11}(n)=\mathbb{I}_{\widehat{\mathcal{K}}_{\sigma_1}} \otimes \pi_{\sigma_1}(n) \forall n \in  N$, we find by identification that $V_\nu(n)=\mathbb{I}_{\widehat{\mathcal{K}}_{\sigma_1}}$ for all $n\in N$. 

It can be proven that $V_\nu$ is projective:
\begin{equation*}
V_\nu(g) V_\nu(h')=\omega(g',h') V_\nu(g'h'),
\end{equation*}
where $\omega$ satisfies the cocycle condition:
$$ \omega(k'g',h') \omega(k',g')=\omega(k',g'h') \omega(g',h').$$
Since $\omega(g'n,h'm)=\omega(g',h')$ for all $n,m\in N$, $V_\nu(g')$ is actually a representation of $G'/N$.
Finally we can write:
\begin{equation}\label{eq:simdec}
 \tilde{T}_{ij}(g)\equiv
\Big\{
	\begin{array}{ll}
		V_\nu(\check{g}_i^{-1}g \check{g}_j)\otimes C_\nu(\check{g}_i ^{-1}g \check{g}_j)  & \mbox{if } \check{g}^{-1}_i g \check{g}_j\in G' \\
		0 & \mbox{otherwise } 
	\end{array}.
\end{equation}
See Fig. \ref{Aresymblock2}(c) for a diagrammatic representation. 

\begin{figure}[ht!]
\begin{center}
\includegraphics[scale=0.7]{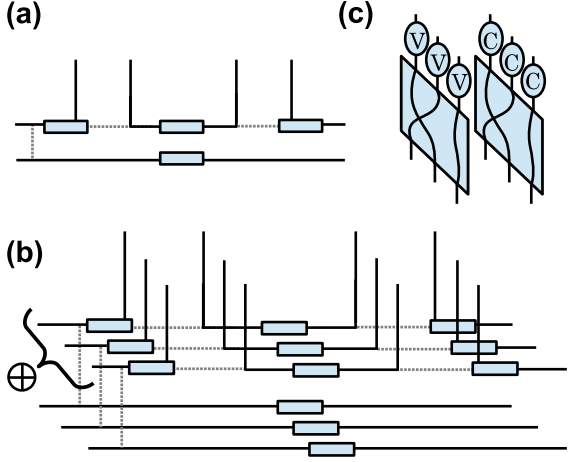}
\caption{ (a) Diagrammatic representation of a term appearing in the block $A^{\rm res}_G(\nu,\mu)$; see Eq.(\ref{eq:diares}). The grey dashed line is meant to indicate that the four boxes represent the operator $\mathbb{I}_{\ell_\nu} \otimes \mathbb{I}_{d_\rho} \otimes \mathbb{I}_{d_\sigma} \otimes \mathbb{I}_{\ell_\mu}$ corresponding to \emph{fixed} values of $\rho, \sigma$ in the sum. (b) Diagrammatic representation of the \emph{whole sum} Eq.(\ref{eq:diares}): each set of four tensors in a same plane parallel to the sheet relate to a same term of the sum, i.e. a same value of the pair $(\rho,\sigma)$. The $\oplus$ symbol is meant to mark the summation over all possible values of $(\rho,\sigma)$. (c) Decomposition of an irrep of $G$, described by the Eq.(\ref{eq:simdec}). We omit to represent the dependence of $V$ and $C$ on $g,\nu, i, j$, see Eq.(\ref{eq:simdec}).} 
\label{Aresymblock2}
\end{center}
\end{figure}
The matrix $\tilde{\Pi}_\nu(g)$ defined as (\ref{eq:pi_nu_cliff}), with $T_{ij}$ replaced with $\tilde{T}_{ij}$ maps $\widehat{\mathcal{K}}_{\sigma_i} \otimes \widehat{\mathcal{H}}_{\sigma_i}$ to $\widehat{\mathcal{K}}_{\sigma_j} \otimes \widehat{\mathcal{H}}_{\sigma_j}$ bijectively according to $gg_j=g_i g'$. If we further define $\psi_{ij}(g) \equiv \check{g}_i^{-1} g \check{g}_j$, we are in a position to specify the action of $\tilde{\Pi}_\nu(g)$,
\begin{align}
\bra{l^{(\sigma_j)},q^{(\sigma_j)}} \tilde{\Pi}_\nu(g) \ket{l^{(\sigma_i)},q^{(\sigma_i)}}=&
\bra{l^{(\sigma_j)}} V_\nu(\psi_{ij}(g))\ket{l^{(\sigma_i)}}  \notag \\
\times &\bra{q^{(\sigma_j)}} C_\nu(\psi_{ij}(g)) \ket{q^{(\sigma_i)}}, \notag
\end{align}
\begin{figure}[ht!]
\begin{center}
\includegraphics[scale=0.6]{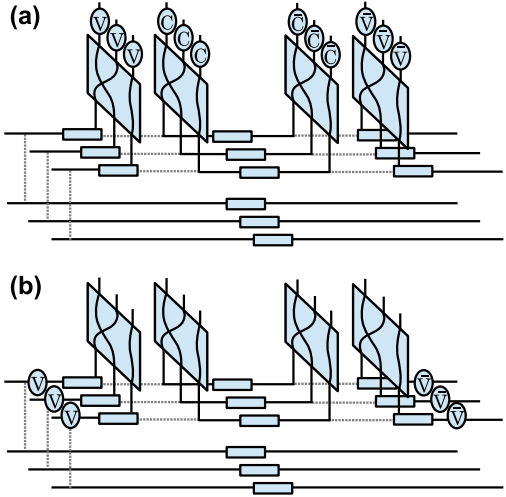}
\caption{ (a) Action of the irrep operator on the tensor according to Eq.(\ref{eq:symresfor1}). (b) Final result of the action on the restricted tensor given by Eq.(\ref{eq:actionblock}).} 
\label{Aresymblock}
\end{center}
\end{figure}

and to apply the operator corresponding to two associate irreps $\nu$ and $\mu$ over $A^{\rm res}_G(\nu,\mu)$:

\begin{widetext}
\begin{equation}\label{eq:symresfor1}
\begin{split}
\Big(&\tilde{\Pi}_\nu(g)\otimes \tilde{\Pi}^{\dagger}_\mu(g)\Big)_p A^{\rm res}_G(\nu,\mu)=
\sum_{\sigma_i(\nu) \sim \rho_i(\mu)}  \frac{1}{d_{\sigma_i}}  
\sum_{\substack{ q^{(\rho_j)}\\l^{(\sigma_j)},l^{(\rho_j)} }}  
\sum_{\substack{ q^{(\sigma_i)}\\l^{(\sigma_i)},l^{(\rho_i)} }} 
 \bra{q^{(\rho_j)}}C^{\dagger}_\mu(\psi_{ij}(g)) \mathbb{I}_{d_{\rho_i}} C_\nu(\psi_{ij}(g))\ket{q^{(\rho_j)}}\times \\
&\hspace{3cm}|l^{(\sigma_i)},q^{(\sigma_i)} ) \langle  l^{(\sigma_j)},q^{(\rho_j)}| \bra{l^{(\sigma_i)}} V_\nu(\psi_{ij}(g))\ket{l^{(\sigma_j)}}\otimes \bra{l^{(\rho_j)}}V^{\dagger}_\mu(\psi_{ij}(g))\ket{l^{(\rho_i)}} 
 |l^{(\rho_j)},q^{(\rho_j)}\rangle  ( l^{(\rho_i)},q^{(\sigma_i)}|,
\end{split}
\end{equation}
where $ \bra{q^{(\rho_j)}}C^{\dagger}_\mu(\psi_{ij}(g)) \mathbb{I}_{d_{\rho_i}} C_\nu(\psi_{ij}(g))\ket{q^{(\rho_j)}}=1 $ since $C_\nu=C_\mu$ for associate irreps $\nu$ and $\mu$ \cite{Clifford}. Then
\begin{equation}
\begin{split}
\Big(&\tilde{\Pi}_\nu(g)\otimes \tilde{\Pi}^{\dagger}_\mu(g)\Big)_p A^{\rm res}_G(\nu,\mu)=\\
&\sum_{\sigma_i(\nu) \sim \rho_i(\mu)}  \frac{1}{d_{\sigma_i}}  
\sum_{\substack{ q^{(\rho_j)}\\l^{(\sigma_i)},l^{(\rho_i)} \\l^{(\sigma_j)},l^{(\rho_j)} }}  
 |l^{(\sigma_i)},q^{(\sigma_i)} ) \langle  l^{(\sigma_j)},q^{(\rho_j)}|  \bra{l^{(\sigma_i)}} V_\nu(\psi_{ij}(g))\ket{l^{(\sigma_j)}} 
  \bra{l^{(\rho_j)}}V^{\dagger}_\mu(\psi_{ij}(g))\ket{l^{(\rho_i)}}  |l^{(\rho_j)},q^{(\rho_j)}\rangle  ( l^{(\rho_i)},q^{(\sigma_i)}|= \\
  &\sum_{\sigma_j(\nu) \sim \rho_j(\mu)}  \frac{1}{d_{\sigma_j}}  
\sum_{\substack{ q^{(\rho_j)}\\l^{(\sigma_i)},l^{(\rho_i)} \\l^{(\sigma_j)},l^{(\rho_j)} }}   
(l^{(\sigma_i)}| V_\nu(\psi_{ij}(g))|l^{(\sigma_j)})
 |l^{(\sigma_i)},q^{(\sigma_i)} ) \langle  l^{(\sigma_j)},q^{(\rho_j)}| \otimes 
   |l^{(\rho_j)},q^{(\rho_j)}\rangle  ( l^{(\rho_i)},q^{(\sigma_i)}| (l^{(\rho_j)}|V^{\dagger}_\mu(\psi_{ij}(g)) | l^{(\rho_i)}) .\notag
\end{split}
\end{equation}
\end{widetext}

Therefore 
\begin{align}\label{eq:actionblock}
\Big(&\tilde{\Pi}_\nu(g)\otimes \tilde{\Pi}^{\dagger}_\mu(g)\Big)_p A^{\rm res}_G(\nu,\mu)=  \\
&A^{\rm res}_G(\nu,\mu)\left( P_\nu(g)\bigoplus V_\nu(\psi_{ij}(g))\otimes P_\mu(g) \bigoplus\bar{V}_\mu(\psi_{ij}(g))\right)_v, \notag
\end{align}
where $P_\nu(g)$ represents the permutation part of the irrep $\nu$ given by the induced representation of $G'$ (see Eq.(\ref{eq:indurep})) mapping $\widehat{\mathcal{K}}_{\sigma_i} \otimes \widehat{\mathcal{H}}_{\sigma_i}$ to $\widehat{\mathcal{K}}_{\sigma_j} \otimes \widehat{\mathcal{H}}_{\sigma_j}$ and where $V_\nu(\psi_{ij}(g))$ is the projective representation of $G'/N$, appearing in the decomposition of  Eq.(\ref{eq:VC}), acting on $\widehat{\mathcal{K}}_{\sigma_j}$. The diagrams for this action are shown in Fig. \ref{Aresymblock}.

We now consider the operators
\begin{align}
S_{\nu,\mu}(g)&=\left(\bigoplus_{\gamma \neq \nu,\mu} \mathbb{I}_{m_{\gamma}}\otimes \mathbb{I}_{d_{\gamma}} \oplus \mathbb{I}_{m_{\nu}}\otimes \tilde{\Pi}_\nu(g) \oplus \mathbb{I}_{m_{\mu}}\otimes \tilde{\Pi}_\mu(g)
\right) \notag\\ 
& \otimes \left(\bigoplus_{\gamma \neq \nu,\mu} \mathbb{I}_{m_{\gamma}}\otimes \mathbb{I}_{d_{\gamma}} \oplus \mathbb{I}_{m_{\nu}}\otimes \tilde{\Pi}^{\dagger}_\nu(g) \oplus \mathbb{I}_{m_{\mu}}\otimes \tilde{\Pi}^{\dagger}_\mu(g)\notag
\right)
\end{align}
and
\begin{align}\label{eq:sinop}
S_{\nu}(g)&=\left(\bigoplus_{\gamma \neq \nu,\mu} \mathbb{I}_{m_{\gamma}}\otimes \mathbb{I}_{d_{\gamma}} \oplus \mathbb{I}_{m_{\nu}}\otimes \tilde{\Pi}_\nu(g) 
\right) \notag\\ 
& \otimes \left(\bigoplus_{\gamma \neq \nu,\mu} \mathbb{I}_{m_{\gamma}}\otimes \mathbb{I}_{d_{\gamma}} \oplus \mathbb{I}_{m_{\nu}}\otimes \tilde{\Pi}^{\dagger}_\nu(g)
\right).
\end{align}
In virtue of Eq.(\ref{eq:actionblock}), these operators correspond to an on-site global symmetry of the state constructed with $A^{\rm res}_G$:
 
$$S^{\otimes L}_{\nu,\mu}(g)|\mathcal{M}(A^{\rm res}_G) \rangle=|\mathcal{M}(A^{\rm res}_G) \rangle.$$\\

\begin{tcolorbox}
The action of the symmetry operator of the group $G$ on each block is a conjugation by a projective representation ($V(g')$ on the virtual d.o.f.) of the group $ G'/N$ which can be extended to the group $G/N$ as an induced representation carrying intrinsically the pattern of permutation-action between blocks. The group $G'$ is defined as the elements of $G$ leaving one (chosen) block invariant under the permutation action of the operator for these elements. The normal subgroup $N$ of $G$ corresponds to the local invariance of the tensor and it encodes the splitting of irreducible blocks under the permutation action of the symmetry.
\end{tcolorbox}
\vspace{5mm}

This structure fits nicely with the one done in \cite{phasesMPS} explained in Section III.A. The group associated to the physical symmetry is $G$ in both cases but in our construction the $N$-injectivity of our \textsf{MPS} reveals an effective $G/N$ representation of the symmetry in the virtual d.o.f. In the case we are dealing with, degenerate ground states, the role of the subgroup $H\subset G$ is played by the quotient $G'/N \subset G/N$.

\subsection{An example: classification for $\mathbf{G_{sym}= \mathbb{Z}_2 }$}

As an illustration of the analysis of the previous subsection, we consider the case where $G_{\rm sym}=\mathbb{Z}_2$. According to \cite{phasesMPS}, the phases will be given solely by permutations between the blocks forming the tensor of the \textsf{MPS} in consideration, since the second cohomology group $H^2(\mathbb{Z}_2,U(1))$ is trivial (see Appendix \ref{ap:coho}). There are two phases: one where $\mathbb{Z}_2$ is represented trivially at the virtual level, and another where $\mathbb{Z}_2$ is faithfully represented, by an appropriate permutation, at the virtual level. This permutation will be the product of disjoint transpositions acting on blocks with the same size. Thus the number of disjoint transpositions characterizes the phase, i.e., the way in which the symmetry acts. 

In order to obtain a nontrivial permutation we have to take the operator associated with the non-trivial semidirect product. The number of disjoint transpositions of this permutation will be given by the number of irreps of the extension in the decomposition of the operator that realizes the symmetry. 

The degeneracy of the ground state manifold is given by the block structure of the matrices and will depend on the group $N$; see Eq.(\ref{eq:blockdesc}). For our purposes, we study the case of the abelian group $N=\mathbb{Z}_n$, with $n$ odd. It is known that all possible extensions of these two groups are semidirect products \cite{Rotman}. Then, we have the different extensions $G=\mathbb{Z}_n \rtimes_{\rho} \mathbb{Z}_2$  (see Appendix \ref{ap:ext}) resulting in the direct product $\mathbb{Z}_n \times \mathbb{Z}_2$ and the dihedral group $D_{2n}$ as the non-trivial semidirect product. The latter is built choosing the inverse automorphism $\rho_1(g)=-g, \; \forall g\in \mathbb{Z}_n$. We will write $g^{-1}$, $n-g$ or $-g$ indistinctly.

The left regular representation of $D_{2n}$ decomposes as (we only consider elements of the normal subgroup):
\begin{equation}
L_{(k,0)}^{D_{2n}}  \cong \left(1\oplus 1 \bigoplus_m \Pi_m(k,0)\oplus \Pi_m(k,0) \right),\notag
\end{equation}
where $\Pi_m$, $m=1,\cdots, (n-1)/2$, are the bidimensional irreps of $D_{2n}$ \cite{Simon} :
\beq
\Pi_m(k,0)=\left(\begin{array}{cc}q^{ km}& 0 \\ 0 & q^{- km} \end{array}\right); \; q=e^{\frac{2\pi i}{n}}, \; k=0,\cdots, n-1. \notag
\eeq
This form for the representation is analogous to Eq.(\ref{eq:Irdec}).

One easily checks that
 $$\Pi_m(k,0)\oplus \Pi_m(k,0)=P( \Pi_m(k,0)\otimes \mathbb{I}_2) P,$$
 where $ P=1\oplus \sigma_x\oplus1$  and
$$\Pi_m(k,0)\otimes \mathbb{I}_2= \left(\begin{array}{cccc}q^{km}& 0& 0& 0 \\ 0 & q^{km}& 0& 0\\ 0 & 0&q^{-km}&  0 \\ 0 & 0& 0&q^{-km} \end{array}\right).$$ 
A bit more explicitly, 
\begin{equation*}
\begin{split}
 \Pi_m(k,0)\otimes \mathbb{I}_2=q^{km} (|m,0\rangle \langle m,0|+|m,1\rangle \langle m,1|)+\\
 q^{-km} (|-m,0\rangle \langle -m,0|+|-m,1\rangle \langle -m,1|).
\end{split}
\end{equation*}
So, up to local unitary equivalence, 
\begin{equation*}
\begin{split}
L_{(k,0)} \cong  \sum_{u=0}^{1}\bigg [ |0,u\rangle \langle 0,u| + \sum_{m=1}^{(n-1)/2}  & \big( q^{km} |m,u\rangle \langle m,u|+\\
&q^{-km} |-m,u\rangle \langle -m,u|\big)  \bigg]
\end{split}
\end{equation*}
After decomposing $L^{\dagger}(k,0)=L(-k,0)$ similarly, we get
\begin{equation*}
\begin{split}
L_{(-k,0)} \cong  \sum_{u=0}^{1}\bigg [ |0,u\rangle \langle 0,u| + \sum_{m=1}^{(n-1)/2}  & \big( q^{-km} |m,u\rangle \langle m,u|+\\
&q^{km} |-m,u\rangle \langle -m,u|\big)  \bigg]
\end{split}
\end{equation*}
Putting all this together, we find
\begin{widetext}
\begin{equation}
\begin{split}
 A^{\rm res}_{D_{2 n}} & \cong \sum_{k=0}^{n-1}  \Bigg\{\sum_{u=0,1}\bigg [  |0,u ) \langle 0,u| +\sum_{m=1}^{(n-1)/2}  \big( q^{km} |m,u) \langle m,u|+q^{-km} |-m,u) \langle -m,u|\big)  \bigg] \Bigg\} \otimes\\
&\otimes \Bigg\{ \sum_{u=0,1}\bigg [ |0,u\rangle ( 0,u| +\sum_{m=1}^{(n-1)/2}  \big( q^{-km} |m,u\rangle ( m,u|+q^{km} |-m,u\rangle ( -m,u| \big)  \bigg] \Bigg\}\frac{1}{n}.\notag
\end{split}
\end{equation}
\end{widetext}
Using $\frac{1}{n}\sum_k e^{2\pi k(i-j)/n}=\delta_{i,j}$ we get the matrix 
\begin{equation}
A^{ {\rm res} \{(m,b)(m',b')\} }_G=\delta_{m,m'}|m,b)(m',b'|, \notag
\end{equation}
where $\{(m,b)(m',b')\}$ label the physical indices. For each value of the physical index, the virtual matrices, of size $2n \times 2n$, of the tensor has a diagonal structure with $n$ two-dimensional blocks related to the $n$ irreps of $\mathbb{Z}_n$. The first one is denoted by the label $m=0$ and the others are grouped in pairs, those which labels $\pm m$, related to the $(n-1)/2$ bidimensional irreps of $D_{2n}$. 

We denote such a pair by $(m)$. The tensor given by Eq.(\ref{eq:blockdesc}) is also diagonal in terms of the irreps of $D_{2n}$, i.e. the associate irreps of $D_{2n}$ are a single irrep. If we fix one block of the pair $(m)$, say $+m$, 
we obtain four different matrices
\begin{equation}
A^{ {\rm res} \{(+m,b)(+m,b')\} }_G = |+m,b)( +m,b'|\equiv B_{b,b'}^{+m}, \notag
\end{equation}
where $[B_{b,b'}^{+m}]_{\alpha,\alpha'}=\delta_{b,\alpha}\delta_{b',\alpha'}$. These matrices span the whole space of $2\times 2$ matrices, so each block is injective \cite{MPSrep}. Now we are going to act on our tensor with different symmetry operators at the physical level, and we will recover the permutations of the blocks of the matrices. The exchange will be between the blocks $\pm m$ belonging to the pair $(m)$. The symmetry operators are the irreps of the non-trivial extension, evaluated at the elements belonging to the non-trivial coset of $G$ by $N$, $\{ (g,1)|g\in N \}$, acting on each pair $(m_0)$ .

The  $(n-1)/2$ two-dimensional irreps of $D_{2n}$ in the coset $\{ (k,1)|k\in \mathbb{Z}_n \}$  take the form
$$\Pi_m(k,1)=\Pi_m(k,0)\Pi_m(0,1)=\left(\begin{array}{cc}0&q^{ km} \\ q^{- km} & 0 \end{array}\right),$$
where $\Pi_m(0,1)$ is nothing but the Pauli matrix $\sigma_x$ which does not depend on $m$. For simplicity, we will deal only with the case of one block in detail reaching a single transposition. The operator associated to one block $(m_0)$, is analogy with Eq.(\ref{eq:sinop}), and is given by
\begin{equation}
\begin{split}
S_{m_0}(l,l')=&[( \Pi_{m_0}(l,1)\otimes\mathbb{I}_2)  \otimes \mathbb{I}_{\rm rest})] \otimes \\
&[(\bar{\Pi}_{m_0}(l',1) \otimes \mathbb{I}_2 ) \otimes \mathbb{I}_{\rm rest}].\notag
\end{split}
\end{equation}
The left-hand side of the previous operator in the selected basis can be expressed as
\begin{equation*}
\begin{split}
 \Pi_{m_0}(l,1)\otimes\mathbb{I}_2= \sum_{u=0}^{1} \big[ &q^{ lm_0} |m_0,u\rangle \langle -m_0,u|+\\
& q^{-lm_0} |-m_0,u\rangle \langle m_0,u| \big] ,\notag
\end{split}
\end{equation*}
acting on the pair $(m_0)$ and the identity operator in the rest of the blocks. Therefore, if we act with this operator on the tensor, we notice that all blocks with $m\neq \pm m_0$ are not affected, but the pair with $(m)=(m_0)$ changes as  
\begin{equation*}
\begin{split}
\Pi_{m_0}(k,0)\Pi_{m_0}(l,1)=\Pi_{m_0}(k+l,1) \quad {\rm and}\\
\Pi_{m_0}(l',1)\Pi_{m_0}(-k,0)=\Pi_{m_0}(k+l',1)
\end{split}
\end{equation*}
on each side of the tensor product respectively. Let us analyze the action of the operator $S_{m_0}(l,l')$ looking at the virtual matrices of the modified tensor $\tilde{A}^{ {\rm res} }_G=S_{m_0}(l,l') {A}^{ {\rm res} }_G$:
\begin{align}
\tilde{A}&^{ {\rm res} \{(+m,b)(+m,b')\} }_G \notag \\
 =&\bigg\{ \begin{array}{lcc}   |+m,b)(+m, b'| & {\rm if} &+m\notin (m_0)
\\
q^{(l-l') m_0}|-m,b)(-m,b'|  & {\rm if} &+m\in (m_0) \end{array},\notag
\end{align}
where only the non-zero elements are written. That is:
\begin{equation}
 \tilde{B}_{b,b'}^{+m}
 =\bigg\{ \begin{array}{lcc}   B_{b,b'}^{+m} & {\rm if} &+m\notin (m_0)
\\
 B_{b,b'}^{-m} & {\rm if} &+m\in (m_0) \end{array},\notag
\end{equation}
As a consequence, we obtain that the two blocks of the matrices, associated with the pair $(m_0)$ are exchanged:
\begin{equation}
\begin{tikzpicture}
                        \matrix (m) [matrix of math nodes,left delimiter=(,right delimiter=)] {
                                B^{+m_0}       & 0 \\
                                0       & B^{-m_0}     \\
                };
                        \draw[<->] ($(m-1-1.center) + (0.1,-0.1)$) to ($(m-2-2.center) + (-0.45,0.1)$);
\end{tikzpicture} \notag
\end{equation}
This permutation between the blocks $+m_0$ and $-m_0$ is nothing but the single transposition that we were looking for. The action of the operator does not depend on the element of $\mathbb{Z}_n$ in the set $\{ (l,1)|l\in\mathbb{Z}_n \}$, the non-trivial coset $[1]\neq[e]\cong \mathbb{Z}_n$, so the result is uniquely determined by the quotient group $\mathbb{Z}_2$ (in general $G_{sym} \cong G/N$). The action of the operator using elements of the subgroup $\{ (g,0)|g\in\mathbb{Z}_n \}$ is trivial because it does not permute the blocks of the virtual matrices. In this example, the multiplicity of the irreps of $N$ in each irrep of $G$ is one, then the projective representation $V$ in Eq.(\ref{eq:VC}) does not play any role here. Instead, the only non-trivial action for this case is a permutation carried out by the induced representation of Eq.(\ref{eq:indurep}). Eq.(\ref{eq:actionblock}) translates as:
\beq
\left(S_{m_0}(g)\right )_p{A}^{ {\rm res} }_G={A}^{ {\rm res} }_G  \left(  (\sigma_x\otimes \mathbb{I}_2)_{m_0}\otimes(\sigma_x\otimes \mathbb{I}_2)_{m_0}   \right )_v,\notag
\eeq
where $g$ represents the non-trivial element of the group $\mathbb{Z}_2$. We can interpret this result as a symmetry breaking phase since the blocks exchanged correspond to linearly independent states of the ground subspace. When we take the operator from the trivial extension, we find that the symmetry is in a non-equivalent phase, characterized by a non-symmetry breaking pattern in the ground subspace.

In order to recover the other permutations, related to disjoint transpositions, we just act with the operator created by adding the different irreps associated  to the two interchangeable blocks. The operator associated to the pairs $(m_0),\cdots, (m_i)$ is given by
\begin{equation}
\begin{split}
( &\Pi_{m_0}(l_0,1)\oplus \cdots \oplus  \Pi_{m_i}(l_i,1))\otimes\mathbb{I}_2)  \otimes \mathbb{I}^{\rm rest}] \otimes \\
 [( &\bar{\Pi}_{m_0}(l_0,1)\oplus \cdots \oplus  \bar{\Pi}_{m_i}(l_i,1))\otimes\mathbb{I}_2)  \otimes \mathbb{I}^{\rm rest}] .\notag
\end{split}
\end{equation}
This operator carries out the transposition between the blocks $\pm m_0, \dots, \pm m_i$ in the virtual matrices. Again, the action is independent from the element $l_0,\cdots, l_i$, so it is uniquely determined by the element of the quotient group just as in the single transposition case. Therefore, we have recovered all the possible phases with symmetry group $\mathbb{Z}_2$ and degenerate ground state for \textsf{MPS}. \\
It is straightforward to use Eq.(\ref{eq:locsympa}) to show that the parent tensor is left invariant by the symmetry operators.

\section{Conclusions and outlook}\label{sec:conc}

We have studied two classes of \textsf{PEPS} related by anyon condensation (parent and restricted model). We have seen that the local invariance of the first under the action of a group G is broken in the second. Some residual symmetry persists though: the restricted \textsf{PEPS} is left invariant by a smaller local symmetry ($G_{\rm topo} \vartriangleleft G$) plus a global symmetry $(G_{\rm sym} \equiv G/G_{\rm topo})$. This symmetry change, from local to global, is closely related to flux confinement and charge condensation. To get a microscopic understanding of these phenomena, we have analyzed how a background defined by the restricted model is affected by the insertion of (virtual) excitations of the parent model. Besides, we have seen that the residual global symmetry is represented by permutations of particle types within each anyonic sector. Also some cohomological effect has been identified. Similarly, when the model lives on a non-trivial manifold, this residual (global) symmetry both leaves the ground subspace invariant and does not act trivially on it. 

On another hand, Wilson loops (corresponding to unconfined excitations) also leave the ground subspace invariant and act non-trivially on it. This coincidence leads us to believe that the two types of operators might be related. We have also discussed why the theory of group extensions is a promising route in modelling the behavior of the anyons under a symmetry. 

Next, we have investigated \textsf{MPS} analogues of our findings; combining the symmetry reduction discussed above with classical results in the theory of group representations \cite{Clifford}, we have been able to re-derive all possible representations of an on-site global symmetry at the virtual level. 

The approach for charge condensation studied here could also be applied to charge confinement using the different tensor network realizations of the same quantum phase described in \cite{BoconTNS}. In that case the restricted tensor would act as a {\it flux condensator} for the parent model. In \cite{ShadowsAnyons,CondTNS,newNorbert} anyon condensation has been numerically studied in the framework of \textsf{PEPS} for different topological orders without symmetries. The authors of these works performed a local parametrized perturbation on the tensor and successfully identified the condensed and confined anyons pattern in the properties of the fixed point of the transfer operator. In contrast we have here studied analytically pairs of phases, corresponding to the extreme points of an anyon condensation process, modeled with \textsf{PEPS}. We have focused on discrete gauge theories (quantum doubles) which has allowed us to analyze the behavior of local/global symmetries in both phases through the  condensation. A complete description of confinement and condensation for quantum doubles in terms of groups algebras can be found in \cite{Bombin}. The complete study of the global symmetry action on anyons and its relation to fusion is left for future work. 

We point out that topological phases with global symmetries, the so-called Symmetry Enriched Topological (SET) phases, has been fully classified in \cite{SETQ}.  SET phases has been also been studied in \cite{Ran} using tensor networks and in \cite{Fiona}  giving solvable model to represent each phase (also see \cite{Tarantino} for the relation to our work). We wonder if the reduction to a global symmetry is a general feature of anyon condensation in gauge theories. Were it the case, enriching an anyonic system with extra global symmetries might allow to detect its topological content locally. One future line of research will be to study how our findings carry over to a more general class of \textsf{PEPS}, with more complex global symmetry and anyon representations, such as those discussed \cite{MPOtop}. Finally, it would be worth parallelling  our work in the context of fermions.

\section*{Aknowledgements}

We warmly thank Norbert Schuch for inspiring discussions and for providing us with an early copy of \cite{newNorbert}. The authors acknowledge support from MINECO (grant MTM2014- 54240-P), from Comunidad de Madrid (grant QUITEMAD+- CM, ref. S2013/ICE-2801), and the European Research Council (ERC) under the European Union's Horizon 2020 research and innovation programme (grant agreement No 648913). DPG acknowledges support from the John Templeton Foundation through grant $\#48322$. The opinions expressed in this publication are those of the authors and do not necessarily reflect the views of the John Templeton Foundation.
This work has been partially supported by ICMAT Severo Ochoa project SEV-2015-0554 (MINECO).

\appendix
\section{Group cohomology} \label{ap:coho}

In this appendix we describe the group $H^2(G,U(1))$ that classifies the nonequivalent projective representations of $G$. A projective representation $D$ of a group $G$ is a mapping from this group to the $GL(\mathcal{H})$ group, where $\mathcal{H}$ denotes the vector space where the representation acts, such that
\begin{equation} \label{2co}D(g_1)D(g_2)=\omega(g_1,g_2)D(g_1g_2), \; g_1,g_2 \in G,\end{equation}
where $\omega:G\times G\to U(1)$, which is called cocycle, satisfies
$$\omega(g_2,g_3)\omega(g_1,g_2g_3)=\omega(g_1,g_2)\omega(g_1g_2,g_3),$$
$$\omega(g,e)=\omega(e,g)=1, \forall g \in G.$$
The first one is the so-called 2-cocycle condition and comes from the application of (\ref{2co}) to the associative condition $[D(g_1)D(g_2)]D(g_3)=D(g_1)[D(g_2)D(g_3)]$. Note that if $\omega(g_1,g_2)=1 \; \forall g_1,g_2 \in G$ we recover a linear representation.  Suppose that we choose a different pre-factor for the representation matrices $D'(g)=c(g)D(g)$ where  $c:G \to U(1)$, so the different cocycles are related through:$$\omega'(g_1,g_2)=\frac{c(g_1)c(g_2)}{c(g_1g_2)}\omega(g_1,g_2).$$ This also shows that the cocycle for a linear representation, also called coboundary, can be written as $\omega(g_1,g_2) = c(g_1)c(g_2)/c(g_1g_2)$. We regard $U'(g)$ and $U(g)$ as equivalent projective representations if the associated cocycles are related by a pre-factor. Such an equivalent relation forms the abelian group $H^2(G,U(1))$, called the second cohomology group of $G$.\\

For our purpose we have to obtain $H^2(\mathbb{Z}_2, U(1))$, this task has been done in \cite{SPTWen} and there, it has been demonstrated that it is equal to the trivial group $\{ e \}$. 
\section{Group extensions}\label{ap:ext}

An extension of a group $Q$ is a group $E$, together with a surjective homomorphism $\pi:E\rightarrow Q$. Let $N$ denote the kernel of $\pi$. It is clear that $N$ is a normal subgroup of $E$. We say that the group $E$ with the homomorphism $\pi$ is an extension of $Q$ by $N$ \cite{Rotman, Morandi}. An extension is encoded in the following short exact sequence:
$$1\rightarrow  N  \stackrel{i}{\rightarrow} E\stackrel{\pi}{\rightarrow} Q \rightarrow 1,$$
where $i$ is an inclusion map. $Q$ is isomorphic to the quotient group $E/N$ ($Q\cong E/N$).\\

In the case where $N$ is an abelian group, given an extension $E$ of $Q$ by $N$ and the homomorphism $\pi$, the extension is characterized by two maps: (i) a homomorphism $\rho:Q  \rightarrow  {\rm Aut}(N)$ and (ii) a cocycle $\omega:Q\times Q\rightarrow N$ which satisfies 
\beq
\omega(g,h)\omega(gh,k)=\rho_g(\omega(h,k))\omega(g,hk).\notag
\eeq
These maps are defined as follows. Given $g$, we pick a pre-image $e_g\in E$ such that $\pi(e_g)=g$, and we construct $\rho_g:n\mapsto e_g n e^{-1}_g$ and $\omega(g,h)=e_g e_h e^{-1}_{gh}$. There is some arbitrariness in this choice of a pre-image: $e_g$ and $e_g k$ obviously lead to the same map $\rho_g$ for any $k \in N$. This arbitrariness partially persists when considering the cocyle $\omega$: the two pre-image choices $e_g$ and $e_g k$ lead to two cocycles that might be different but will belong to the same second cohomology class $H^2(Q,U(1))$.

$E$, as a set, can be expressed as the cartesian product $N\times Q$ with the rule for multiplication:
\begin{equation}\label{eq:carprod}
(a,g)(b,h)=(a\rho_g(b)\omega(g,h),gh).
\end{equation}
The product $(a,e)(b,e)=(ab,e)$ generates the normal subgroup $N$ and the product $(e,g)(e,h)=(\omega(g,h),gh)$ generates the group $Q$ after quotienting by $N$.  
If there exists a homomorphism $\phi:Q\rightarrow E$ such that $\pi \circ \phi = \mathbb{I}_Q$, we say that the group extension {\it splits} and it is associated with the semidirect product $E=N \rtimes_{\rho} Q$. If such a homomorphism $\phi$ exists, the cocycle $\omega$ is trivial, i.e. $H^2(Q,U(1))=1$.
Two extensions are said to be equivalent if there is a isomorphism $\sigma:E\rightarrow E'$ such that the following diagram commutes:

\beq \label{eq:extdia}
\begin{array}{lllllllll}
1 & \longrightarrow  &N &\stackrel{i}{\longrightarrow} & E& \stackrel{\pi}{\longrightarrow}
& Q& \longrightarrow  & 1 \\
&  & \downarrow \,{\rm Id} &  & \downarrow \,\sigma &  & \downarrow \,{\rm Id} &  &  \\
1 & \longrightarrow  & N & \stackrel{i'}{\longrightarrow} &E' & \stackrel{\pi'}{\longrightarrow}
& Q & \longrightarrow  & 1
\end{array}
\eeq

The conditions are $\sigma(i(n))=i'(n) \; \forall n\in N$ and $\pi'(\sigma(g))=\pi(g) \; \forall g\in Q$. $E$ and $E'$ are isomorphic if the diagram (\ref{eq:extdia}) commutes, but the converse need not be true. An important result is that if two extensions are equivalent then the action $Q\rightarrow {\rm Aut}(N)$ is the same for both extensions, and the cocycles describing the two extensions are in the same class of cocycles in $H^2(Q,N)$.\\

To deal with the non-abelian case, two maps $\omega$ and $\rho$ are again constructed. But the map $\rho:Q\rightarrow {\rm Aut}(N)$ need not be a group homomorphism now. In fact it satisfies
$$\rho_g \rho_h={\rm Inn}(\omega(g,h)) \rho_{gh},$$
where Inn$(n)$ denotes the inner automorphism $m\mapsto n m n^{-1}: n,m \in N$. $\omega(g,h)$ is defined as in the abelian case. The map $\omega$ also satisfies the generalized cocycle condition $\omega(g,h)\omega(gh,k)=\rho_g(\omega(h,k))\omega(g,hk)$. However the maps seen as $\rho:Q\rightarrow {\rm Out}(N)$ is a group homomorphism where the outer automorphism group is defined as ${\rm Out}(N)={\rm Aut}(N)/{\rm Inn}(N)$. The group $E$ can also be written as Eq.(\ref{eq:carprod}), the maps $n\mapsto (n,1)$ and $g\mapsto (1,g)$ are again group homomorphisms and the extension equivalence is again defined as the commutation of the diagram (\ref{eq:extdia}). The extensions are now classified by $H^2(Q,Z(N))$ where $Z(N)$ is the center of $N$. This is because the inner automorphisms are invariant under the multiplication by an element of $Z(N)$. 

In the main text, we have called $G_{\rm topo}\equiv N$ and $G_{\rm sym}\equiv Q$, the extension group $E$ as $G$ and we take as normal subgroup of $G$ $\{ (g,e) |\; g\in G_{\rm topo} \} \cong G_{\rm topo}$ whose choice, in general, is not unique. So the reduction of the tensor is achieved by associating the elements of the group $G_{\rm sym}$ to the identity. We can take $L_{(g,e)}^G= L_g^{G_{\rm topo}}\otimes\mathbb{I}_{G_{\rm sym}}$ for any extension group that we choose. This is because acting on the group algebra basis ($\mathbb{C}[G]=\mathbb{C}[G_{\rm topo}])\otimes \mathbb{C}[G_{\rm sym}]$: 
\begin{equation}
\begin{split} 
L_{(k,e)}^G|n,h\rangle&=|k\rho_{e}(n) \omega(e,h),h\rangle= \\
 &=|kn,h\rangle= (L_k^{G_{\rm topo}}\otimes L_e^{G_{\rm sym}})|k,h\rangle,\notag
\end{split}
\end{equation}
where we have used that $\omega(e,h)=e$ and $\rho_{e}=\mathbb{I}$ because this choice does not change the class of the extension.

\section{Dyonic quasiparticles}\label{ap:dyonic}

This section is devoted to study dyonic quasiparticles of the quantum double model, $D(G)$, \cite{Preskillnotes} in the \textsf{PEPS} formalism \cite{PEPSdescp}. These quasiparticles complete the characterization of the topological low-energy sector of these models, together with pure fluxes and pure charges that have been analyzed in the main text. We first show the representation of the virtual operator corresponding to a dyon and we describe its principal topological properties. Then (Appendix \ref{ap:dyonic3}), we show how that operator is created physically according to the original model \cite{Kitaev} using a certain \textsf{PEPS} representation of the ground state. We finally (Appendix \ref{ap:dyonic2}) address the question of how the parent dyons behave when they are placed in a background of restricted tensors.\\

A dyonic excitation is a quasiparticle containing a flux part and a compatible charge part in $D(G)$ . For abelian groups, this quasiparticle is just a combined object of a pure charge and a pure flux (see its $G$-isometric \textsf{PEPS} representation in \cite{newNorbert}). For general finite groups, a dyon is associated to an irrep of the normalizer of a given conjugacy class of $G$. Given $h\in G$ we denote the normalizer subgroup of this element as $N_h=\{ n\in G | nh=hn\}$. The normalizer of another element in the conjugacy class of $h$, $h^g\equiv ghg^{-1}\in C[h]$, is $N_{h^g}=gN_hg^{-1}$. So the normalizers of the elements of a conjugacy class are all isomorphic, and the expression $N_{C[h]}$ is meaningful. We can decompose the group $G$ in left cosets of $N_h$ with representatives $k_1=e,k_2,\cdots,k_\kappa$ where $\kappa=|G|/|N_h|$. A relation between these cosets and elements of the conjugacy class can be given by $h_j=k_jh k^{-1}_j$. \\

A dyonic quasiparticle, associated to an irrep $\alpha$ of $N_{C[h]}$, can be expressed as an operator acting on the virtual d.o.f of the \textsf{PEPS} representation of the $D(G)$. We point out that these quasiparticles are always created in pairs (see Appendix \ref{ap:dyonic3} for details), as a dyon-antidyon composite, but here we have focused on one particle of the pair. A dyon corresponds to a string of right regular representations $R_h$ (corresponding to the flux part) placing the following operator (corresponding to the compatible charge part) at the extremity of the string:
\begin{equation}
\label{dyonend}
D^m_\alpha \equiv \sum_{n \in N_{h}} \chi^{N_{C[h]}}_\alpha(m n) \sum^\kappa_{j=1} |k_jn\rangle\langle k_jn|,
\end{equation}
where we have chosen the element $h$ as the representative of the conjugacy class, $\chi_\alpha$ is the character of the irrep $\alpha$ of $N_{C[h]}$ and $m\in N_h$ corresponds to the internal state of the charge associated to the quasiparticle (see Fig. \ref{Dyonpic}).\\

\begin{figure}[ht!]
\begin{center}
\includegraphics[scale=0.33]{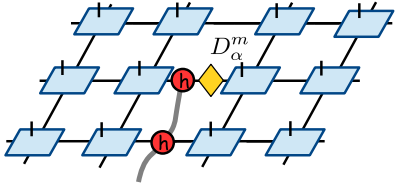}
\caption{The virtual representation of the dyonic quasiparticle is depicted. The red dots represent the operator $R_h$ acting on the virtual d.o.f. of the tensor network. The yellow rhombus corresponds to the operator $D^m_\alpha$ of (\ref{dyonend}).}
\label{Dyonpic}
\end{center}
\end{figure}
 
Let us now show some of the properties of this quasiparticle representation:
\begin{itemize}

\item Self-braiding: the effect of a half exchange of the dyon and its antiparticle or equivalently a $2\pi$ rotation of one dyon. This operation corresponds to braiding the string of the quasiparticle with (\ref{dyonend}) at its end in the following way: 

\begin{figure}[ht!]
\begin{center}
\includegraphics[scale=0.3]{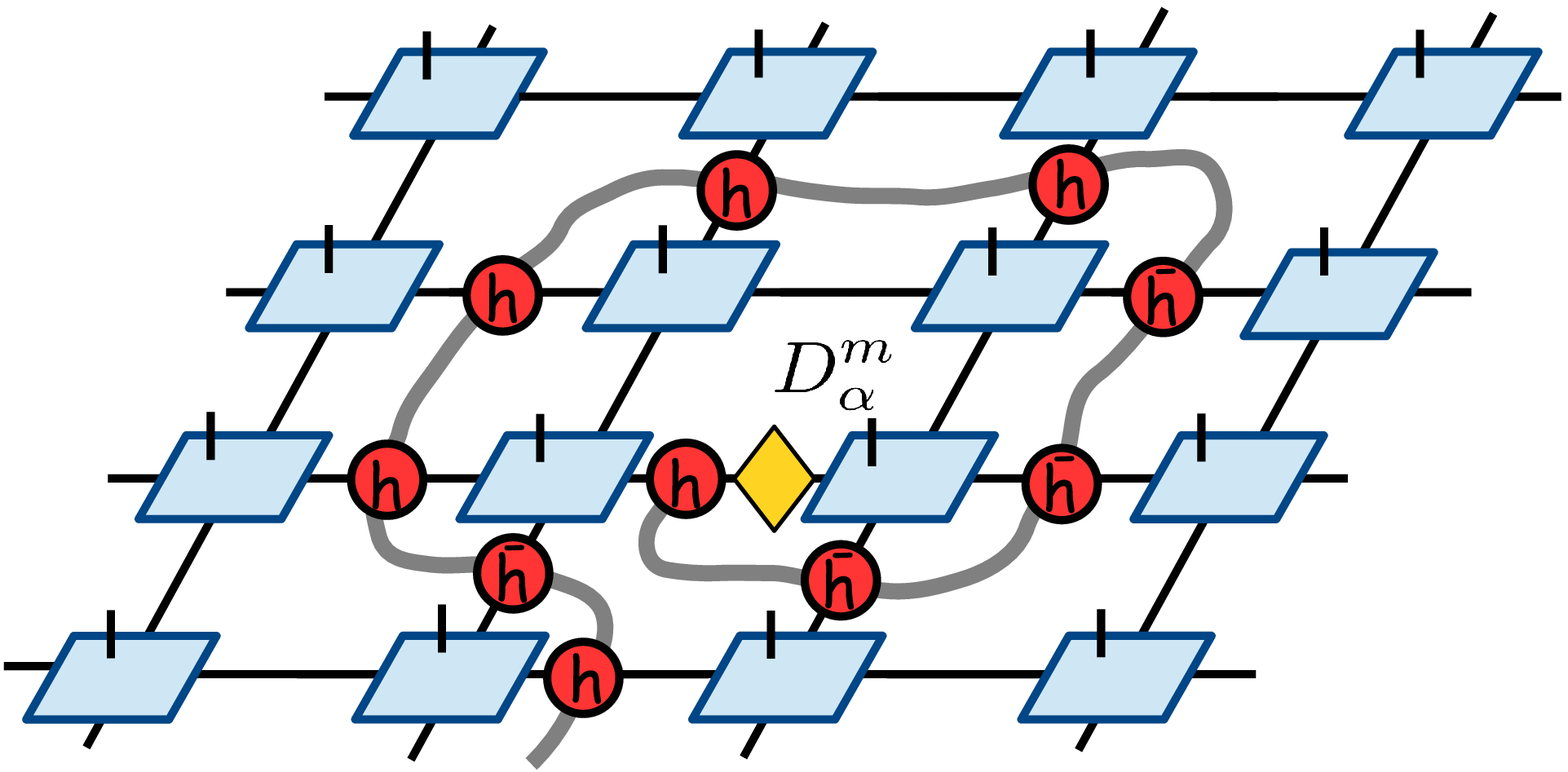}
\end{center}
\end{figure}

which is equivalent to \\

\begin{figure}[ht!]
\begin{center}
\includegraphics[scale=0.3]{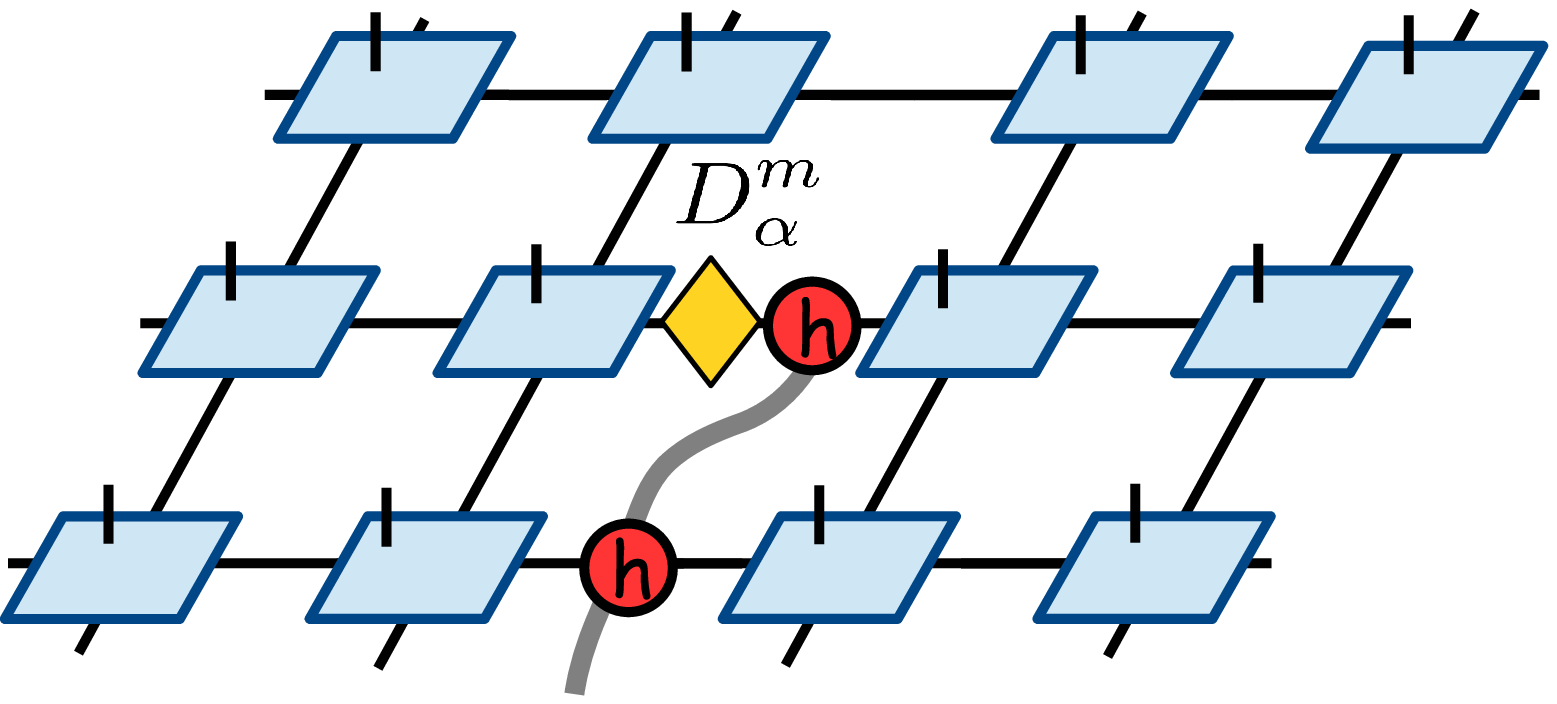}
\end{center}
\end{figure}

In order to complete the whole $2\pi$ spin we express $ D^m_\alpha R_h$ as $R_h (R^{\dagger}_h D^m_\alpha R_h )$; the effect of this operation is the conjugation by $R_h$ in the charge part of the dyon. Since $h$ is central in $N_h$ the matrix representation of $h$ is a multiple of the identity so  $R^{\dagger}_h D^m_\alpha R_h= \chi^{N_{C[h]}}_\alpha(h)D^m_\alpha$. The corresponding topological spin is $\chi^{N_{C[h]}}_\alpha(h)/d_\alpha$, where $d_\alpha$ is the dimension of the irrep $\alpha$.

\item Braiding with $g\in N_h$: this operation corresponds to the conjugation by $R_g$ over the string and over (\ref{dyonend}). The flux part remains invariant because $g^{-1}hg=h$ and the charge part transforms as $R^{\dagger}_g D^m_\alpha R_g= D^{g^{-1}m}_\alpha$. This change of the internal state of the charge part can be measured using interferometric experiments \cite{PEPSdescp}.

\item Braiding with $g\notin N_h$: the string gets conjugated $h\to h^{\bar{g}} $ and the charge part gets also conjugated $R^{\dagger}_g D^m_\alpha R_g$. We write $\chi_\alpha$ instead of $\chi^{N_{C[h]}}_\alpha$ from now on so the conjugation action is given by:
$$ \sum_{n \in N_h} \chi_\alpha(m n) \sum^\kappa_{j=1} |k_jng\rangle\langle k_jng|.$$
In order to operate with this expression we rewrite $k_jng=   \tilde{k}_{x_j}    \; (  \tilde{k}^{-1}_{x_j} \; g \;  g^{-1}k_jg)\;  g^{-1}ng$, where we have just inserted identities and the element $\tilde{k}_{x_j}$. To define this element let us denote the representatives of the left cosets of $G/N_{h^{\bar{g}}}$ as $\tilde{k}_j=g^{-1} k_jg$ with the relation $\tilde{h}_j = \tilde{k}_j \;h^{\bar{g}}\; \tilde{k}^{-1}_j $. We now denote with the index $x_j\in [1,\cdots,\kappa]$ the element corresponding to $\tilde{h}_{x_j}= \tilde{k}_{x_j}  h^{\bar{g}}\tilde{k}^{-1}_{x_j}=g \tilde{h}_j g^{-1} $. We can prove that $\tilde{n}_{x_j} \equiv \tilde{k}^{-1}_{x_j} g\tilde{k}_j   $ belongs to $N_{h^{\bar{g}}}$ and $R^{\dagger}_g D^m_\alpha R_g$ equals

\begin{align}\label{eq:braidyon}
& \sum_{n \in N_h} \chi_\alpha(mn) \sum^\kappa_{j=1} \ket{\tilde{k}_{x_j}  \tilde{n}_{x_j}  g^{-1}ng} \bra{\tilde{k}_{x_j}  \tilde{n}_{x_j}  g^{-1}ng}  \notag\\
=& \sum_{\tilde{n} \in N_{h^{\bar{g}}} } \chi_\alpha(mg\tilde{n}g^{-1}) \sum^\kappa_{j=1}
\ket{\tilde{k}_{x_j}  \tilde{n}_{x_j}  \tilde{n}} \bra{\tilde{k}_{x_j}  \tilde{n}_{x_j}  \tilde{n}}  \notag \\
=& \sum^\kappa_{j=1} \sum_{\tilde{n} \in N_{h^{\bar{g}}} } \chi_\alpha(mg \tilde{n}^{-1}_{x_j}  \tilde{n} g^{-1}) \ket{ \tilde{k}_{x_j} \tilde{n}} \bra{ \tilde{k}_{x_j} \tilde{n}} \notag\\
=& \sum^\kappa_{j=1} \sum_{\tilde{n} \in N_{h^{\bar{g}}} } \chi_\alpha( g^{-1}mg \tilde{n}^{-1}_{x_j} \tilde{n} ) \ket{ \tilde{k}_{x_j} \tilde{n}} \bra{ \tilde{k}_{x_j} \tilde{n}} .
\end{align}

This action is analogous to the symmetry transformations of the quantum double algebras \cite{Dijkgraaf,Mark}.

\end{itemize}

\subsection{Creation of dyons}\label{ap:dyonic3}

A composite dyon-antidyon excitation is created by acting with certain combination of operators, the so called ribbon operators \cite{Kitaev,Bombin}, over the ground state of $D(G)$. This ground state can be constructed using $G$-isometric {\sf PEPS} tensors \cite{PEPSdescp}. Here we have obtained the virtual representation of the ribbon operator corresponding to the composite dyon-antidyon excitation. In order to do so we apply that operator over the physical indices of a tensor network. Analyzing the virtual indices of the boundary we obtain the desired equivalence between physical and virtual operator.

The ribbon operator we chose acts in four edges of the square lattice with the orientation illustrated in Fig. \ref{orientation}.
\begin{figure}[ht!]
\begin{center}
\includegraphics[scale=1]{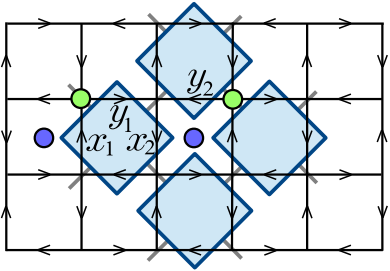}
\caption{ The operator of Eq.(\ref{eq:ribbop}) acts on four adjacent edges, denoted as $x_1,y_1,x_2,y_2$, and involves three vertices and three plaquettes. This operator depends on the orientation of each edge and we take the one represented by the arrows. The green and blue points identified the vertices and plaquettes excited respectively.}
\label{orientation}
\end{center}
\end{figure}

The operator can be written as
\begin{equation}\label{eq:ribbop}
\mathcal{O}_{\alpha}\equiv \frac{d_{\alpha}}{|N_h|} \sum_{n\in N_h} \bar{\chi}_\alpha(n)\sum_{i,j=1}^{\kappa}  F^{\bar{h}_i,k_in\bar{k}_j},
\end{equation}
where the $k_i$'s are the representatives of the left cosets of $G$ by $N_h$ and the operator $ F^{h,g}$ acts over the four chosen edges as follows (see Fig. \ref{orientation} for clarification):
$$F^{h,g}\ket{x_1,y_1,x_2,y_2}= \delta_{g,x_1\bar{x}_2}\ket{x_1\bar{h},y_1,\bar{y}_1h y_1x_2,y_2}. $$

The ground state of the quantum double of $G$ can be constructed with the following tensor \cite{PEPSdescp}:
$$K= \sum_{l,r,s,p\in G}\ket{p\bar{l},l\bar{r},r\bar{s},s\bar{p}} |p,l)(r,s|.$$
This tensor has a the following virtual invariance:
\begin{align} \label{eq.Rregular}
&K\big[R_g\otimes R_g\otimes R^{\dagger}_g\otimes R^{\dagger}_g\big]_v =  \\
&\sum_{l,r,s,p\in G}\ket{p\bar{l},l\bar{r},r\bar{s},s\bar{p}} \big[R_g\otimes R_g |p,l)(r,s| R^{\dagger}_g \otimes R^{\dagger}_g\big]=K, \notag
\end{align}
$\forall g \in G$ which endows the state with topological properties.
We now express the edges involved in the action of the operator of Eq.(\ref{eq:ribbop}) in its tensor network representation:
\begin{equation}\label{eq:physten}
 \mathcal{P}(K)\equiv \sum_{l,r,s,p,t\in G}\ket{p\bar{l},l\bar{r},r\bar{s},s\bar{p},t\bar{r}} |p,l)(s,t|,
\end{equation}
for a driagrammatic representation see Fig. \ref{fig:indices}.
\begin{figure}[ht!]
\begin{center}
\includegraphics[scale=0.8]{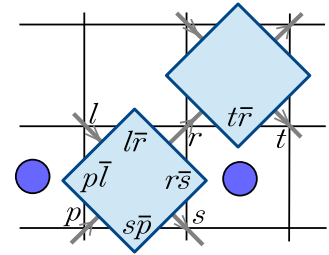}
\caption{Tensor network representation of the physical system involved in the creation of a dyon. The blue dots are depicted for comparison with Fig. \ref{orientation} and the virtual index $r$ is depicted for clarification.}
\label{fig:indices}
\end{center}
\end{figure}
The creation operator of the dyon acts on the tensor network representation as follows:
\begin{align}
&\frac{|N_h|}{ d_{\alpha}}\mathcal{O}_{\alpha}  \mathcal{P}(K)=\notag \\
 \sum_{\substack{ n\in N_h   \\  i,j=1,\cdots,\kappa  \\  l,r,s,p,t\in G }} &\delta_{k_in\bar{k}_j,l\bar{t}} \;\bar{\chi}_\alpha(n) \; \ket{p\bar{l}h_i,l\bar{r},r\bar{l}\; \bar{h}_il\bar{s},s\bar{p},t\bar{r}}  |p,l)(s,t|.\notag
 \end{align}
We now can relabel the indices ($s'=s\bar{l} h_i l$ and $p'=p\bar{l} h_i l$) to obtain the action on the virtual d.o.f.:
\begin{align}\label{eq:creadyon}
\sum_{\substack{  i,j=1,\cdots,\kappa  \\  l,r,s,p,t\in G }} &\bar{\chi}_\alpha(\bar{k}_i l \bar{t} k_j)  \ket{p\bar{l},l\bar{r},r\bar{s},s\bar{p},t\bar{r}}  |p\bar{l}\;\bar{h}_il,l)(s\bar{l}\;\bar{h}_il,t| \notag \\
&=\sum_i\mathcal{F}_i \circ\mathcal{C}_i[ \mathcal{P}(K)],
\end{align}
 where the operator $\sum_i\mathcal{F}_i \circ\mathcal{C}_i$ acts purely on the virtual d.o.f of $\mathcal{P}(K)$ and its components are defined as follows:
\begin{align}\label{eq:opedyon}
&\mathcal{F}_i\big[| p,l)(s,t| \big]\equiv \sum_{g\in G} R^{\dagger}_{\bar{g}\;\bar{h}_i g}\otimes |g)(g| \big[| p,l)(s,t|\big]R_{\bar{g}\;\bar{h}_i g}\otimes \mathbb{I},\notag \\
&\mathcal{C}_i \big[| p,l)(s,t| \big]\equiv \notag \\
&\sum_{n,m\in N_h} \bar{\chi}_\alpha(n\bar{m})\; \mathbb{I}\otimes|k_in)(k_i n|\big[| p,l)(s,t|\big] \mathbb{I}\otimes \sum_j^\kappa |k_jm)(k_j m|, \notag
\end{align}
where $\mathcal{F}_i$ and $\mathcal{C}_i$ can be regarded as the flux and charge part of the dyon respectively.
We can represent diagramatically the virtual operator corresponding to Eq.(\ref{eq:creadyon}) as follows:
$$\sum_{\substack{  i=1,\cdots,\kappa  \\ g\in G \\n,m\in N_h }} \bar{\chi}_\alpha(n\bar{m})\;
\parbox[c]{0.3\textwidth}{ \includegraphics[scale=0.9]{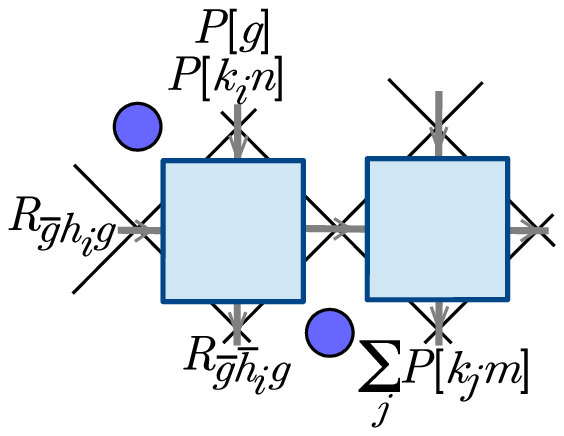}},$$
where $P[a]=\ket{a}\bra{a}$. If we only consider the one particle of the composite dyon-antidyon, the virtual representation we are dealing with is equivalent to a string of $R_g$ operators ended with the operator (\ref{dyonend}).

We point out that the tensor $K$ has the invariance described in Eq.(\ref{eq.Rregular}) because of the clockwise direction, of the edges contained in $K$ (see Fig. \ref{orientation}), chosen \cite{PEPSdescp}. A counterclockwise direction would give rise to a tensor with the virtual invariance represented by $L_g$ instead of $R_g$, which would be unitary equivalent to the $G$-isometric \textsf{PEPS} of Eq.(\ref{eq:gisopeps}). This relation connects the tensor $K$, obtained in \cite{PEPSdescp}, used in this section with the convention used through the main text.

\subsection{Symmetry reduction with dyons}\label{ap:dyonic2}
We now turn to the situation analyzed in this work where a parent excitation is placed on a background of restricted tensors. We also study the effect of the symmetry operators over the dyons of the restricted phase. 

If we place a parent dyon with the flux part corresponding to a conjugacy class not in $G_{\rm{topo}}$, the chain cannot be moved freely and the the dyon is confined. Let us analyze the more involved case where the parent dyon is unconfined and its flux and charge parts split. Let $C^G[h]=\{h_i, i=1,\cdots, \kappa \}$ be a conjugacy class of $G$ which is in $G_{\rm topo}$. This conjugacy class can be decomposed in conjugacy classes of $G_{\rm topo}$: $C^G[h]= \cup_j C^{G_{\rm topo}}[h_j]$ where $j$ only runs over in the indices corresponding to the elements $h_i$ with disjoint conjugacy classes of $G_{\rm topo}$. It is clear that any two of these conjugacy classes of $G_{\rm topo}$ can be related by conjugation with an element of $G$ and vice versa. This fact is what is causing the splitting of the flux part of an unconfined parent dyon and the action of the symmetry over the flux part of a dyon of the restricted model. 

Take now a representative element $h_j$ of $C^G[h]$ and denote its normalizer in $G$ as $N^G_{h_j}=\{n\in G | nh_j=h_jn\}$. Trivially $N^{G_{\rm topo}}_{h_j}=\{k\in G_{\rm topo} | kh_j=h_jk\}$ is a subgroup of $ N^G_{h_j}$. It is also normal: $(nkn^{-1})h_j=h_j(nkn^{-1})$ $\forall n\in N^G_{h_j}$ and $\forall k\in N^{G_{\rm topo}}_{h_j}$. Therefore $N^{G_{\rm topo}}_{gh_jg^{-1}}$ is normal in $N^G_{gh_jg^{-1}}$. By Clifford's Theorem \cite{Clifford} the irreps of $N_{C^G[h]}$ will decompose into an equal weight superposition of irreps of $N_{C^{G_{\rm topo}}[h]}$, all of them related by conjugation. This describes the splitting of the charge part of an unconfined parent dyon and also describes qualitatively the action of the symmetry over the charge part of a dyon of the restricted model.

In order to compute quantitatively the effect of the symmetry operator over the charge part of a restricted dyon let us denote $g=(q,[z])$ with $q\in G_{\rm topo}$ and $[z] \in G_{\rm sym}$. As a set, $G$ can be viewed as the cartesian product of $G_{\rm topo}$ and $G_{\rm sym}$ (see Apendix \ref{ap:ext}). We will write down the case with trivial cocycle for the sake of simplicity. Now $N_h$ will denote the normalizer of $h$ in $G_{\rm topo}$, $\chi_\beta$ the character of the irrep $\beta$ of the subgroup $N_{C[h]}$ of $G_{\rm topo}$ and now $k_j$ will denote the representatives of the right cosets of $G_{\rm topo}$ by $N_h$ ( $k_1=e,k_2,\cdots,k_\xi$ where $\xi=|G_{\rm topo}|/|N_h|$). With this notation the charge part of the restricted dyon, corresponding to the conjugacy class $C[h]$ of $G_{\rm topo}$ and the irrep $\beta$ of $N_{C[h]}$, can be associated to the following operator (at the end plaquette of a string of $L_h$ corresponding to the flux part):  
\beq \label{eq:restdyon}
D^m_\beta\equiv \sum_{n \in N_{h}} \chi_\beta(m n)\sum_{[y]\in G_{\rm sym}} \sum^\xi_{j=1} \ket{n k_j ,[y]}\bra{n k_j ,[y]},
\eeq
where $m$ belongs to $N_h$. The action of the symmetry over this operator is the conjugation by $L_g$ with $g\in G$. 
Applying $L_g$ to the basis elements in Eq.(\ref{eq:restdyon}) we end up with $L_g|nk_j,[y]\rangle=|q\rho(nk_j),[z][y]\rangle$, where $\rho_g(n)=gng^{-1}$  which goes from $N_{C[h]}$ to $N_{C[h^g]}$ with $C[h]\neq C[h^g]$ if $g\in G-G_{\rm topo}$. If $g\in G_{\rm topo}$ the action will be equivalent to the corresponding analysis in Eq.(\ref{eq:braidyon}). Now we rewrite $q\rho_g(nk_j)$ as $q\rho_g(n)q^{-1}\; \big[q\rho_g(k_j)q^{-1}\; q \; \tilde{k}^{-1}_{x^g_j}\big]\; \tilde{k}_{x^g_j}$. To define $\tilde{k}_{x^g_j}$ let us denote the representatives of the cosets of $G_{\rm topo}/N_{h^{gq}}$ as $\tilde{k_j}=qg k_jg^{-1}q^{-1}$ with the relation $\tilde{h}_j = \tilde{k}_j \;h^{gq}\; \tilde{k}^{-1}_j $.  We now denote with the index $x^g_j\in [1,\cdots,\xi]$ the element corresponding to $\tilde{h}_{x^g_j}= \tilde{k}_{x^g_j}  h^{gq} \tilde{k}^{-1}_{x^g_j}=q \tilde{h}_j q^{-1} $. We can prove that $\tilde{n}_{x^g_j} \equiv \tilde{k}_j q  \tilde{k}^{-1}_{x^g_j}$ belongs to $N_{h^{qg}}$. Denoting $\tilde{n}\equiv \rho_{qg}(n)$ the conjugation by the inverse of $p$ as $\rho_{\bar{p}}(\cdot)$ and $P[a]=\ket{a}\bra{a}$, $L_g D^m_\beta L^\dagger_g$ equals
\begin{align}
&\sum_{\tilde{n} \in N_{h^{qg}}} \chi_\beta(m \rho_{\bar{qg}}(\tilde{n})) \sum_{[y]\in G_{\rm sym}} \sum^\xi_{j=1} P[\tilde{n}\tilde{n}_{x^g_j} k_{x^g_j},[zy]]  \notag \\
=&\sum^\xi_{j=1} \sum_{\tilde{n} \in N_{h^{qg}}} \chi_\beta(m \rho_{\bar{qg}}(\tilde{n} \tilde{n}^{-1}_{x^g_j})) \sum_{[y]\in G_{\rm sym}}   P[\tilde{n} k_{x^g_j},[y]] \notag \\
=&\sum^\xi_{j=1} \sum_{\tilde{n} \in N_{h^{gq}}} \chi_\beta(\rho_{\bar{qg}}[ \rho_{qg}(m) \tilde{n} \tilde{n}^{-1}_{x^g_j}])\sum_{[y]\in G_{\rm sym}}  P[ \tilde{n} k_{x^g_j},[y]]\notag \\
=&\sum^\xi_{j=1} \sum_{\tilde{n} \in N_{h^{gq}}} \chi_{\beta'}( \rho_{qg}(m) \tilde{n} \tilde{n}^{-1}_{x^g_j})\sum_{[y]\in G_{\rm sym}}   P[ \tilde{n} k_{x^g_j},[y]], \notag
\end{align}
where we have denoted as $\beta'$ the irrep of the normalizer of $h^{qg}$, with $h^{qg} \notin C^{G_{\rm topo}}[h]$, which is obtained by conjugation of the irrep $\beta$. This action corresponds to a permutation of the charge part of the dyon compatible with the corresponding permutation in the flux part.

\end{document}